\documentclass[12pt]{article}
\makeatletter \oddsidemargin  -.1in \evensidemargin -.1in
\newcommand{\singlespacing}{\let\CS=\@currsize\renewcommand{\baselinestretch}{1}\tiny\CS}
\newcommand{\oneandahalfspacing}{\let\CS=\@currsize\renewcommand{\baselinestretch}{1.25}\tiny\CS}
\newcommand{\doublespacing}{\let\CS=\@currsize\renewcommand{\baselinestretch}{1.35}\tiny\CS}

\newtheorem{rule-def}[theorem]{Rule}

\RequirePackage[dvips]{graphicx} \textheight 22.5cm
\usepackage{multirow}
\usepackage{amsmath}
\usepackage{amssymb}
\usepackage{latexsym}
\begin{document}

\title {\bf Peristaltic Pumping of Blood Through Small Vessels of Varying Cross-section }
\author{\small J.C.Misra$^1$\thanks{Email
address: {\it misrajc@rediffmail.com (J. C. Misra)}},
~~~S. Maiti$^2$\thanks{Email address: {\it
somnathm@cts.iitkgp.ernet.in (S. Maiti)}}~  \\
\it$^1$Professor, Department of Mathematics,\\ Institute of Technical
Education and Research,\\ Siksha O Anusandhan University, Bhubaneswar-751030, India\\
$^2$\it School of Medical Science and Technology $\&$ Center for
Theoretical Studies, \\Indian Institute of Technology, Kharagpur-721302, India \\}
\date{}
\maketitle \noindent \doublespacing

\begin{abstract}
 The paper is devoted to a study of the peristaltic motion of blood in
 the micro-circulatory system. The vessel is considered to be of
 varying cross-section. The progressive peristaltic waves are taken to
 be of sinusoidal nature. Blood is considered to be a Herschel-Bulkley
 fluid. Of particular concern here is to investigate the effects of
 amplitude ratio, mean pressure gradient, yield stress and the power
 law index on the velocity distribution, streamline pattern and wall
 shear stress. On the basis of the derived analytical expression,
 extensive numerical calculations have been made. The study reveals
 that velocity of blood and wall shear stress are appreciably affected
 due to the non-uniform geometry of blood vessels. They are also
 highly sensitive to the magnitude of the amplitude ratio and the
 value of the fluid index. \\ \it Keywords: {\small Non-Newtonian
   Fluid, Retrograde Flow, Wall Shear Stress, Trajectory.}
\end{abstract}

\section{Introduction}
Peristaltic transport is well known to physiologists as a natural
mechanism of pumping materials in the case of most physiological
fluids. Apart from physiological fluids, some other fluids also
exhibit peristaltic behaviour. Peristalsis usually occurs when the
flow is induced by a progressive wave of area contraction/expansion
along the length of the boundary of a fluid-filled distensible
tube. Usefulness of studies on peristaltic flow has been discussed in
detail in our earlier publications (Misra et
al. \cite{Misra1,Misra2,Misra3,Misra4,Misra5,Misra6,Misra7,Misra8,Misra9},
Maiti and Misra \cite{Maiti1,Maiti2}) and in some other references
\cite{Guyton,Jaffrin}.
\begin{center}
\begin{tabular}{|l l|}\hline
{~\bf Nomenclature} &~ \\
~~$a_0 $ & Half-width of the channel at the inlet\\
~~$b$ & Wave amplitude\\
~~$H $ & Vertical displacement of the wall \\
~~$n$ & Flow index number\\
~~$n_1 $ & Reciprocal of n \\
~~$P $ & Fluid pressure\\
~~$\bar{Q} $ & Flux at axial location\\
~~~~$t $  & Time\\
~~$X,Y $ & Rectangular Cartesian co-ordinates\\
~~$U,V$ & Velocity components in X,Y directions respectively\\
~~$\delta $ & Wave number\\
~~$\Delta p$ & Pressure difference between the channel ends\\
~~$\lambda $ & Wave length of the travelling wave motion of the wall\\
~~$\Lambda $ & A parametric constant\\
~~$\mu $ & Blood viscosity\\
~~$\nu $ & Kinematic viscosity of blood\\
~~$\phi$ & Amplitude ratio \\
~~$\rho$ & Density of blood \\
~~$\tau_0$ & Yield stress of blood\\
~~$\tau_h$ & Wall shear stress\\
\hline
\end{tabular}
\end{center}
The phenomenon of peristalsis plays an important role in the
functioning of heart-lung machine, blood pump machine and dialysis
machine. Fung and Yih \cite{Fung1} presented a theoretical analysis of
peristaltic transport primarily with inertia-free Newtonian flows
driven by sinusoidal transverse waves of small
amplitude. Investigation of peristaltic motion in connection with
functions of different physiological systems such as the ureter, the
gastro-intestinal tract, the small blood vessels and other glandular
ducts was first made by Shapiro et al. \cite{Shapiro}. They presented
a closed form solution for an infinite train of waves for small
Reynolds number flow. Their study was, however, restricted to cases
where the wave length is long and the wave amplitude is
arbitrary. Conditions for the the presence of physiologically
significant phenomena of trapping and reflux were also suggested by
them. Jaffrin and Shapiro \cite{Jaffrin} as well as Srivastava and
Srivastava \cite{Srivastava1} summarized the early literatures on
peristaltic transport. Some of the recent studies on peristaltic
transport were discussed by Usha and Rao \cite{Usha1}, Mishra and Rao
\cite{Mishra}, Yaniv et al. \cite{Yaniv}, Jimenez-Lozano et al
\cite{Jimenez-Lozano1}, Wang et al. \cite{Wang} and Hayat et
al. \cite{Hayat}.

Some important theoretical analyses on different aspects of blood flow
were carried out in a systematic manner by Misra et
al. \cite{Misra10,Misra11,Misra12}. Attempt to consider the complex
rheology of various physiological fluids was made in several studies
\cite{Usha2,Takabatake,Pozrikidis,Jimenez-Lozano2,Bhargava,Bohme,Srivastava2,Provost,Chakraborty}. The
non-Newtonian behaviour of blood mainly owes to the presence of
erythrocytes in whole blood. In the case of blood, such behaviour
starts becoming prominent when the hematocrit rises above 20$\%$. This
particular behaviour plays a dominating role when the hematocrit level
lies between 40$\%$ and 70$\%$ \cite{Rand,Bugliarello,Chien}. Some
other relevant theoretical studies on non-Newtonian fluid flows were
carried out by Masud and Kwack \cite{Masud}, Kwack and Masud
\cite{Kwack} as well as by Anand and Rajagopal \cite{Anand}.

It is known that the flow behaviour of blood in small vessels
(diameter$<$0.02 cm) and at low shear rate ($<20 sec^{-1}$) can be
represented by a power law fluid \cite{Charm1,Charm2}. Merill et
al. \cite{Merrill} pointed out that Casson model holds satisfactory
for blood flowing in tubes of 130-1000$\mu m$. Moreover, Blair and
Spanner \cite{Blair} reported that blood obeys Casson model for
moderate shear rate flows. However, they pointed out that for cow's
blood, Herschel-Bulkley model is more appropriate than Casson
model. 

It is known that physiological organs are by and large non-uniform
ducts \cite{Wiedman,Wiederhielm,Lee}. Several authors
\cite{Srivastava1,Gupta,Srivastava3} made some initial attempts to
perform theoretical studies pertaining to peristaltic transport of
physiological fluids in vessels of non-uniform cross section. These
analyses were mostly restricted to the assumption of either a
Newtonian fluid or a non-Newtonian fluid of Casson/power-law
type. Moreover, in these reports the different flow characteristics
have not been adequately discussed. The strong merit of
Herschel-Bulkley model is that fluids represented by this model
describe very well material flows with a non-linear constitutive
relation depicting the behaviour of shear-thinning/shear-thickening
fluids that are of much importance in the field of biomedical
engineering \cite{Malek}. It is also to be noted that for formulating
a non-Newtonian model of blood, Herschel-Bulkley fluid model is more
general than most other non-Newtonian fluid models. It is worthwhile
to mention that results for a fluid represented by Bingham plastic
model, power law model and Newtonian fluid model can be derived from
those of the Herschel-Bulkley fluid model. Also this model has been
found to yield more accurate results than many other non-Newtonian
models.

In view of the above, we have taken up here a study on the peristaltic
transport of blood in a non-uniform channel, by treating blood as a
Herschel-Bulkley fluid. It is worthwhile to mention that flow through
axisymmetric tubes is qualitatively similar to the case of flow in
channels. On the basis of their theoretical/experimental study on
peristaltic pumping at low Reynolds number, Shapiro et
al. \cite{Shapiro} also confirmed that flow behaviour in the case of
an axisymmetric tube is identical to that in the case of channel
flow. The formulation and analysis presented in the sequel are
particularly suitable for investigating the peristaltic motion in
vessels of small dimensions, e.g. arterioles and venules. Since for
flow of blood through smaller vessels in the micro-circulatory system,
the Reynolds number is low and since the ratio between half-width of
the channel under consideration and the wave length is considered
small, it has been possible to perform the theoretical analysis for
the present problem in a convenient manner, by using the lubrication
theory \cite{Shapiro}. On the basis of the derived analytical expressions,
computational work has been executed keeping a specific situation of
micro-circulation in view, with the main purpose of examining the
distribution of velocity of blood and wall shear stress as well as the
streamline pattern, trajectories of individual fluid particles and
pumping performance. The plots for the computed results give us a
clear idea of qualitative variation of various fluid dynamical
parameters. The results indicate plug flow in the central region,
where most of the erythrocytes are accumulated and non-plug flow in
the peripheral region. The results of the present study are in good
agreement with those reported earlier by previous investigators. The
study bears the potential to explore some important phenomena that are
useful for having a better insight into the flow in the
micro-circulatory system.

The study has an important bearing on the clinical procedure of
extra-corporeal circulation of blood by using the heart-lung machine,
where there is a chance of damage of erythrocytes owing to significant
variation of the wall shear stress. The results are also likely to
find important application in roller pumps and arthro-pumps by which
fluids can be transported in living organs in pathological states.

\section{Formulation and Analysis}
Let us consider the peristaltic motion of blood, by treating it as an
incompressible viscous non-Newtonian fluid. The non-Newtonian
behaviour is considered to be of Herschel-Bulkley type. We shall study
two-dimensional channel flow, width of the channel being
non-uniform. 

\begin{figure}
\centering
\includegraphics[width=3.5in]{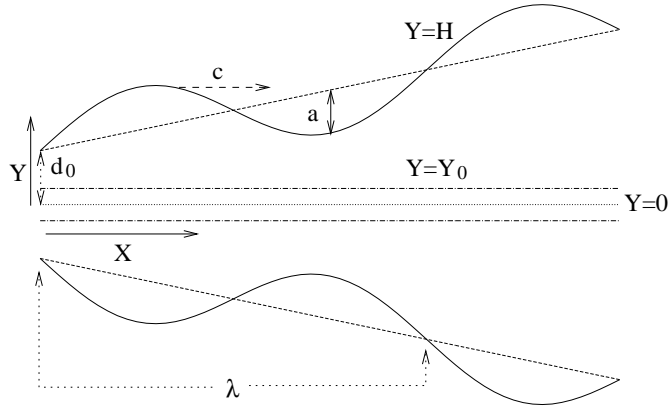}
\caption{A physical sketch of the problem for a tapered channel}
\label{jam_geo4.1}
\end{figure} 
We take (X,Y) as Cartesian coordinates of the location of
a fluid particle, X being measured in the direction of wave
propagation and Y in the normal direction. Let $ Y=H$ and $Y=-H$ be
respectively the upper and lower boundaries of the channel
(cf. Fig. \ref{jam_geo4.1}).  The medium is considered to be
induced by a progressive sinusoidal wave train propagating with a
constant speed $c$ along the channel wall, such that
$H=d(X)+a~\sin(\frac{2\pi}{\lambda}(X-ct)),$ with
d(X)=$d_0+\Lambda$X. $d(X)$ represents the half-width of the channel
at any axial distance X from the inlet, $d_0$ being the half-width at the
inlet, $\Lambda(<$1) a constant whose magnitude depends on the length
of the channel as well as the inlet and outlet dimensions, $a$ the wave
amplitude, $t$ the time and $\lambda$ the wave length.

Under the  assumptions stated above, the basic equations  that govern
the fluid motion are the field equations
\begin{eqnarray*}
\nabla \cdot{\bf V}=0
\end{eqnarray*}
\begin{eqnarray*}
\rm{ and}~~~\rho \frac{d{\bf V}}{dt}=\nabla\cdot{\bf \sigma}+ \rho {\bf f},
\end{eqnarray*}
where ${\bf V}$ is the velocity, {\bf f} the body force per unit mass
and $\rho$ the density of the fluid, while $\frac{d}{dt}$ denotes the
material time derivative. ${\bf \sigma}$ is the Cauchy stress defined
by ${\bf \sigma}=-PI+T$, where $T=2\mu E_{ij}+\eta I S$ and $S=\nabla \cdot
{\bf V}$, $E_{ij}$ being the symmetric part of the velocity gradient
$L$, $E_{ij}=\frac{1}{2}[L+L^T$], $L=\nabla{\bf V}$. \\$-PI$ denotes the
indeterminate part of the stress due to the constraint of
incompressibility; $\mu$ and $\eta$ are viscosity
parameters. The Herschel-Bulkley model gives the combined effect of
Bingham plastic and power-law behavior of a fluid. At low strain rate
( $\dot{\gamma}<\frac{\tau_0}{\mu_0}$ ), the material acts like a
viscous fluid with constant viscosity $\mu_0$. But when the strain
rate increases and the yield stress threshold, $\tau_0$, is reached,
the fluid behavior is described by a power law
\begin{eqnarray*}
\mu=\frac{\tau_0+\alpha\left\{\dot{\gamma}^n-\left(\frac{\tau_0}{\mu_0}\right)^n\right\}}{\dot{\gamma}},
\end{eqnarray*}
 where $\alpha$ and n denote respectively the consistency factors and
 the power law index. $n<1$ and $n>1$ correspond to a shear thinning
 fluid and a shear thickening fluid respectively. If the channel
 length is an integral multiple of the wavelength, the pressure
 difference across the ends of the channel is a constant. The pressure
 p remains constant across any axial station of the channel, when the
 wavelength is large and curvature effects are negligibly small. Since
 in the study, the geometry of wall surface is non-uniform, the flow
 is inherently unsteady in the laboratory frame as well as in the wave
 frame of reference. Disregarding the body forces (i.e. taking f=0),
 and using the Herschel-Bulkley equations, the governing equations of
 the incompressible fluid motion in the micro-vessel in the fixed
 frame of reference may be put in the form
\begin{equation}
\rho \left (\frac{\partial U}{\partial t}+U\frac{\partial U}{\partial
  X}+V\frac{\partial U}{\partial Y}\right )=-\frac{\partial
  P}{\partial X}+\frac{\partial \tau_{XX}}{\partial X}+\frac{\partial
  \tau_{XY}}{\partial Y}
\end{equation}
\begin{equation}
\rho\left (\frac{\partial V}{\partial t}+U\frac{\partial V}{\partial
  X}+V\frac{\partial V}{\partial Y}\right )=-\frac{\partial
  P}{\partial Y}+\frac{\partial \tau_{YX}}{\partial X}+\frac{\partial
  \tau_{YY}}{\partial Y}
\end{equation}
\begin{eqnarray}
with ~\tau_{ij}=2\mu E_{ij}=\mu \left(\frac{\partial U_i}{\partial
  X_j}+\frac{\partial U_j}{\partial X_i}\right)~,\\
\mu=\left\{\begin{array}{r@{\quad : \quad}l}\mu_0~ & for~\Pi\le\Pi_0,
\\ \alpha\Pi^{n-1}+\tau_0\Pi^{-1} & for~\Pi\ge\Pi_0
\end{array} \right.\\
\Pi=\sqrt{2E_{ij}E_{ij}}
\end{eqnarray}
The limiting viscosity $\mu_0$ is considered such that
\begin{equation}
\mu_0=\alpha\Pi_0^{n-1}+\tau_0\Pi_0^{-1}
\label{cnsns_manuscript_mu_0}
\end{equation}
In the analysis that follows, we shall make use of the following
non-dimensional variables:
\begin{eqnarray}
\bar{X}=\frac{X}{\lambda},~~\bar{Y}=\frac{Y}{d_0},~~\bar{U}=\frac{U}{c},~\bar{V}=\frac{V}{c\delta},~ \delta=\frac{d_0}{\lambda},~\bar{P}=\frac{d_0^{n+1}P}{\mu
c^n\lambda},~\bar{t}=\frac{ct}{\lambda},
~h=\frac{H}{d_0},~\phi=\frac{a}{d_0},\nonumber\\~\bar{\tau}_0=\frac{\tau_0}{\mu\left(\frac{c}{d_0}\right)^n},~\bar{\tau}_{YX}=\frac{\tau_{YX}}{\mu\left(\frac{c}{d_0}\right)^n},~\bar{\Psi}=\frac{\Psi}{d_0
c}~~~~~~~~~~~~~~~~~~~~~~~~~~~~~~~~~~~~~~~~~~~~~~~~~~~~~~~~~
\label{jam_non-dimensionalize}
\end{eqnarray}
The equation governing the  flow of the fluid  can now be written in the form (dropping the bars over the symbols)
\begin{equation}
 Re\delta \left (\frac{\partial U}{\partial t}+U\frac{\partial
   U}{\partial X}+V\frac{\partial U}{\partial Y}\right
 )=-\frac{\partial P}{\partial X}+2\delta^2\frac{\partial
   \left(\Phi\frac{\partial U}{\partial X}\right)}{\partial
   X}+\frac{\partial \left(\Phi\left(\frac{\partial U}{\partial
     Y}+\delta^2\frac{\partial V}{\partial X}\right)\right)}{\partial
   Y}
\end{equation}
\begin{equation}
Re\delta^3\left (\frac{\partial V}{\partial t}+U\frac{\partial
  V}{\partial X}+V\frac{\partial V}{\partial Y}\right
)=-\frac{\partial P}{\partial
  Y}+\delta^2\frac{\partial\left(\Phi\left(\frac{\partial U}{\partial
    Y}+\delta^2\frac{\partial V}{\partial X}\right)\right)}{\partial
  X}+\delta^2\frac{\partial \left(\Phi\frac{\partial V}{\partial
    Y}\right)}{\partial Y}
\end{equation}
\begin{eqnarray}
where~~~\Phi=\left|\sqrt{2\delta^2\left\{\left(\frac{\partial U}{\partial
    X}\right)^2+\left(\frac{\partial
    V}{\partial Y}\right)^2\right\}+\left(\frac{\partial U}{\partial
    Y}+\delta^2\frac{\partial V}{\partial
    X}\right)^2}\right|^{n-1}\nonumber\\+\tau_0\left|\sqrt{2\delta^2\left\{\left(\frac{\partial U}{\partial
    X}\right)^2+\left(\frac{\partial
    V}{\partial
    Y}\right)^2\right\}+\left(\frac{\partial U}{\partial
    Y}+\delta^2\frac{\partial V}{\partial Y}\right)^2}\right|^{-1}
\end{eqnarray}
Using the long wavelength approximation ($\delta\ll 1$) and the
lubrication approach \cite{Shapiro,Mishra}, the governing equations
and the boundary conditions describing the flow in the fixed frame of
reference may be rewritten in terms of the dimensionless variables as
\begin{equation}
\frac{\partial P}{\partial X}=\frac{\partial \tau_{YX}}{\partial Y}~,
\label{jam_xmomentum}
\end{equation}
\begin{equation}
\frac{\partial P}{\partial Y}=0
\end{equation}
\begin{eqnarray}
 where~~\tau_{YX}=\left(\tau_0+\left|\frac{\partial U}{\partial Y}\right|^n\right)sgn\left(\frac{\partial U}{\partial Y}\right)~~(cf.~ \cite{Huang}),
\end{eqnarray}
 \begin{eqnarray}
 \Psi=0,~U_Y=\Psi_{YY}=0,~ \tau_{YX}=0~ at~ Y=0;~~U=\Psi_Y=0~ at~ Y=h~~~~~~~~~~~~~~~~~~~~~~~~~~~~~~~~
\label{jam_boundary_condition}
\end{eqnarray}
Thus the pressure gradient $\frac{\partial P}{\partial X}$ is independent of Y. The solution of equation (\ref{jam_xmomentum}) satisfying (\ref{jam_boundary_condition}) is found in the form
\begin{equation}
U(X,Y,t)=\left\{\begin{array}{r@{\quad : \quad}l}\frac{1}{(n_1+1)P_1}\left[(P_1h-\tau_0)^{n_1+1}-(P_1Y-\tau_0)^{n_1+1}\right] & if~~ Y\ge 0 \\ \frac{1}{(n_1+1)P_1}\left[(P_1h-\tau_0)^{n_1+1}-(-P_1Y-\tau_0)^{n_1+1}\right] & if~~ Y<0, \\
\end{array} \right. 
\label{jam_axial_velocity}
\end{equation}
where~$P_1=-\frac{\partial P}{\partial X},~n_1=\frac{1}{n}$.
If the plug flow region be given by $Y=Y_0$,
\begin{eqnarray*}
U_Y=0~~\rm{at}~Y=Y_0
\end{eqnarray*}
\begin{eqnarray*}
\rm{Then}~~Y_0=\frac{\tau_0}{P_1}.
\end{eqnarray*}
h defined in (\ref{jam_non-dimensionalize}) stands for the non-dimensional vertical displacement. If $\tau_{YX}=\tau_h$ at Y=h, we find $h=\tau_h/P_1$. 
\begin{eqnarray}
Now~~~\frac{Y_0}{h}=\frac{\tau_0}{\tau_h}=\tau~(say),~~0<\tau<1~~~~~~~~~~~~~
\end{eqnarray}
The plug velocity is then given by
\begin{eqnarray}
U_p=\frac{(P_1h-\tau_0)^{n_1+1}}{(n_1+1)P_1}~~~~~~~~~~~~~~~~~~~~~~~~~~~~~~
\label{jam_plug_velocity}
\end{eqnarray}
Using the boundary conditions\\
 $~~~~~~~~~~~~~~~~~~~~~~~~~~~~~~~~~~\Psi_p=0~~at~Y=0~\\~~~~~~~~~~~~~~~~~~~~~~~~~~~~and~\Psi=\Psi_p~at~Y=Y_0$, \\and
 integrating (\ref{jam_axial_velocity}) and
 (\ref{jam_plug_velocity}), the stream function $\Psi$ is found to be
 given by
\begin{eqnarray}
\Psi=\left\{\begin{array}{r@{\quad
 : \quad}l}\frac{P_1^{n_1}}{(n_1+1)}\left[Y(h-Y_0)^{n_1+1}-\frac{(Y-\tau_0)^{n_1+2}}{n_1+2}\right]
 & if~~ Y_0\le Y \le
 h \\ \frac{P_1^{n_1}}{(n_1+1)}\left[Y(h-Y_0)^{n_1+1}+\frac{(-Y-\tau_0)^{n_1+2}}{n_1+2}\right]
 & \rm{if}~~ -h\le Y \le -Y_0, \\ \end{array} \right. 
\label{jam_stream_function_main}
\end{eqnarray}
\begin{eqnarray}
and~~\Psi_p=\frac{P_1^{n_1}(h-Y_0)^{n_1+1}Y}{n_1+1}~,~~~~~\rm{if}~~~-Y_0\le Y \le Y_0
\end{eqnarray}
The instantaneous rate of volume flow through each section, $\bar{Q}(x,t)$, is given by
\begin{eqnarray}
\bar{Q}(X,t)=\int_{0}^{Y_0}U_YdY+\int_{Y_0}^{h}U~dY~~~~~~~~~~~~~\nonumber~~~\\=\frac{P_1^{n_1}(h-Y_0)^{n_1+1}(n_1h+h+Y_0)}{(n_1+1)(n_1+2)},~~n_1=\frac{1}{n}
\label{jam_volume_flow_rate}
\end{eqnarray}
\begin{eqnarray}
Now~~ \frac{\partial P}{\partial
X}=-\left[\frac{\bar{Q}(X,t)(n_1+1)(n_1+2)}{(h-Y_0)^{n_1+1}(n_1h+h+Y_0)}\right]^n~~~~~~~~~~~\nonumber~~~~\\=-\left[\frac{\bar{Q}(X,t)(n_1+1)(n_1+2)}{h^{n_1+2}(1-\tau)^{n_1+1}(n_1+1+\tau)}\right]^n
\label{jam_pressure_gradient}
\end{eqnarray}
The average pressure rise per wave length is calculated as
\begin{eqnarray}
\Delta P=-\int_{0}^{1}\int_{0}^{1}\left(\frac{\partial P}{\partial X}\right)dxdt
\label{jam_pressure_rise}
\end{eqnarray}
In the fixed frame of reference, the expression for the
non-dimensional transverse velocity $V$ is found to be given by
\begin{eqnarray}
V(X,Y,t)=\left[\left\{\frac{n_1}{\bar{Q}^n(X,t)}-\frac{(n_1+1)(n_1+2)}{h(1-\tau)(n_1+1+\tau)}\right\}\left\{h(1-\tau)^{n_1+1}Y-\frac{h^2}{n_1+2}\left(\frac{Y}{h}-\tau\right)^{n_1+2}\right\}\right.\nonumber\\\left.+(n_1+1)(1-\tau)^n_1Y\right]\times\frac{h^n_1P_1^{n_1}\{\Lambda\lambda/d_0+2\pi\phi\cos(2\pi(X-t))\}}{n_1+1}~~~~~~~~~~~~~~
\end{eqnarray}

It may be noted that if we put $n=1,~\tau=0,~\Lambda=0$ in equations
(\ref{jam_axial_velocity}), (\ref{jam_stream_function_main}),
(\ref{jam_volume_flow_rate}) and (\ref{jam_pressure_gradient}), the
expressions reduce to those reported earlier in \cite{Shapiro}. Our
results also tally with those of \cite{Lardner}, when the eccentricity
of the elliptical motion of cilia tips is set equal to zero in their
analysis for a Newtonian fluid flowing through a uniform
channel. Moreover, when $n=1~and~Y_0=0$, the expression for the
pressure gradient given by (\ref{jam_pressure_gradient}) reduces to
that obtained by Gupta and Seshadri \cite{Gupta} for the peristaltic
motion of a Newtonian fluid having constant viscosity. It may be noted
that since the right hand side of equation (\ref{jam_pressure_rise})
cannot integrated in closed form, for non-uniform/uniform geometry,
for further investigation of our problem, we had to resort to the use
of appropriate softwares, as mentioned in the next section.

\section{Computational Results and Discussion}
The instantaneous rate of volume flow
$\bar{Q}(X,t)$ has been assumed to be periodic in (X-t)
(\cite{Srivastava1,Gupta,Srivastava3}), so that $\bar{Q}$ appearing equation
(\ref{jam_pressure_rise}) can be expressed as
\begin{eqnarray}
\bar{Q}^n(X,t)=Q^n+\phi \sin 2\pi (X-t)~,
\end{eqnarray}
$Q$ being the time-averaged flow flux.
Due to complexity of the problem, it has not been possible to find the
expression for the average pressure rise, $\Delta P$ given by
(\ref{jam_pressure_gradient}). It has been computed numerically by
using the software Mathematica.

In this section, on the basis of the present study, we shall obtain
theoretical estimates of different physical quantities that are of
relevance to the physiological problem of blood flow in
micro-circulatory system. For this purpose, we have used the following
data valid in the physiological range
(\cite{Guyton,Srivastava1,Barbee,Fung2}): $d_0=10$ to $60\mu m$,
$\phi=0.1$ to $0.9$, $\frac{d_0}{\lambda}=0.01$ to $0.02$, $\Delta
P=-300$ to $50$, $\tau=0.0$ to $0.2$, $Q=0$ to $2$,
$n=\frac{1}{3}$ to $2$. Unlike other studies on non-uniform geometry,
the value of $\Lambda$ has been so chosen that for converging tubes
(e.g. arterioles), the width of the outlet of one wave length is
$25\%$ less than that of the inlet, while in the case of diverging
tubes (e.g. venules), the width of the outlet of one wave length is
$25\%$ more than that of inlet.
 
It is important to mention that the results presented in the sequel
for shear thinning and shear thickening fluids are quite relevant for
the study of blood rheology. Normal blood usually behaves like a shear
thinning fluid for which with the increase in shear rate, the
viscosity decreases. It has been mentioned in \cite{Fung2,White,Xue}
that in the case of hardened red blood cell suspension, the fluid
behaviour is that of a shear thickening fluid for which with the fluid
viscosity is enhanced due to an elevation of the shear rate.

\begin{figure}
\centering
\includegraphics[width=3.5in]{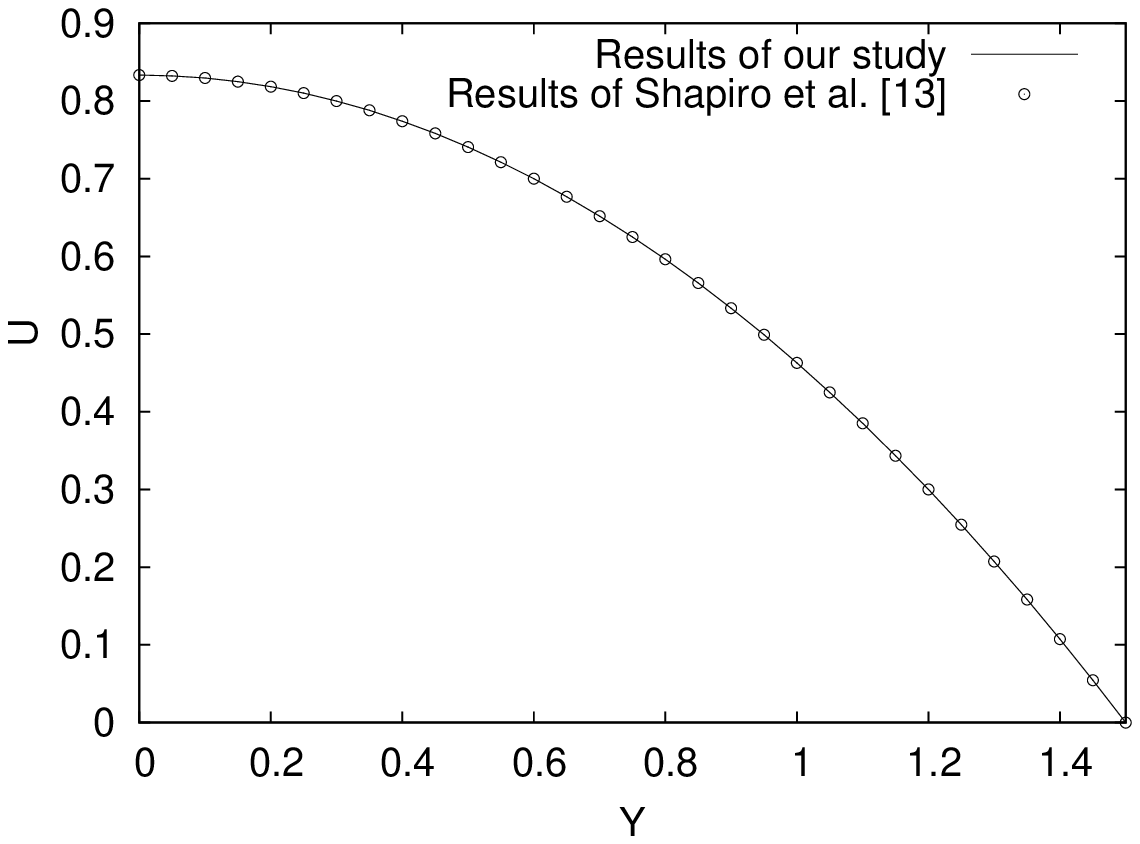}\includegraphics[width=3.5in]{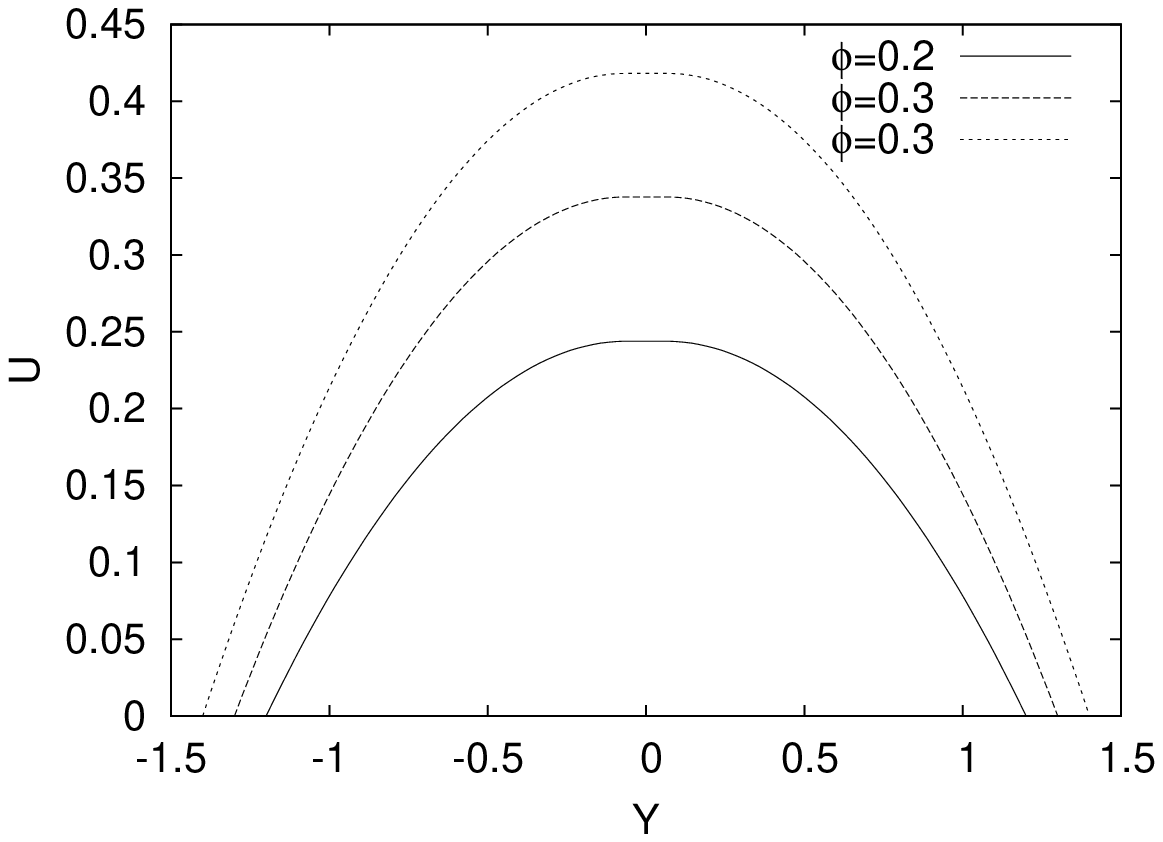}\\
$~~~~~~~~~~~~~~~~~~~~~~~~~~~~~~~~~~~~~(a)~~~~~~~~~~~~~~~~~~~~~~~~~~~~~~~~~~~~~~~~~~~~~~~~~~~~~~~~~~~~~~~~~~~~~(b)~~~~~~~~~~~~~~~~~~~~$
\includegraphics[width=3.5in]{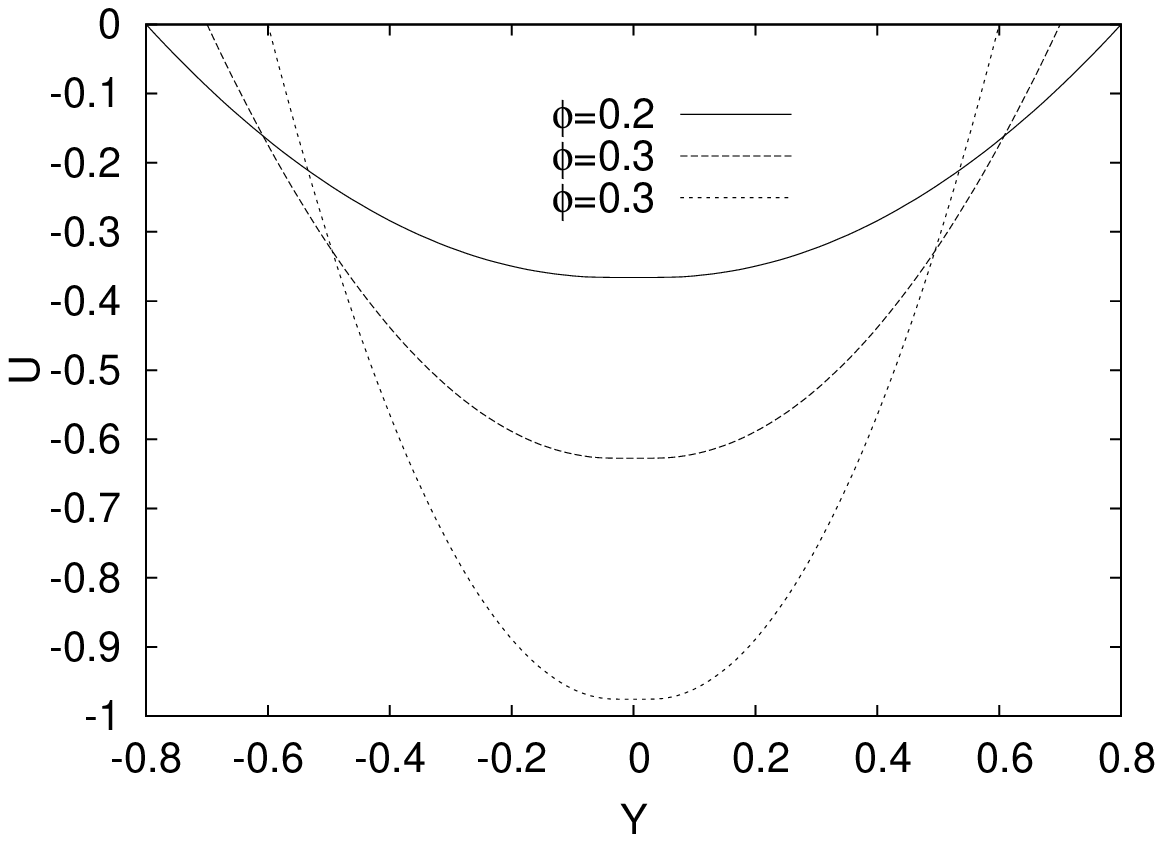}\\
(c)
\caption{Variation of axial velocity in the vertical direction (a) at
  x=0.5, t=0.25 $(\Lambda=0,~\phi=0.5,~\tau=0,~n=1,~\Delta P=0)$ (b)
  for different values of $\phi$ at wave crest when $\Lambda=0$,
  $n=1$, $\tau=0$, $Q=0$ (c) for different values of $\phi$ at wave trough when
  $\Lambda=0$, $n=1$, $\tau=0$, $Q=0$}
\label{jam_velocompare}
\end{figure} 

\begin{table}
\centering

\begin{tabular}{c|c|c|c|}
\cline{2-4}
& $\phi$ & Our results & Results in [25] \\ \hline
\multicolumn{1}{|c|}{\multirow{3}{*}{Wave Crest}} & 0.2 & 0.244 & 0.25 \\ \cline{2-4}\multicolumn{1}{|c|}{} 
 & 0.3 & 0.338 & 0.344 \\ \cline{2-4}
\multicolumn{1}{|c|}{}
 & 0.4 & 0.418 & 0.427 \\ \cline{1-4}
\multicolumn{1}{|c|}{\multirow{3}{*}{Wave Trough}} & 0.2 & -0.366 &
-0.375 \\ \cline{2-4}\multicolumn{1}{|c|}{}
 & 0.3 & -0.627 & -0.635 \\ \cline{2-4}
\multicolumn{1}{|c|}{}
 & 0.4 & -0.976 & -0.984 \\ \cline{1-4}
\end{tabular}
\caption{Comparison of the present study with those of Takabatake and
  Ayukawa [25] for small values of $Re$ and $\delta$ ($n=1$,
  $\tau=0$, $\Lambda=0$, $Q=0$)}
\end{table}

\subsection{Velocity Distribution}

\begin{figure}
\includegraphics[width=3.8in]{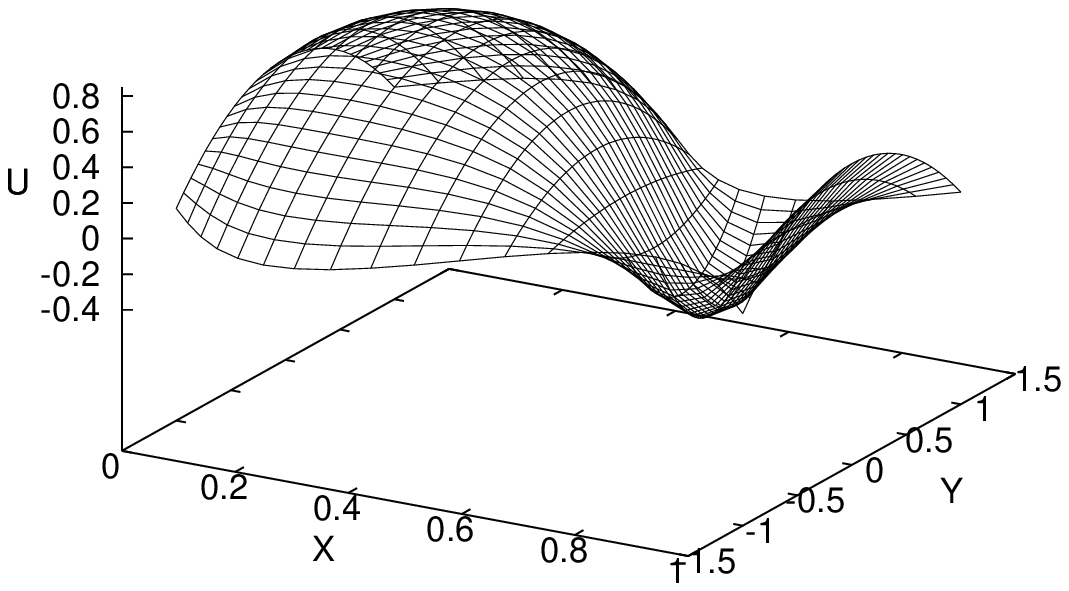}\includegraphics[width=3.8in]{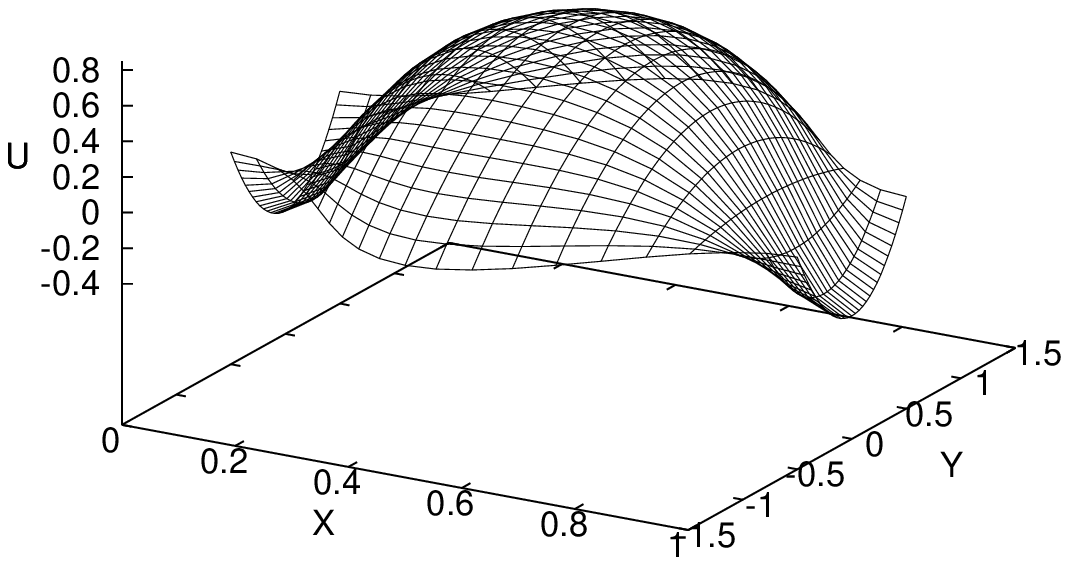}
\\$~~~~~~~~~~~~~~~~~~~~(a)~~\rm{for}~t=0.0~~~~~~~~~~~~~~~~~~~~~~~~~~~~~~~~~~~~~~~~~~~~~(b)~~\rm{for}~t=0.25~~~~~~~~~~~~~~~~~~~~~~~~~~~~~~~~$\\
\includegraphics[width=3.8in]{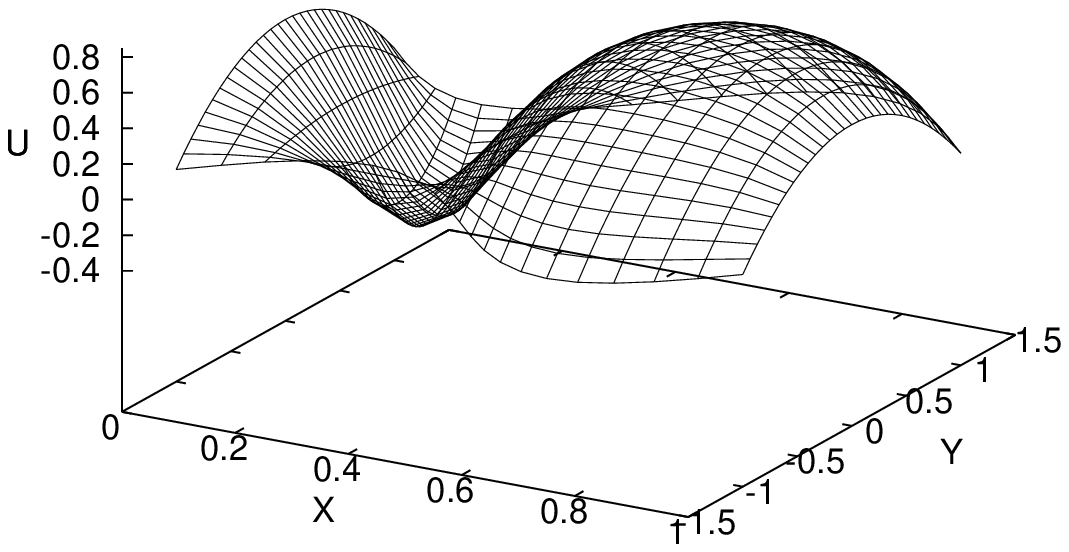}\includegraphics[width=3.8in]{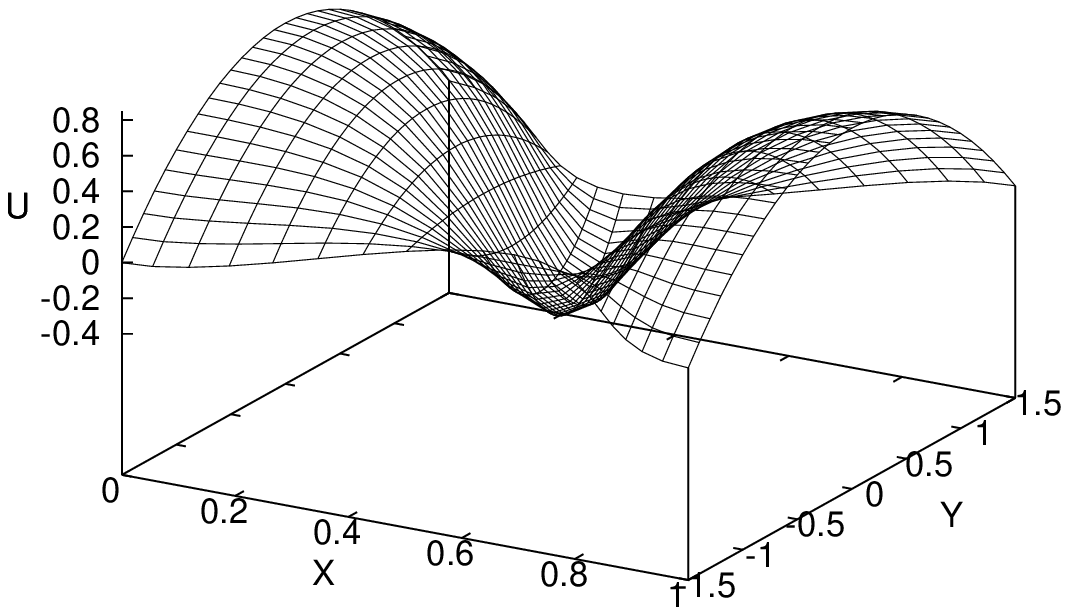}
\\$~~~~~~~~~~~~~~~~~~~~(c)~~\rm{for}~t=0.5~~~~~~~~~~~~~~~~~~~~~~~~~~~~~~~~~~~~~~~~~~~~~(d)~~\rm{for}~t=0.75~~~~~~~~~~~~~~~~~~~~~~~~~~~~~~~~$\\
\caption{Aerial view of the velocity distribution at different
instants of time ($n=1,~\Lambda=0,~\Delta P=0,~\tau=0,~\phi=0.5$)}
\label{jam_velo3D4.1.1-4.1.4}
\end{figure}

\begin{figure}
\centering
\includegraphics[width=3.6in]{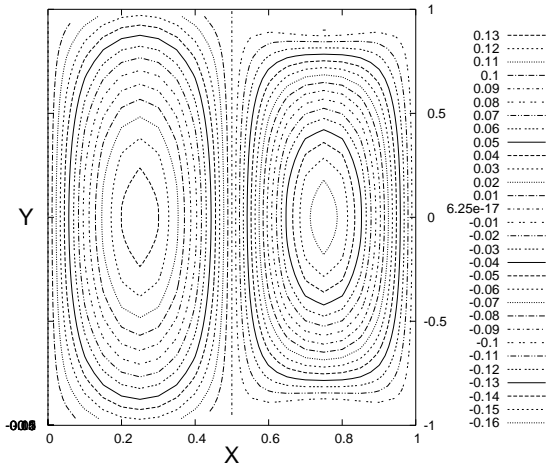}\includegraphics[width=3.6in]{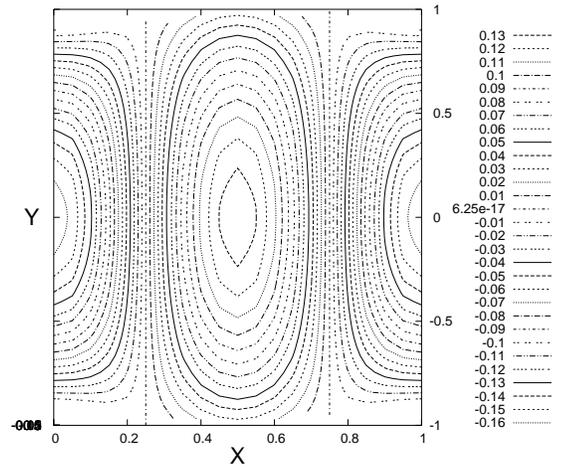}
\\$~~~~~~~~~~~~~~~~~~~~(a)~~\rm{for}~t=0.0~~~~~~~~~~~~~~~~~~~~~~~~~~~~~~~~~~~~~~~~~~~~~(b)~~\rm{for}~t=0.25~~~~~~~~~~~~~~~~~~~~~~~~~~~~~~~~$
\includegraphics[width=3.6in]{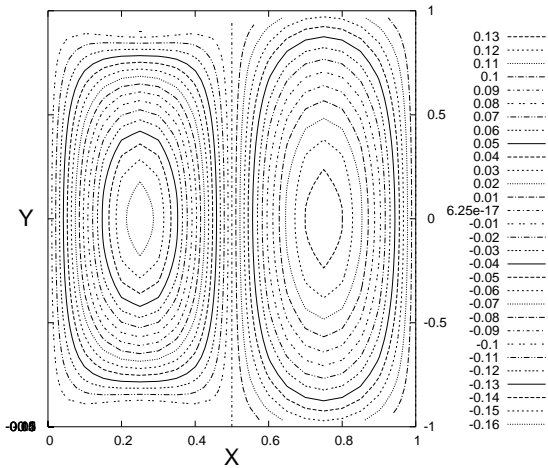}\includegraphics[width=3.6in]{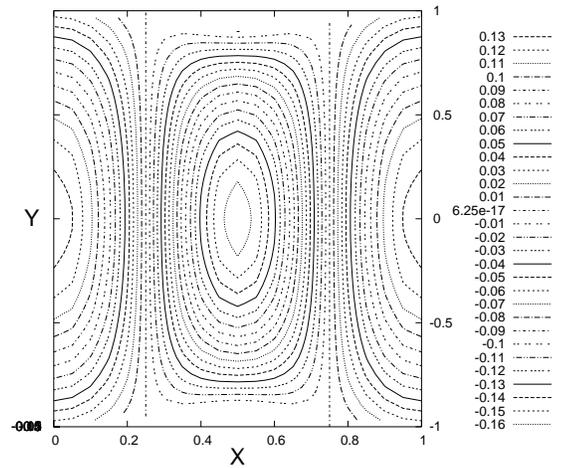}
\\$~~~~~~~~~~~~~~~~~~~~(c)~~\rm{for}~t=0.5~~~~~~~~~~~~~~~~~~~~~~~~~~~~~~~~~~~~~~~~~~~~~(d)~~\rm{for}~t=0.75~~~~~~~~~~~~~~~~~~~~~~~~~~~~~~~~$
\includegraphics[width=3.5in]{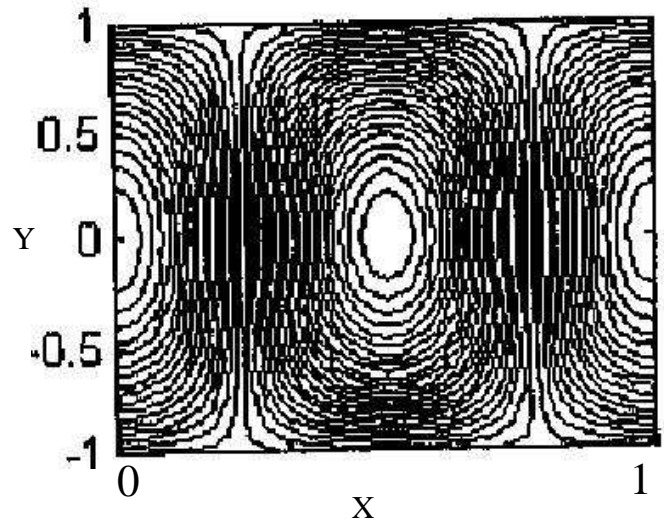}\\
(e)~Contours reproduced from Selvarajan et al. [55]
\caption{Axial velocity contour at different time when $\phi=0.1,~n=1,~\Lambda=0,~\tau=0,~Q=0$}
\label{jam_velocon4.1.1-4.1.4}
\end{figure}

\begin{figure}
\centering
\includegraphics[width=4.5in,height=3.0in]{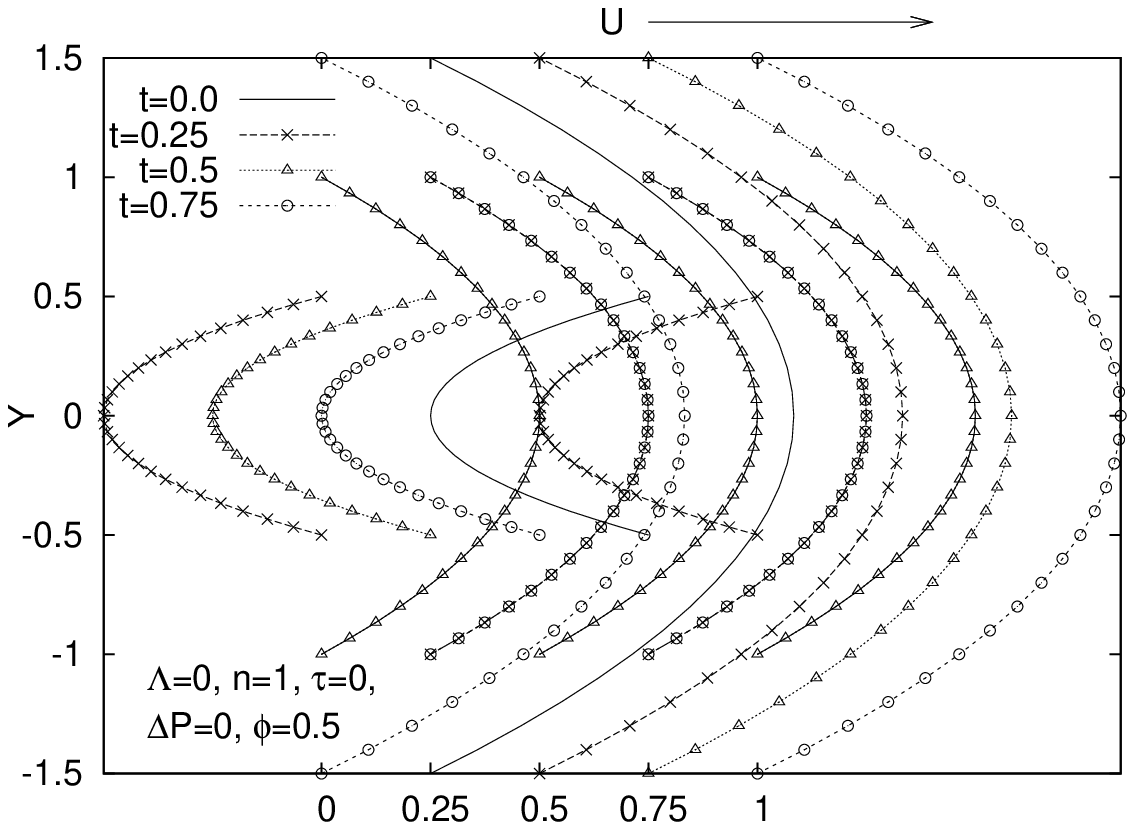}
\caption{Velocity distribution at different instants of time}
\label{jam_velo4.1}
\end{figure}

\begin{figure}
\includegraphics[width=3.6in]{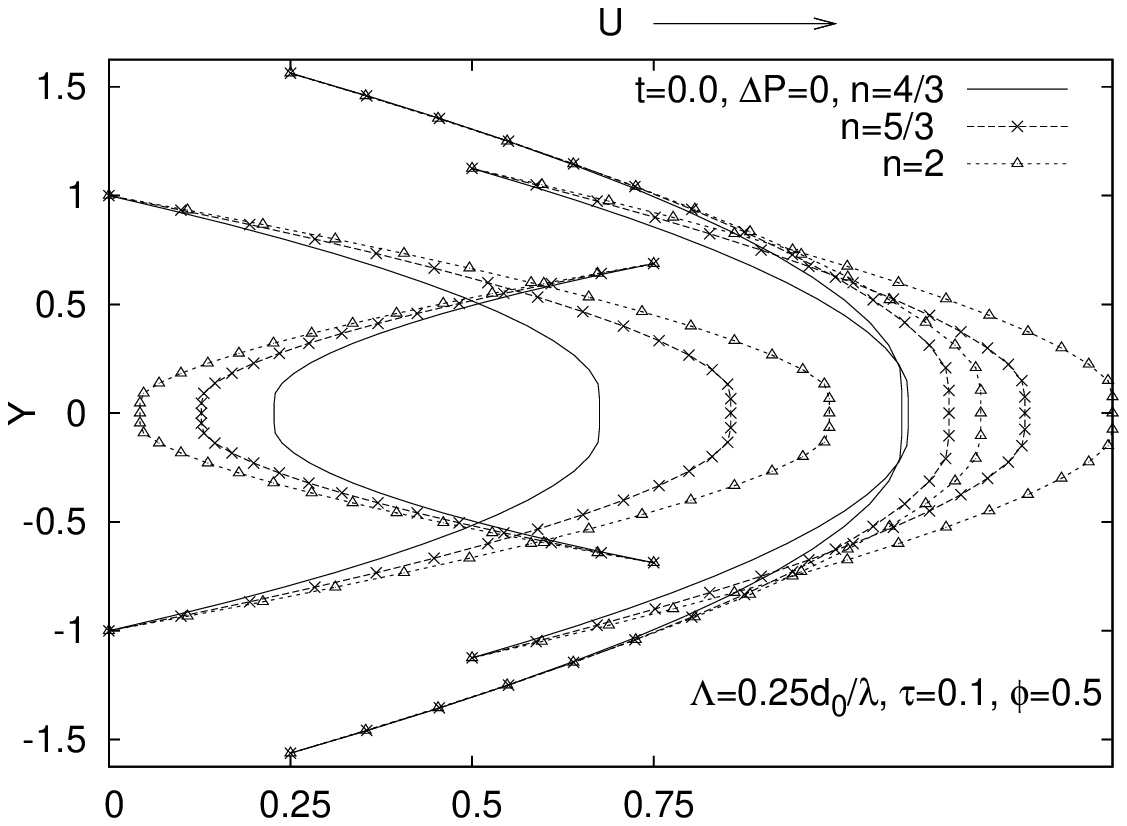}\includegraphics[width=3.6in]{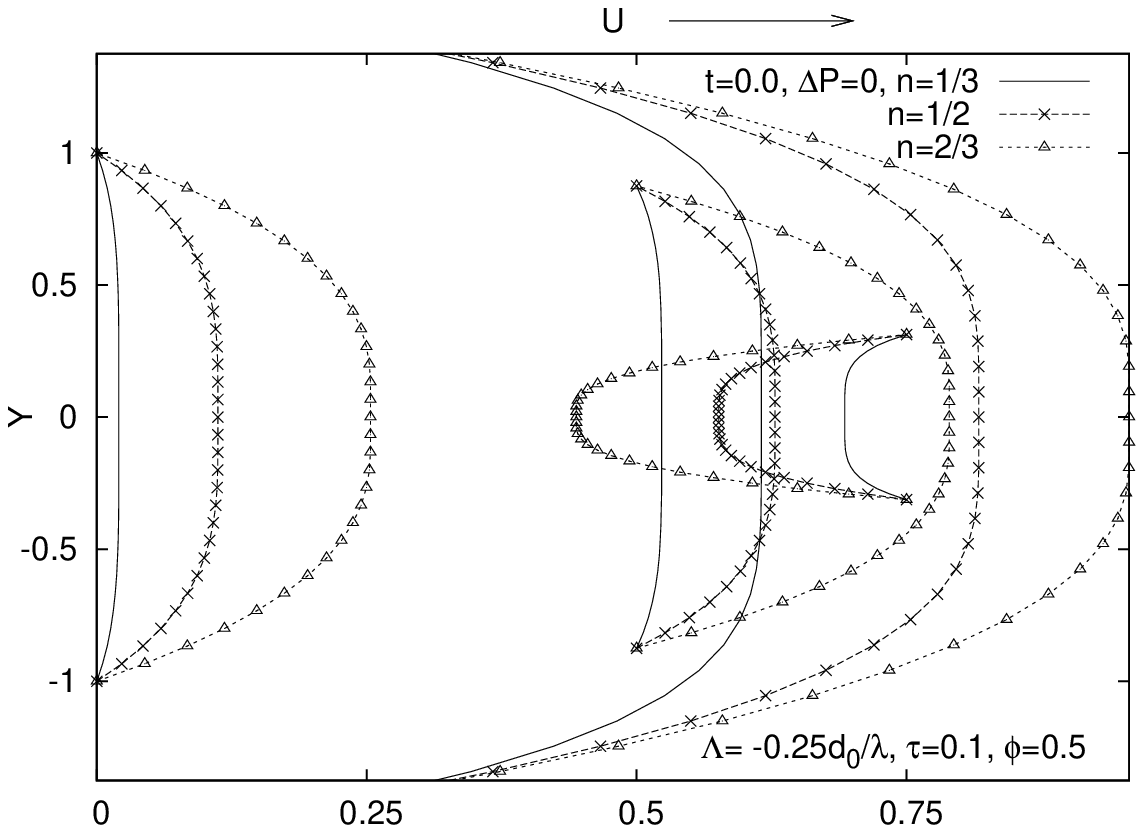}
\\$~~~~~~~~~~~~~~~~~~~~(g) ~~~~~~~~~~~~~~~~~~~~~~~~~~~~~~~~~~~~~~~~~~~~~~~~~~~~~~~~~(h)~~~~~~~~~~~~~~~~~~~~~~~~~~~~~~~~~~~~~~~~~$\\
\includegraphics[width=3.6in]{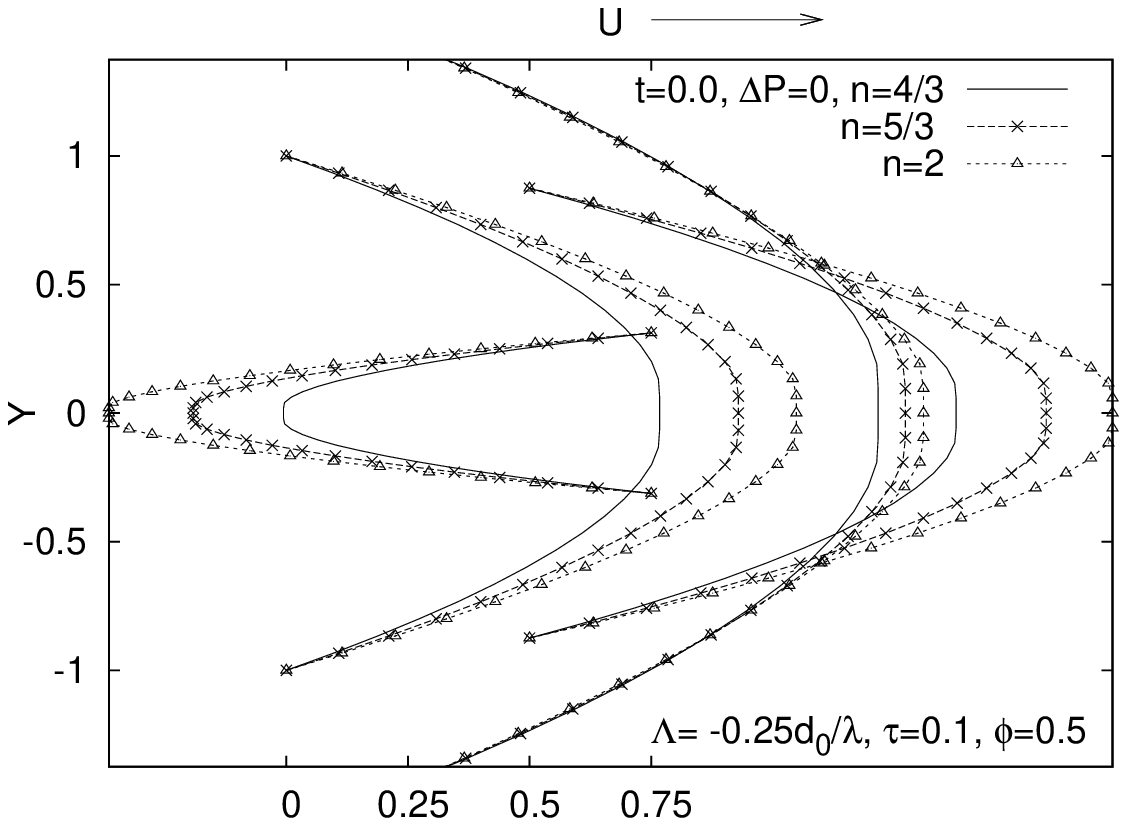}\includegraphics[width=3.6in]{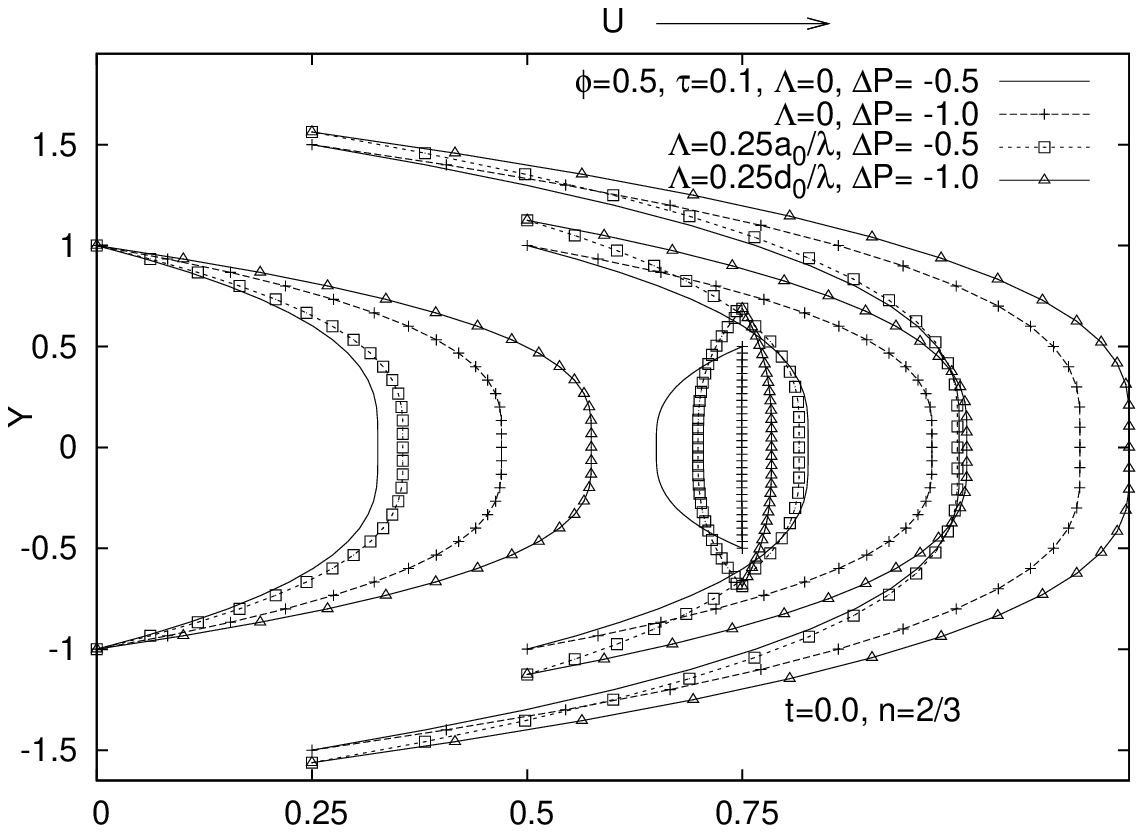}
\\$~~~~~~~~~~~~~~~~~~~~(i) ~~~~~~~~~~~~~~~~~~~~~~~~~~~~~~~~~~~~~~~~~~~~~~~~~~~~~~~~~(j)~~~~~~~~~~~~~~~~~~~~~~~~~~~~~~~~~~~~~~~~~$\\
\includegraphics[width=3.6in]{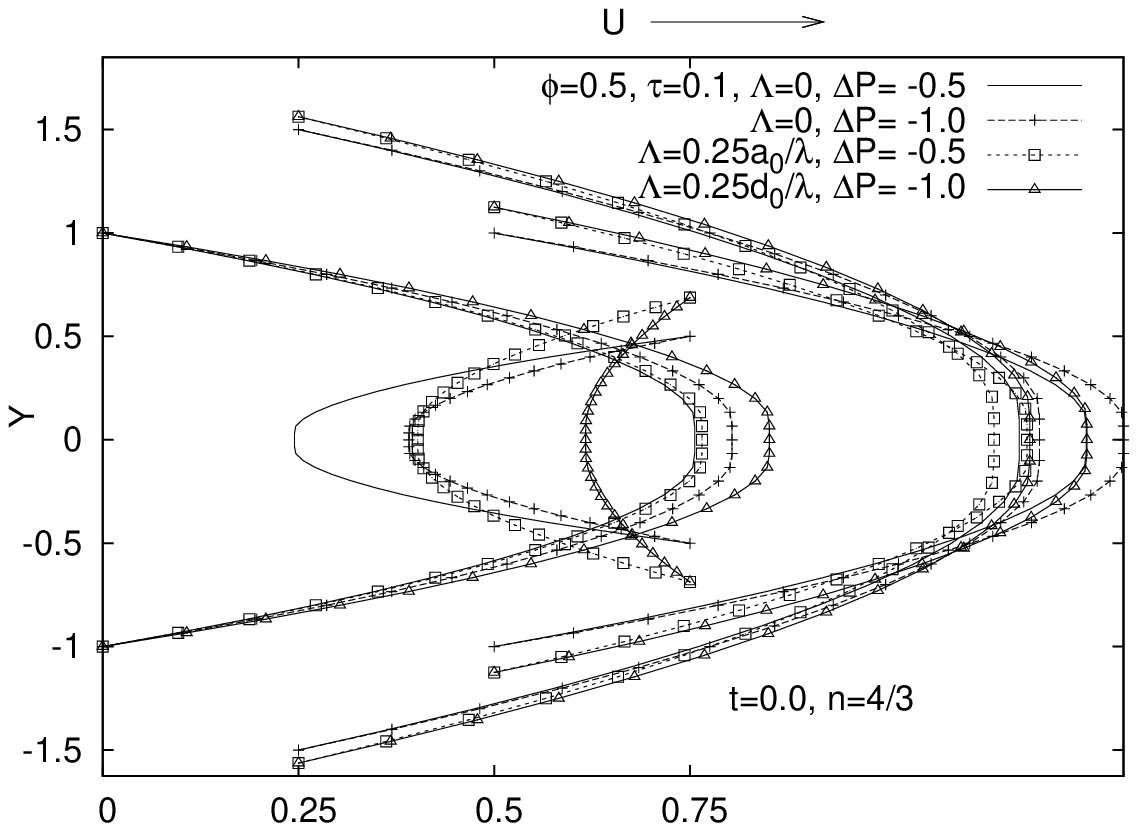}\includegraphics[width=3.6in]{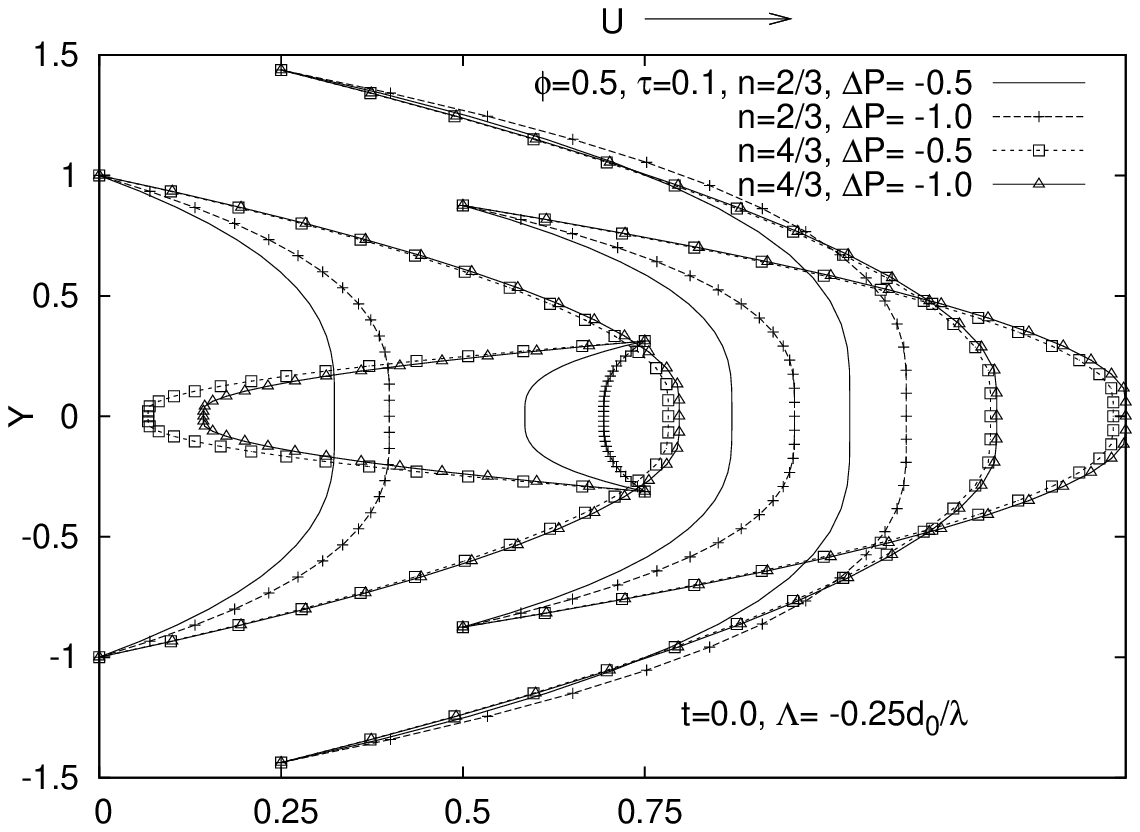}
\\$~~~~~~~~~~~~~~~~~~~~(k) ~~~~~~~~~~~~~~~~~~~~~~~~~~~~~~~~~~~~~~~~~~~~~~~~~~~~~~~~~(l)~~~~~~~~~~~~~~~~~~~~~~~~~~~~~~~~~~~~~~~~~$
\caption{Distribution of velocity in different situations}
\label{jam_velo4.2-4.5.3}
\end{figure}

Plots in Figs. \ref{jam_velocompare}-\ref{jam_velo4.2-4.5.3} give the
distribution of axial velocity in the cases of free pumping, pumping
and co-pumping for different values of the amplitude ratio $\phi $,
flow index number n, $\tau$, $\Lambda$.  Fig. \ref{jam_velocompare}(a)
shows that the results computed on the basis of our study for the
particular case of Newtonian fluid tally well with the results
reported by Shapiro et al. \cite{Shapiro} when the amplitude ratio
$\phi=0.5$. Variation of axial velocity in the vertical direction at
the wave crest and the wave trough are exhibited in
Figs. \ref{jam_velocompare}(b,c) for different values of
$\phi$. In order to compare our results with those reported by
Takabatake and Ayukawa \cite{Takabatake}, we have reproduced their
results alongside the results computed on the basis of the present
study in Table 1. From the tabulated values, one can
observe that axial velocity along the central line at the wave
crest and the wave trough match well with \cite{Takabatake} when the
wave number is small.
Since the velocity profiles and height of the channel change with
time, investigation has been made on the basis of the present study
for the distribution of velocity at an interval of
T/4. Figs. \ref{jam_velo3D4.1.1-4.1.4} depict the aerial view of a few
typical axial velocity distributions for a Newtonian fluid flowing
over a uniform channel. The corresponding velocity contours for the
peristaltic motion corresponding to $\phi=0.1$ are shown in
Figs. \ref{jam_velocon4.1.1-4.1.4}. The velocity contours presented in
Fig. \ref{jam_velocon4.1.1-4.1.4}(b) resemble with those in
Fig. \ref{jam_velocon4.1.1-4.1.4}(e) reported by Selvarajan et
al. \cite{Selvarajan}, although in our plots, there is a slight
increase in area of contour lines at the region of wave
expansion. This is possibly because the curved boundary was not taken
into consideration in \cite{Selvarajan} for calculating the velocity
contours.

Fig \ref{jam_velo4.1} gives the velocity distribution in the plane of
the channel. This figure reveals that at any instant of time, although
there exists a retrograde flow region, the forward flow region is
predominant, the time averaged flow rate being positive. We have also
found that if $Q=0$, the retrograde flow region occupies exactly
half of the one complete wave length, the remaining half being
occupied by the forward mean flow region. For $t=T/4$, our study
reveals that there exists two stagnation points on the axis, one being
located near the trailing end and the other near the leading end. Both
these observations agree with those of Takabatake and Ayukawa
\cite{Takabatake}, who performed a similar study numerically for a
Newtonian fluid. Fig. \ref{jam_velo4.2-4.5.3}(a) shows that when $n=1$
(that is, in the case of a Newtonian fluid) and $\Delta P=0$, in both
the forward flow region and the backward flow region velocity
increases as the value of $\phi$ is raised. One can also have an idea
of the extent by which the value of $\phi$ affects the axial velocity
for shear thinning (cf. Fig. \ref{jam_velo4.2-4.5.3}(b)) and shear
thickening fluids (Fig. \ref{jam_velo4.2-4.5.3}(c)). It is apparent
from these figures that in either region, as $\tau$ increases, the
magnitude of velocity decreases for both types of fluids.

The extent to which the velocity distribution is influenced by the
value of the rheological fluid index 'n' can be observed from
Figs. \ref{jam_velo4.2-4.5.3}(d-i) for uniform/non-uniform
channels. We observe that the the parabolic nature of the velocity
profiles is disturbed due to non-Newtonian effect and that as the
value of 'n' increases, the magnitude of the velocity increases. It is
worthwhile to mention that for a converging channel, the magnitude of
the velocity is greater than that in the case of a uniform channel;
however, for a diverging channel, we have an altogether different
observation.  Figs. \ref{jam_velo4.2-4.5.3}(j-l) depict the influence
of pressure on velocity distribution for shear thinning ($n<1$)/shear
thickening ($n>1$) fluids. While the plots given in
Figs. \ref{jam_velocompare}-\ref{jam_velo4.2-4.5.3}(i) correspond to
the case of $\Delta P=0$ (free pumping),
Figs. \ref{jam_velo4.2-4.5.3}(j-l) have been plotted for the
co-pumping case $\Delta P<0$. For these plots, we consider the case
when $\Delta P=-1$. In this case, for a shear thinning fluid, flow
reversal is totally absent in the case of a uniform/diverging
channel. For a converging channel, however, although there is
reduction in the region of flow reversal, it does not vanish
altogether, irrespective of whether the fluid is of shear thinning or
of shear thickening type.

\begin{figure}
\includegraphics[width=3.6in]{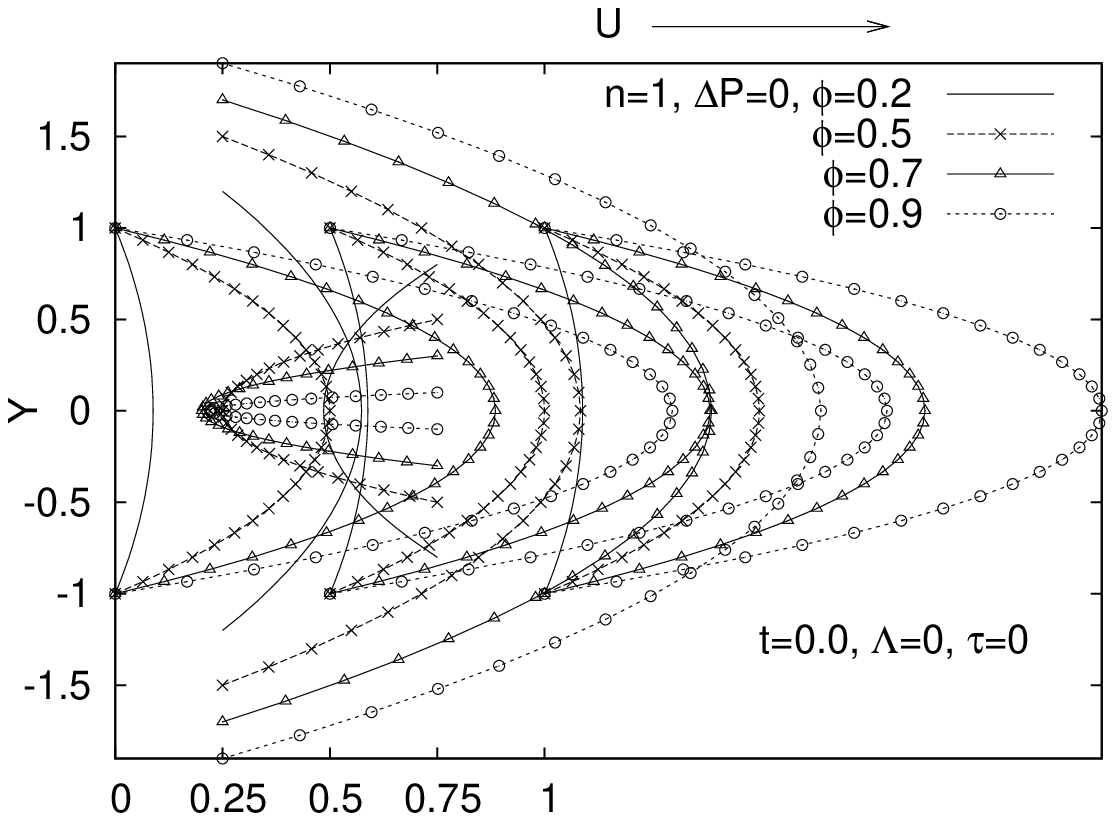}\includegraphics[width=3.6in]{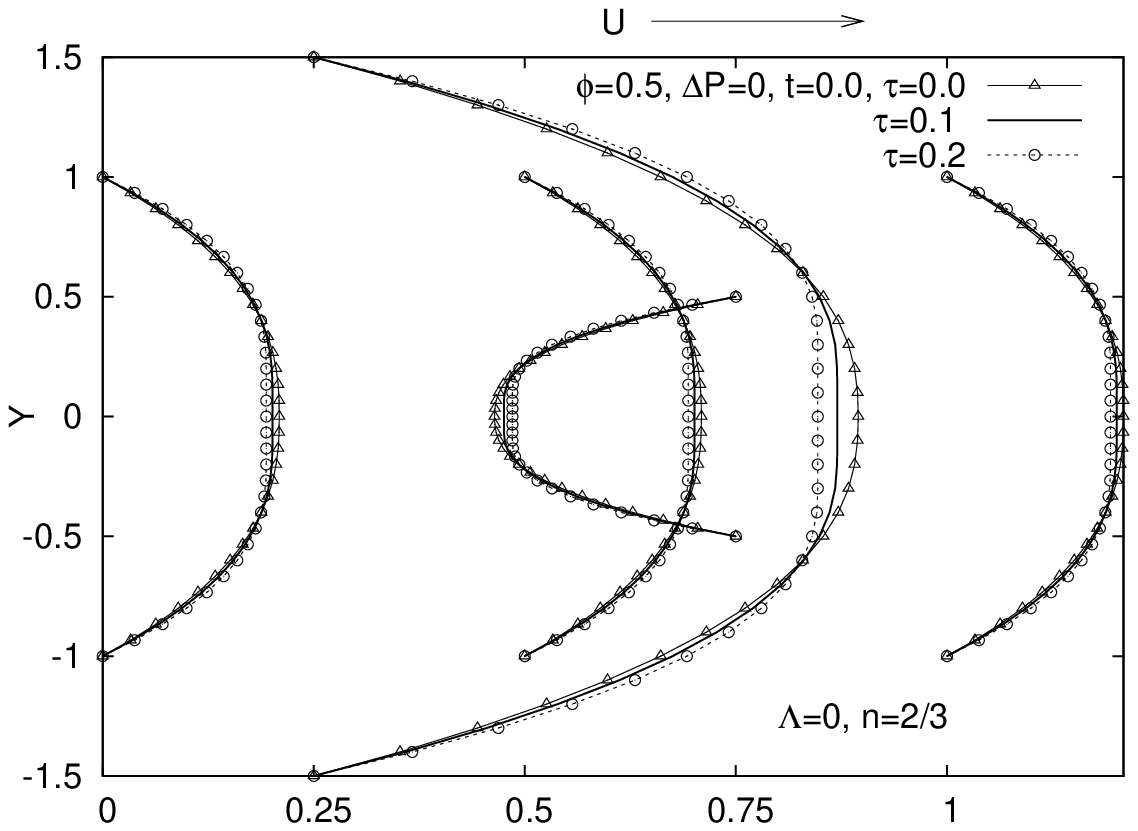}
\\$~~~~~~~~~~~~~~~~~~~~(a) ~~~~~~~~~~~~~~~~~~~~~~~~~~~~~~~~~~~~~~~~~~~~~~~~~~~~~~~~~(b)~~~~~~~~~~~~~~~~~~~~~~~~~~~~~~~~~~~~~~~~~$\\
\includegraphics[width=3.6in]{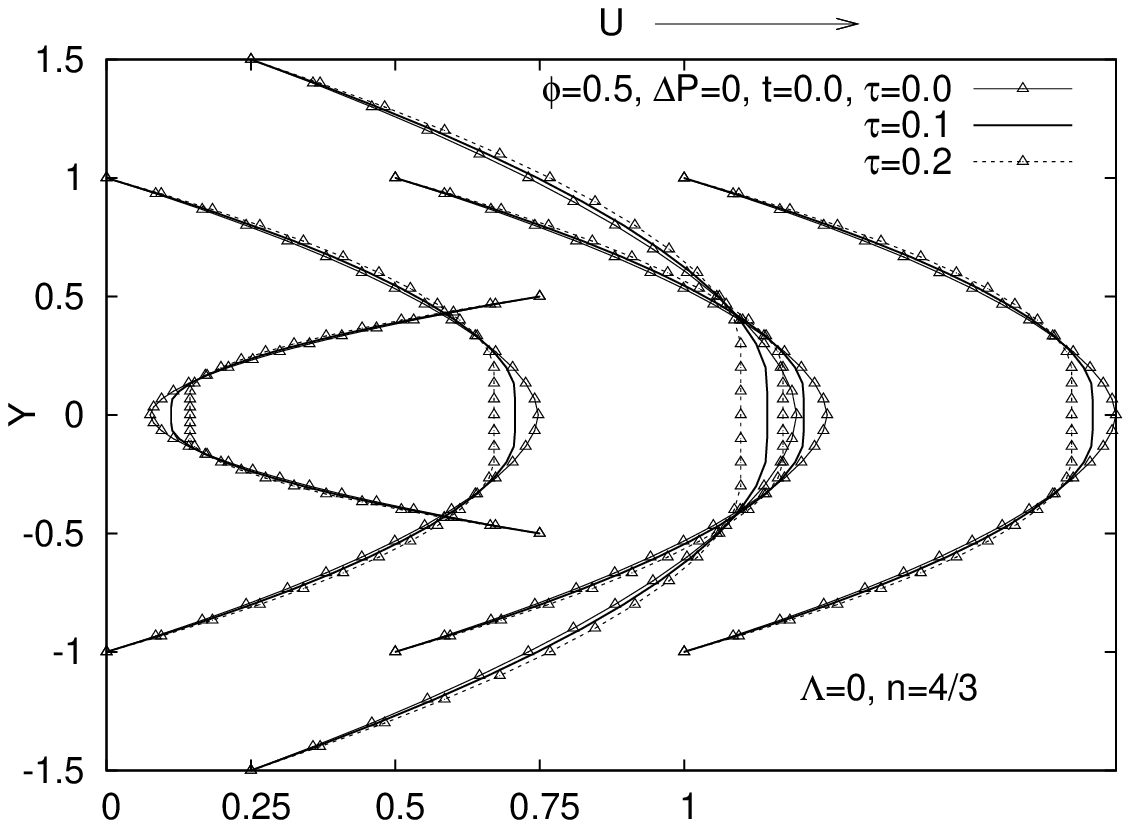}\includegraphics[width=3.6in]{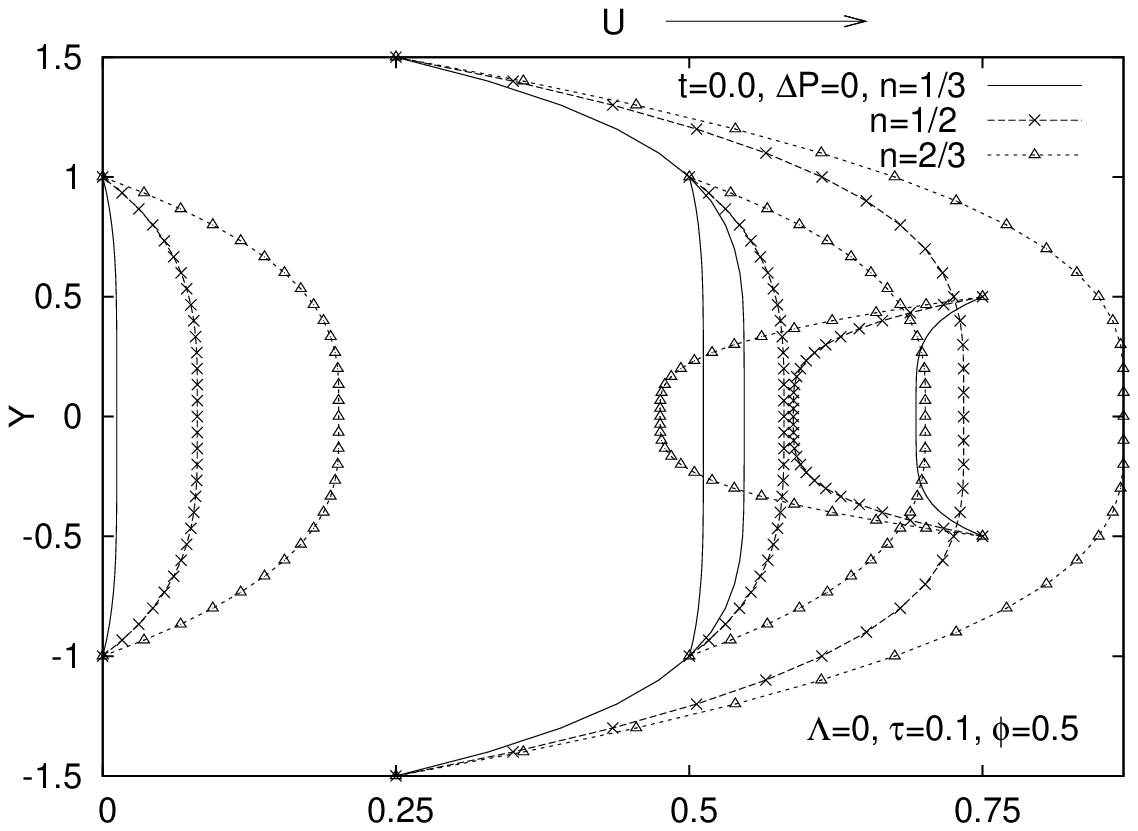}
\\$~~~~~~~~~~~~~~~~~~~~(c) ~~~~~~~~~~~~~~~~~~~~~~~~~~~~~~~~~~~~~~~~~~~~~~~~~~~~~~~~~(d)~~~~~~~~~~~~~~~~~~~~~~~~~~~~~~~~~~~~~~~~~$\\
\includegraphics[width=3.6in]{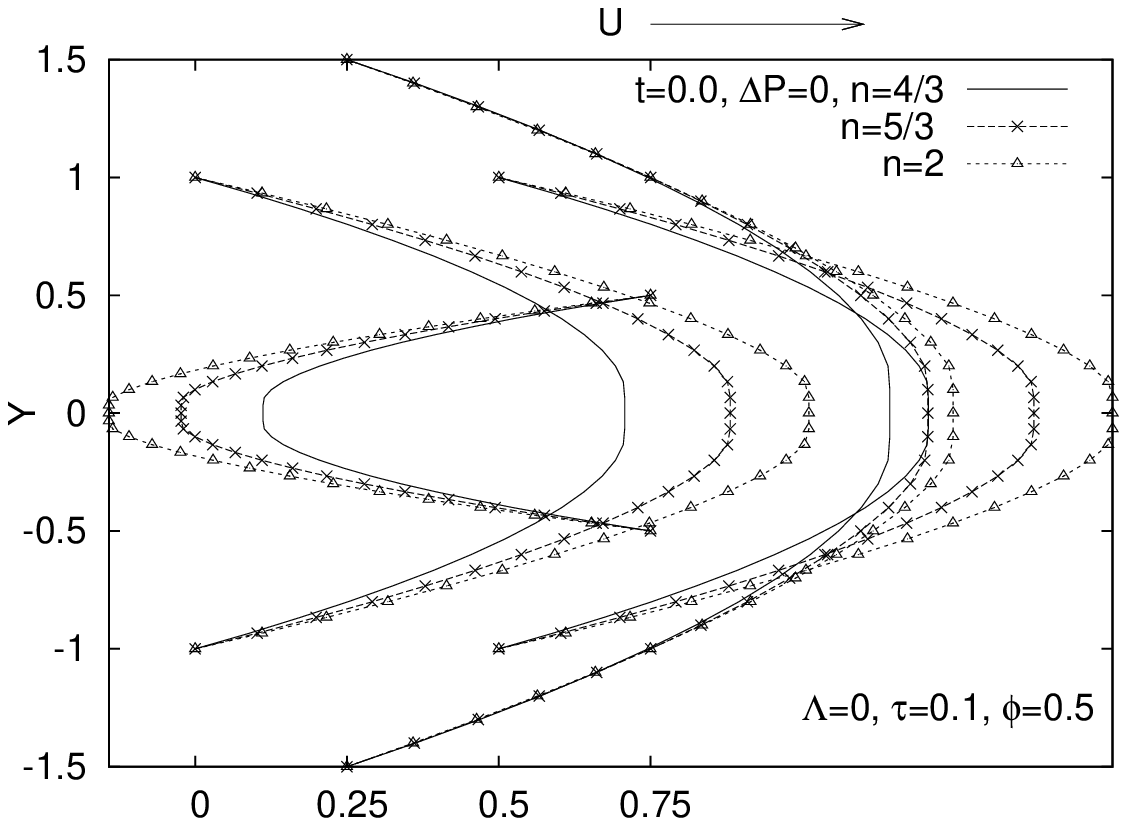}\includegraphics[width=3.6in]{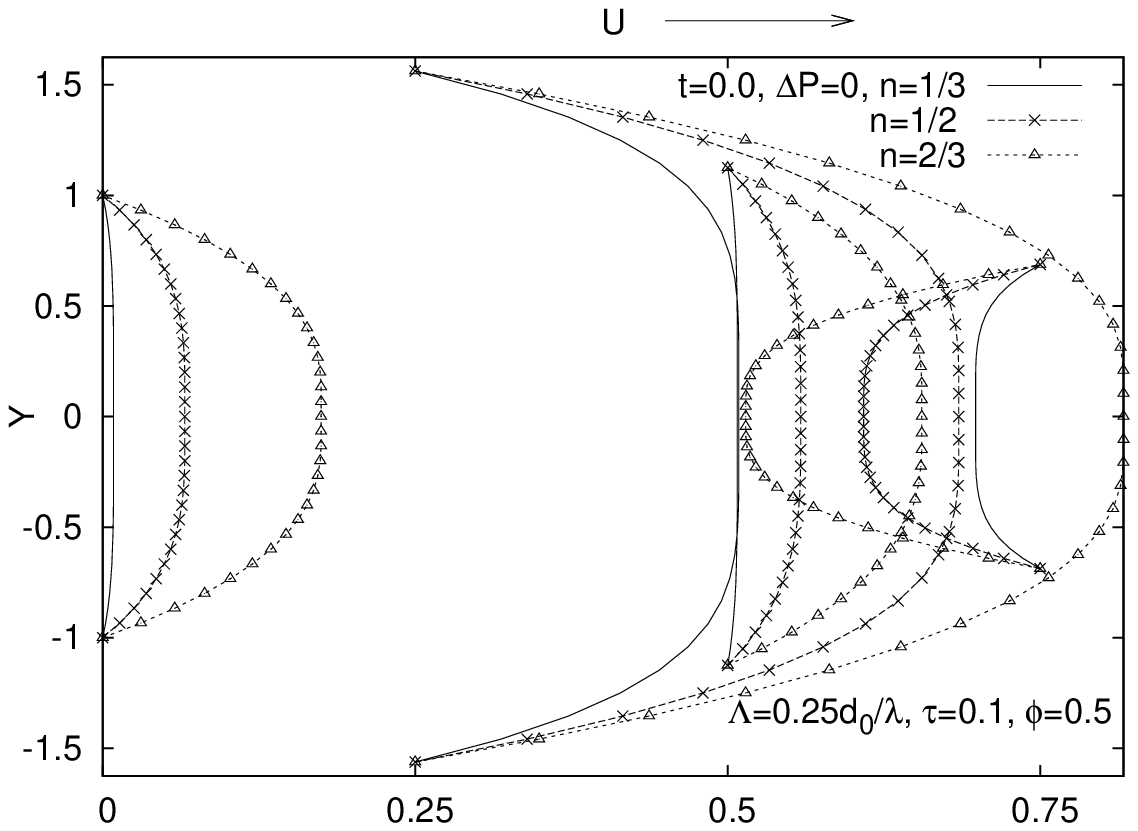}
\\$~~~~~~~~~~~~~~~~~~~~(e) ~~~~~~~~~~~~~~~~~~~~~~~~~~~~~~~~~~~~~~~~~~~~~~~~~~~~~~~~~(f)~~~~~~~~~~~~~~~~~~~~~~~~~~~~~~~~~~~~~~~~~$\\
\end{figure}

\subsection{Pumping Performance}
\begin{figure}
\centering
\includegraphics[width=3.5in,height=1.8in]{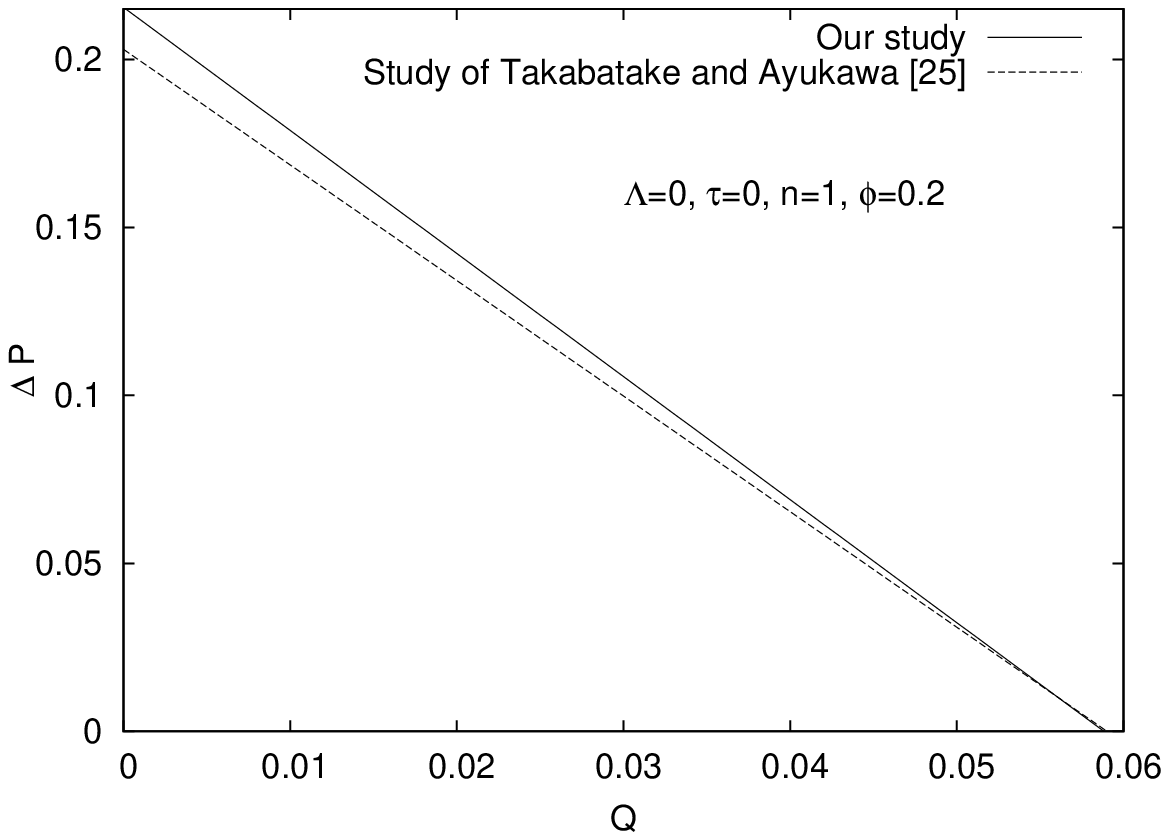}\includegraphics[width=3.5in,height=1.8in]{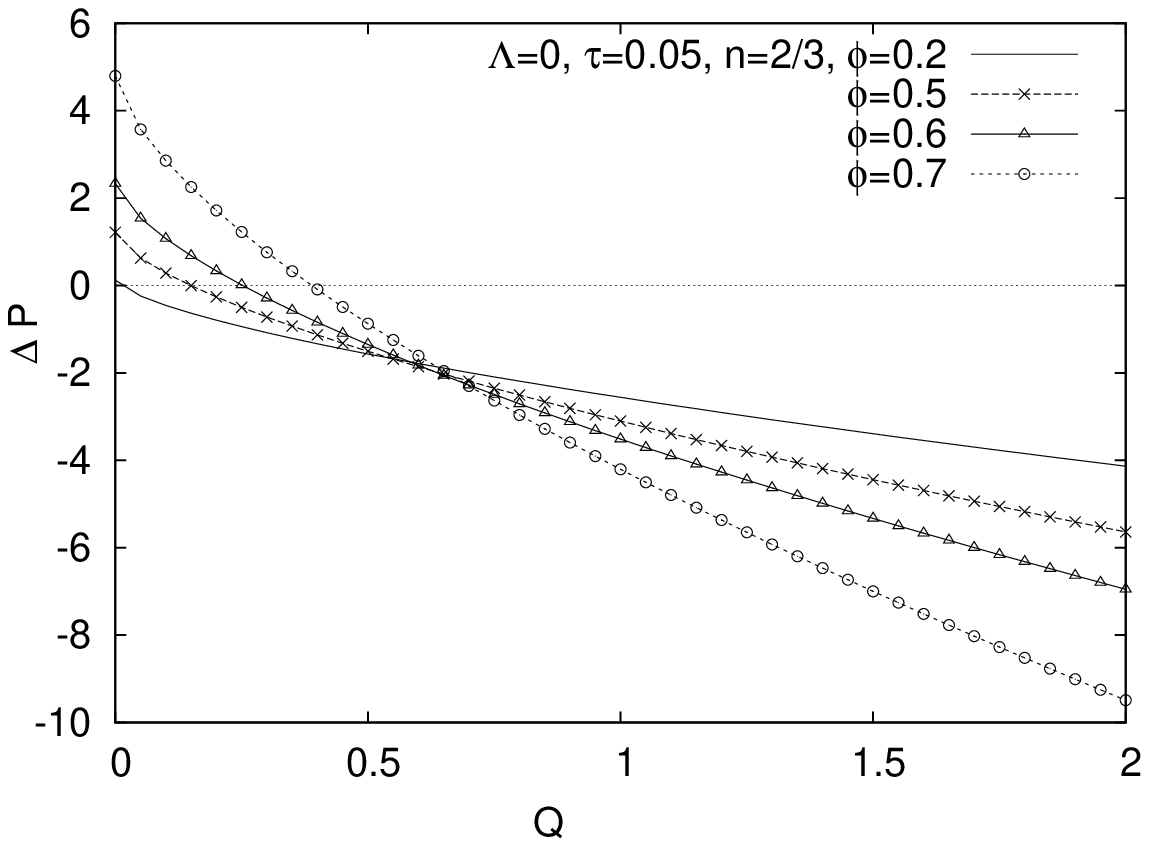}
\\$~~~~~~~~~~~~~~~~~~~~(a) ~~~~~~~~~~~~~~~~~~~~~~~~~~~~~~~~~~~~~~~~~~~~~~~~~~~~~~~~~(b)~~~~~~~~~~~~~~~~~~~~~~~~~~~~~~~~~~~~~~~~~$\\
\includegraphics[width=3.5in,height=1.8in]{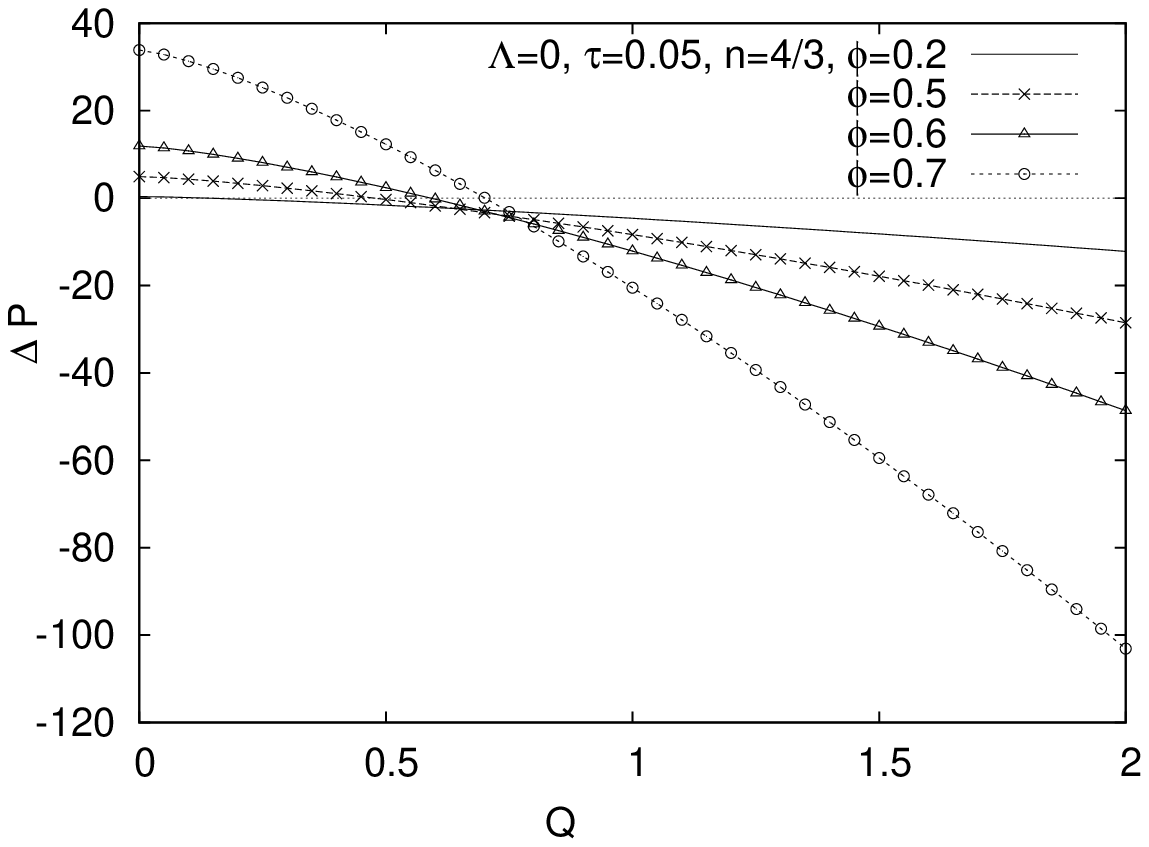}\includegraphics[width=3.5in,height=1.8in]{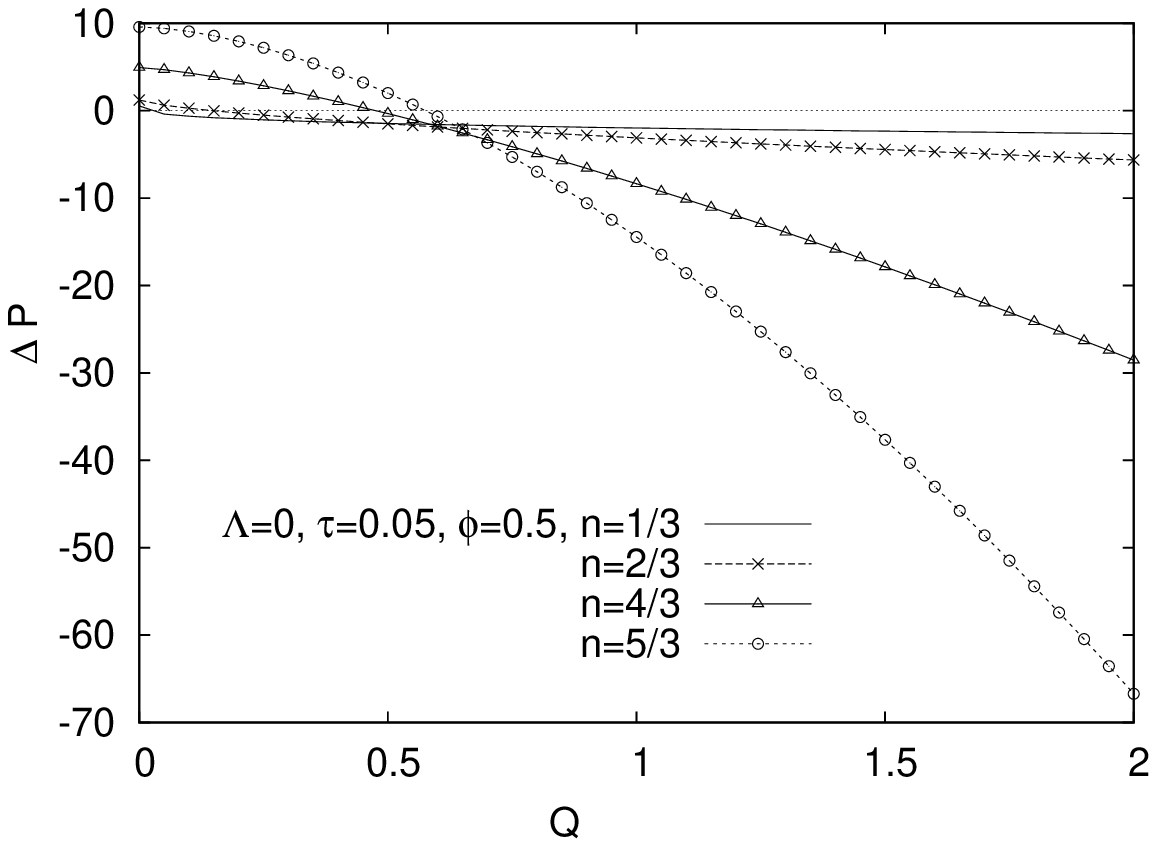}
\\$~~~~~~~~~~~~~~~~~~~~(c) ~~~~~~~~~~~~~~~~~~~~~~~~~~~~~~~~~~~~~~~~~~~~~~~~~~~~~~~~~(d)~~~~~~~~~~~~~~~~~~~~~~~~~~~~~~~~~~~~~~~~~$\\
\includegraphics[width=3.5in,height=1.8in]{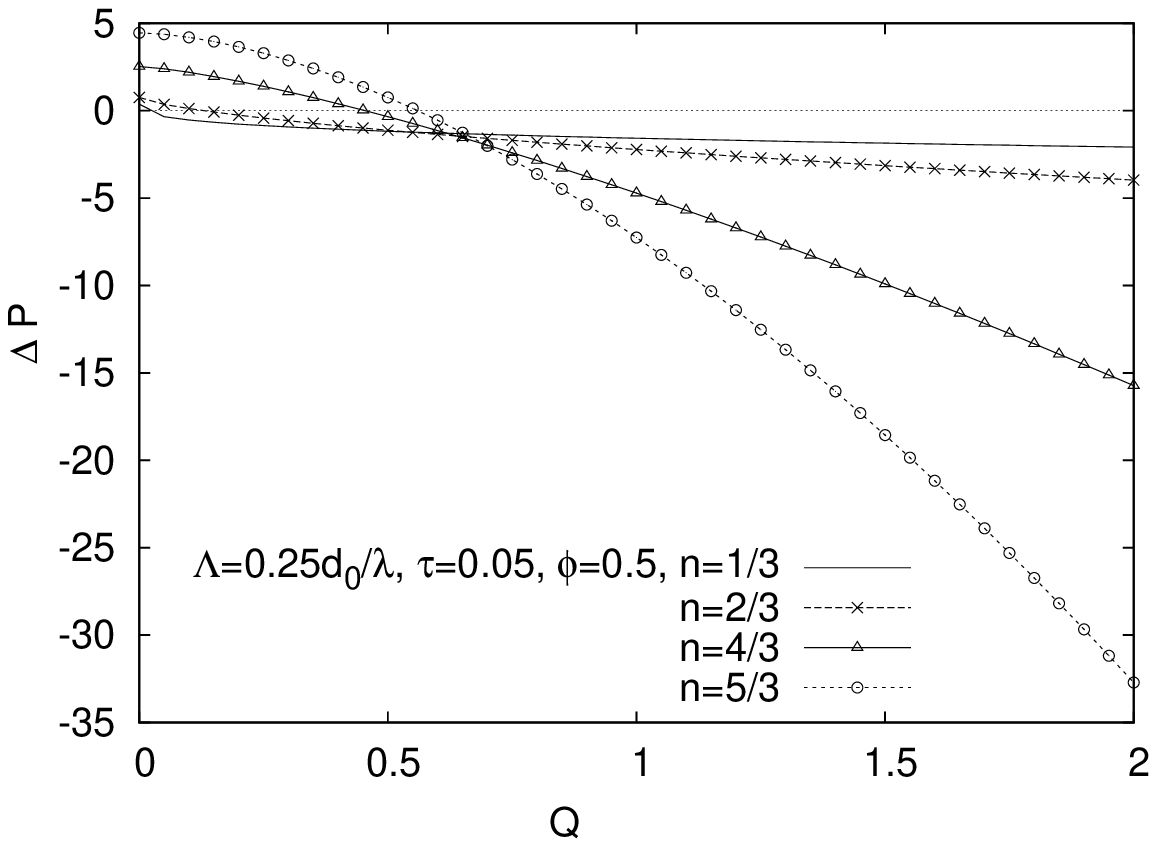}\includegraphics[width=3.5in,height=1.8in]{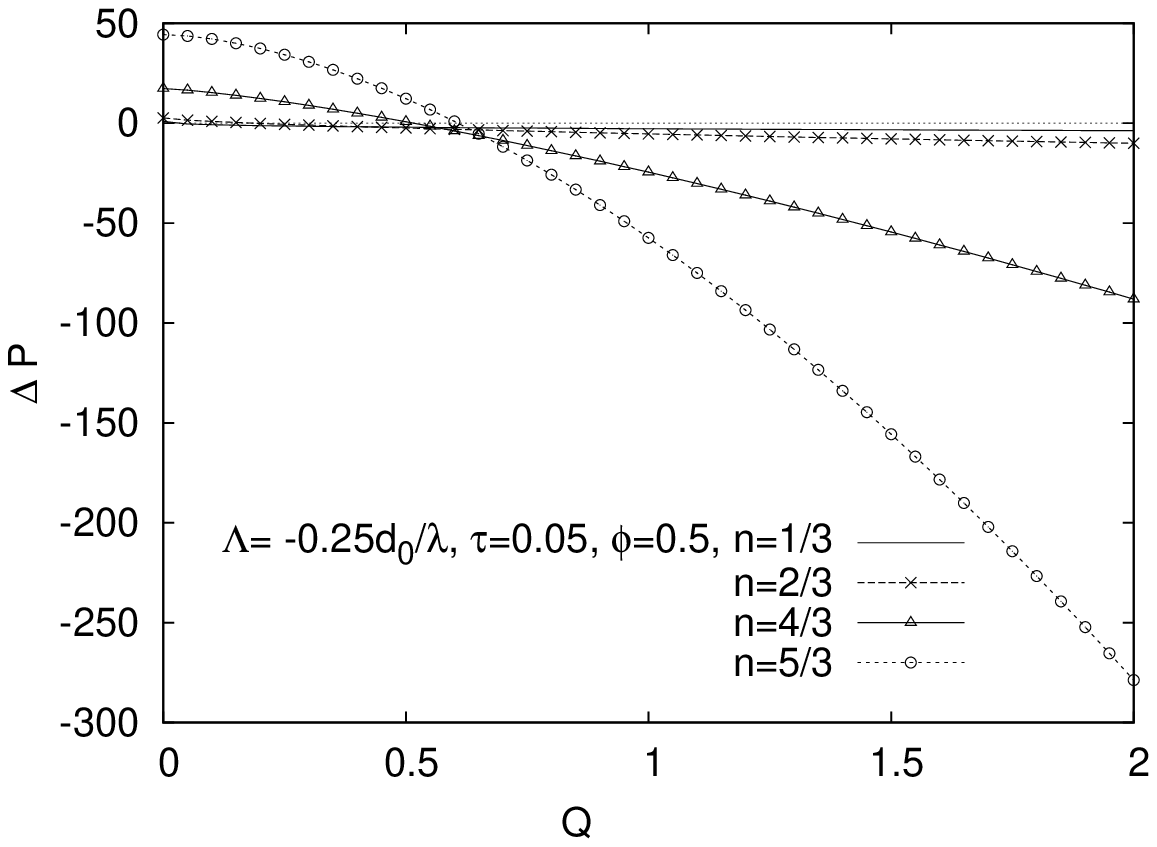}
\\$~~~~~~~~~~~~~~~~~~~~(e) ~~~~~~~~~~~~~~~~~~~~~~~~~~~~~~~~~~~~~~~~~~~~~~~~~~~~~~~~~(f)~~~~~~~~~~~~~~~~~~~~~~~~~~~~~~~~~~~~~~~~~$\\
\includegraphics[width=3.5in,height=1.8in]{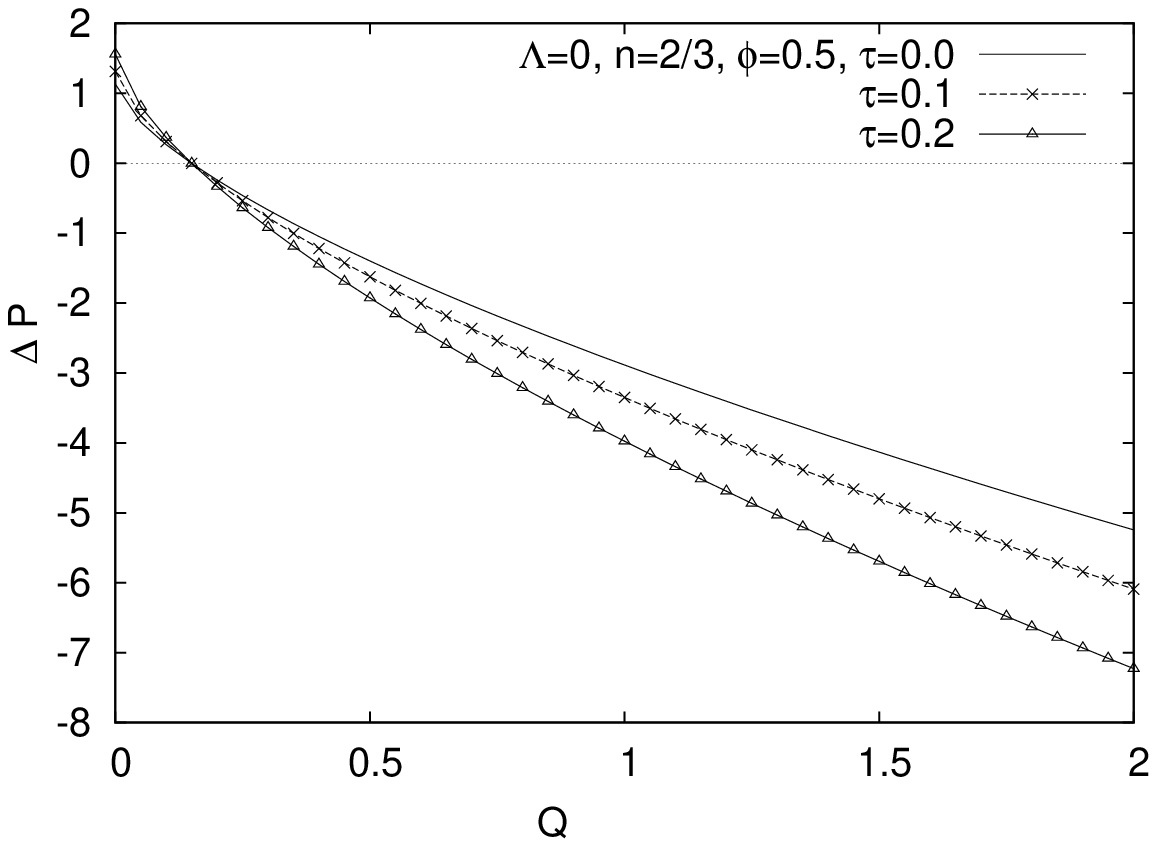}\includegraphics[width=3.5in,height=1.8in]{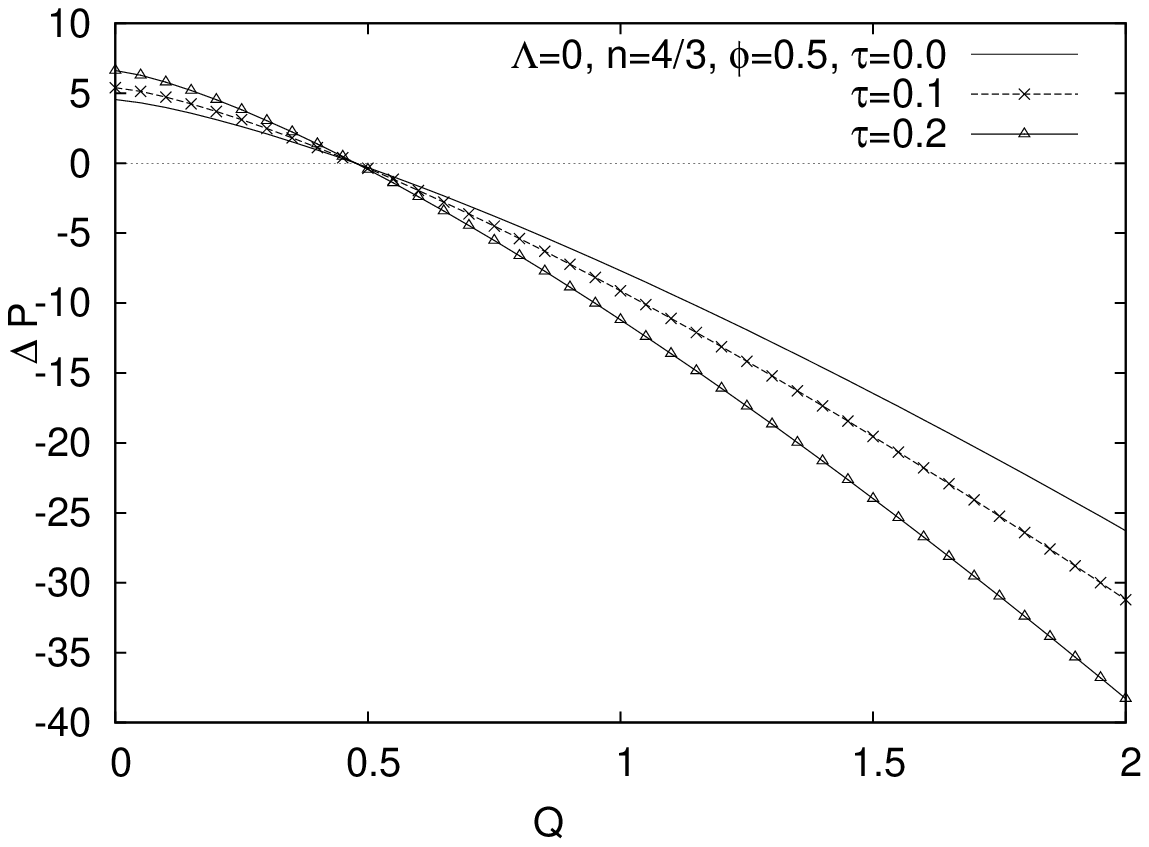}
\\$~~~~~~~~~~~~~~~~~~~~(g) ~~~~~~~~~~~~~~~~~~~~~~~~~~~~~~~~~~~~~~~~~~~~~~~~~~~~~~~~~(h)~~~~~~~~~~~~~~~~~~~~~~~~~~~~~~~~~~~~~~~~~$\\
\caption{Pressure rise versus flow rate}
\label{jam_pump4.1.2.1-4.3.2.4}
\end{figure}

Since peristaltic transport of a fluid is associated with the concept
of mechanical pumping, it is worthwhile to examine the pumping
performance under the purview of the present study. The average
pumping flow rate corresponding to a given pressure against which the
pumping action takes place gives a measure of peristaltic pumping
performance. By using lubrication theory, it has been observed by
Shapiro et al. \cite{Shapiro} that in the infinite tube model, the
flow rate averaged over one wave varies linearly with pressure
difference.  However, the existence of a peripheral layer (Newtonian
fluid), makes the relationship non-linear \cite{Brassur}. The present
study being concerned with the peristaltic transport of a
non-Newtonian fluid, the relationship between the pressure difference
and the mean flow rate is found to be non-linear (cf. Figs
\ref{jam_pump4.1.2.1-4.3.2.4}(b-h)), although the study is based upon
the consideration that the channel length is an integral multiple of
the wave length. This has been the observation both for converging and
diverging channels for all values of the fluid index n ($\ne 1$). In
the Newtonian case, however, the relation is linear (cf.
\ref{jam_pump4.1.2.1-4.3.2.4}(a)) and this is in conformity to the
observation reported by previous investigators \cite{Takabatake,Li}.

The plots in Figs. \ref{jam_pump4.1.2.1-4.3.2.4} show that the mean
flow rate, $Q$ increases as $\Delta P$ decreases. Figs. \ref{jam_pump4.1.2.1-4.3.2.4}(b) and
\ref{jam_pump4.1.2.1-4.3.2.4}(c) reveal that the pumping region
increases with the increase in the value of the amplitude ratio $\phi$
for both shear thinning and shear thickening
fluids. Figs. \ref{jam_pump4.1.2.1-4.3.2.4}(d-f) depict the influence
of the rheological parameter `n' on the pumping performance in
uniform/diverging/converging channels. It is important to note that
for all types of channels, the pumping region ($\Delta P>0$)
significantly increases with the increase in the value of 'n'. In the
co-pumping region ($\Delta P<0$) the pressure rise decreases when
$Q$ exceeds a certain value. From
Figs. \ref{jam_pump4.1.2.1-4.3.2.4}(g) and
\ref{jam_pump4.1.2.1-4.3.2.4}(h), we find that $Q$ is not
significantly affected by $\tau$ for free pumping case. Moreover, when
$Q$ exceeds a certain critical limit (equal to the value of
$Q$ in the free pumping case), pressure rise decreases with the
increase in $\tau$. We further find that for both shear thinning and
shear thickening fluids, pumping region increases with $\tau$
increasing; however, the effect of $\tau$ is greater in the case of
shear thickening fluids.

\begin{figure}
\includegraphics[width=3.6in]{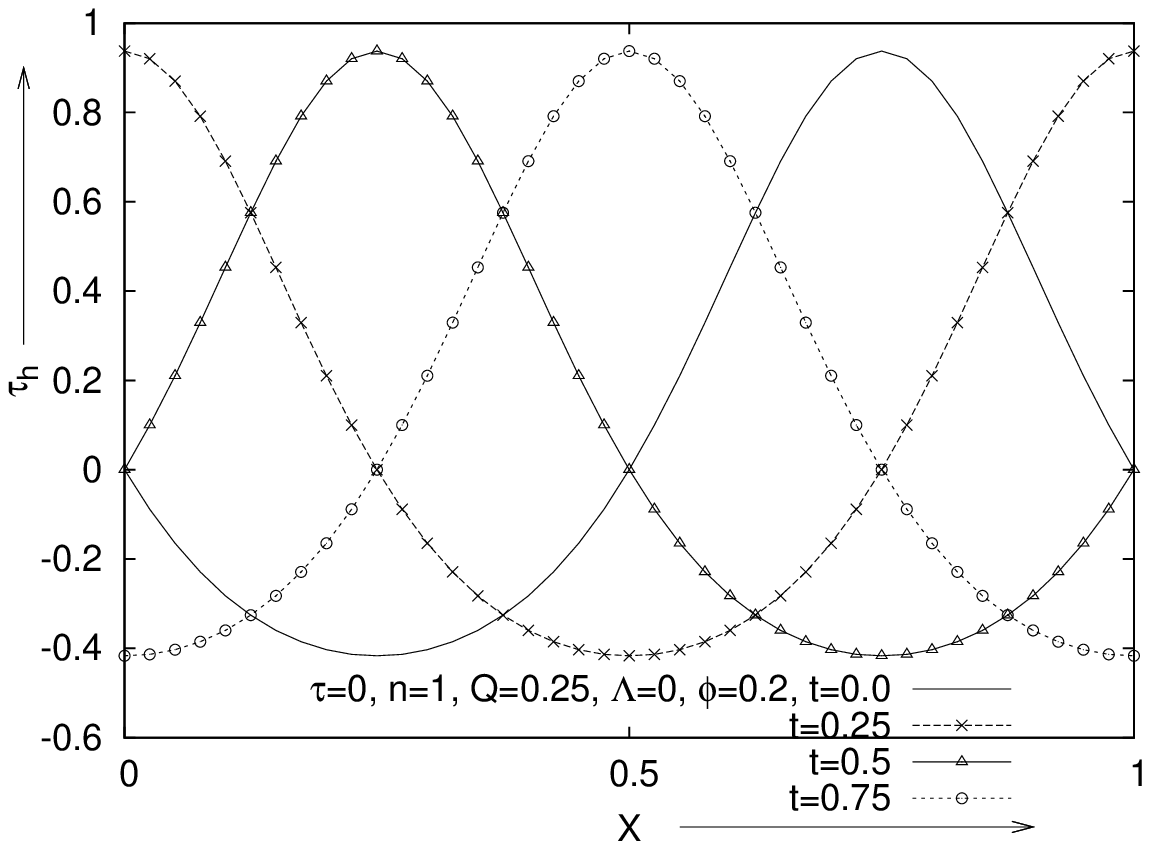}\includegraphics[width=3.6in]{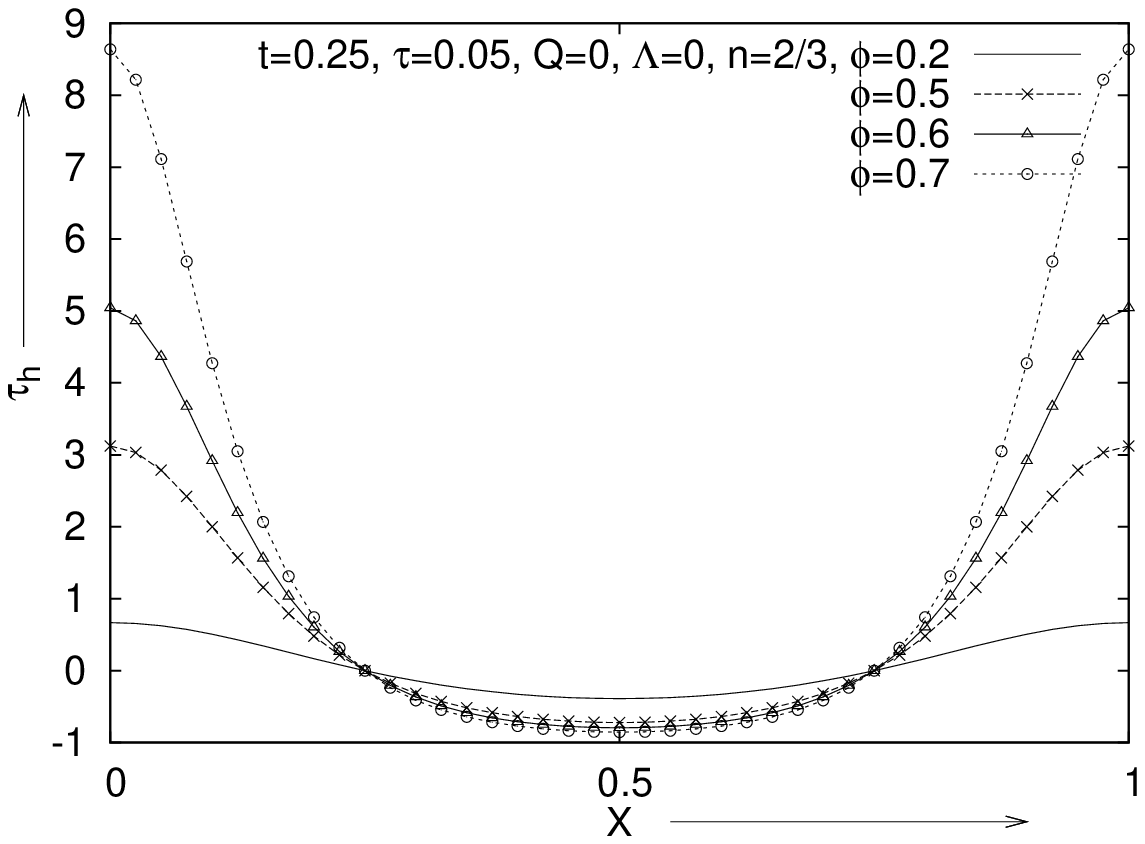}
\\$~~~~~~~~~~~~~~~~~~~~(a) ~~~~~~~~~~~~~~~~~~~~~~~~~~~~~~~~~~~~~~~~~~~~~~~~~~~~~~~~~(b)~~~~~~~~~~~~~~~~~~~~~~~~~~~~~~~~~~~~~~~~~$\\
\includegraphics[width=3.6in]{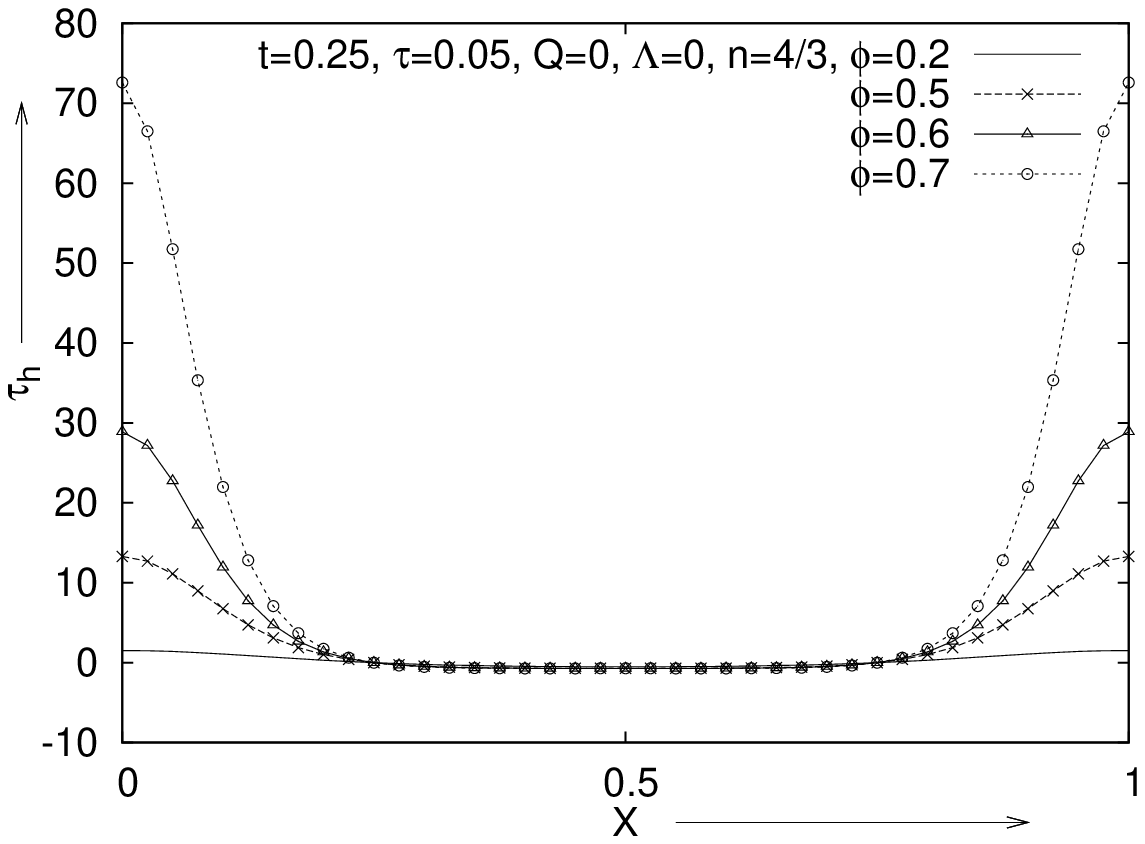}\includegraphics[width=3.6in]{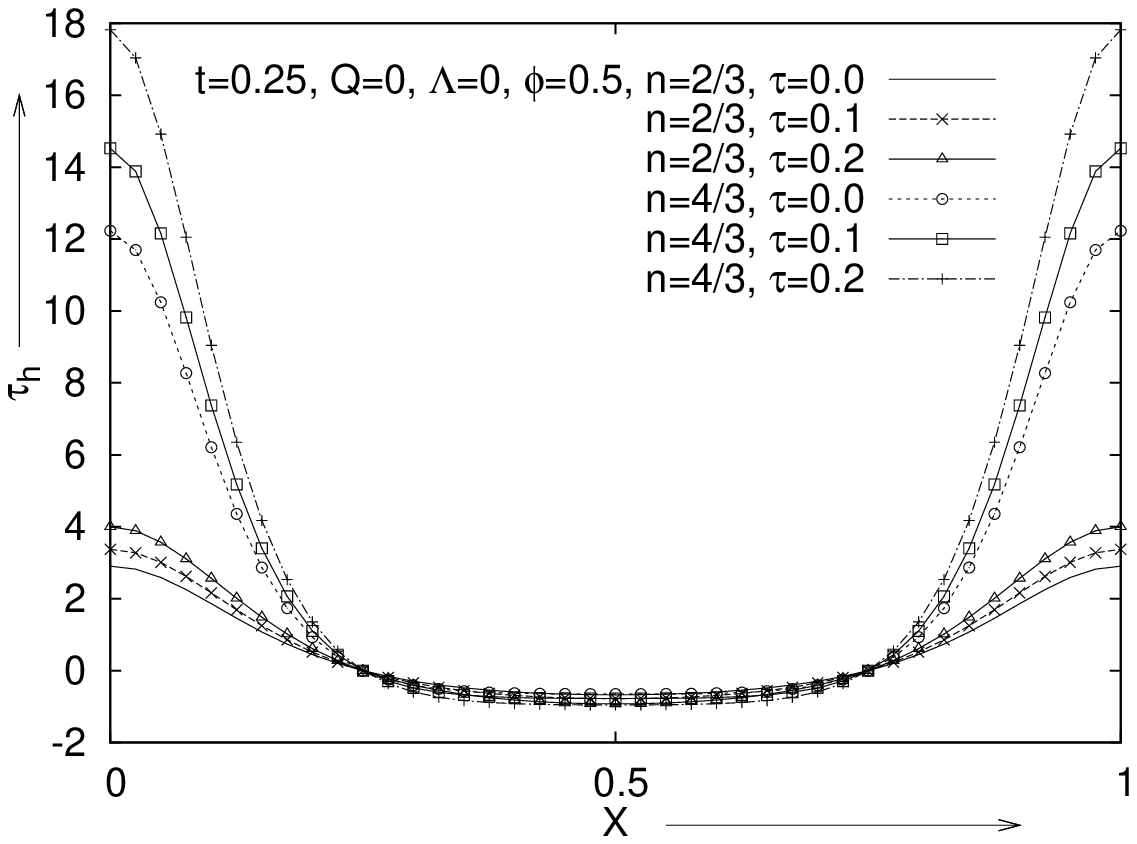}
\\$~~~~~~~~~~~~~~~~~~~~(c) ~~~~~~~~~~~~~~~~~~~~~~~~~~~~~~~~~~~~~~~~~~~~~~~~~~~~~~~~~(d)~~~~~~~~~~~~~~~~~~~~~~~~~~~~~~~~~~~~~~~~~$\\
\includegraphics[width=3.6in]{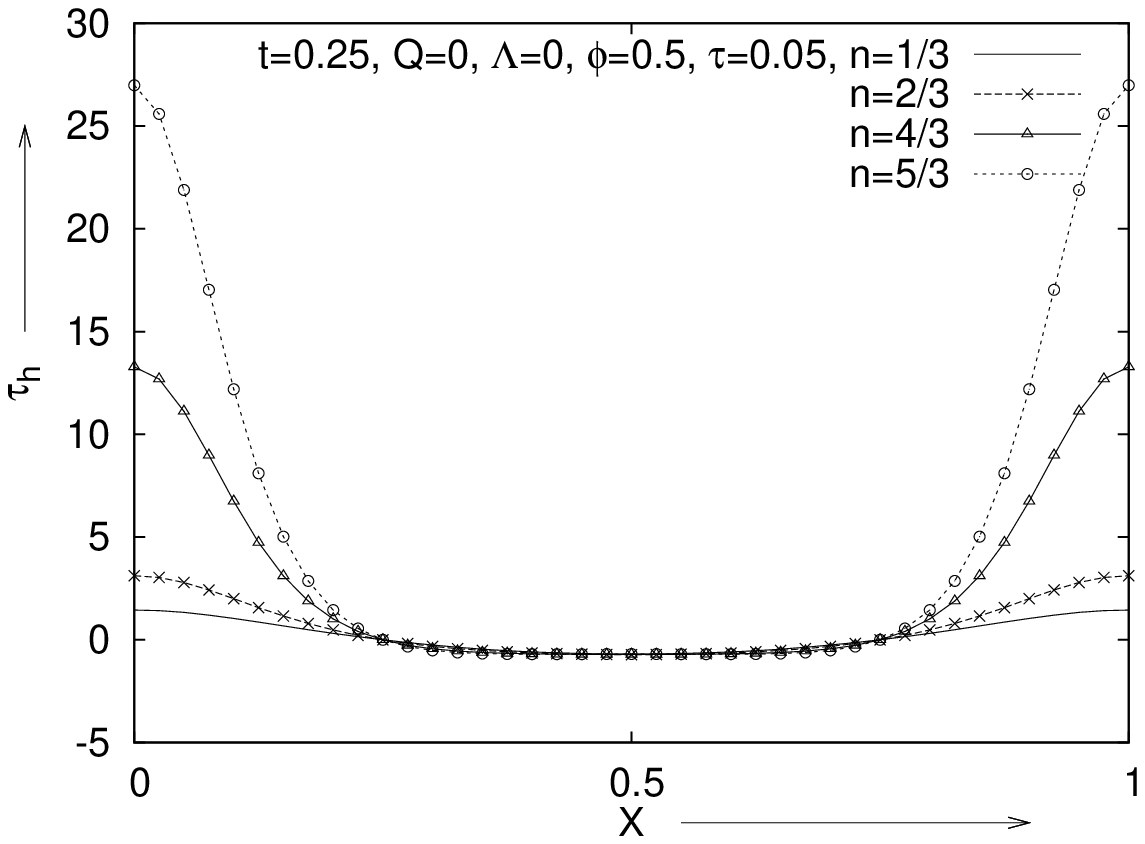}\includegraphics[width=3.6in]{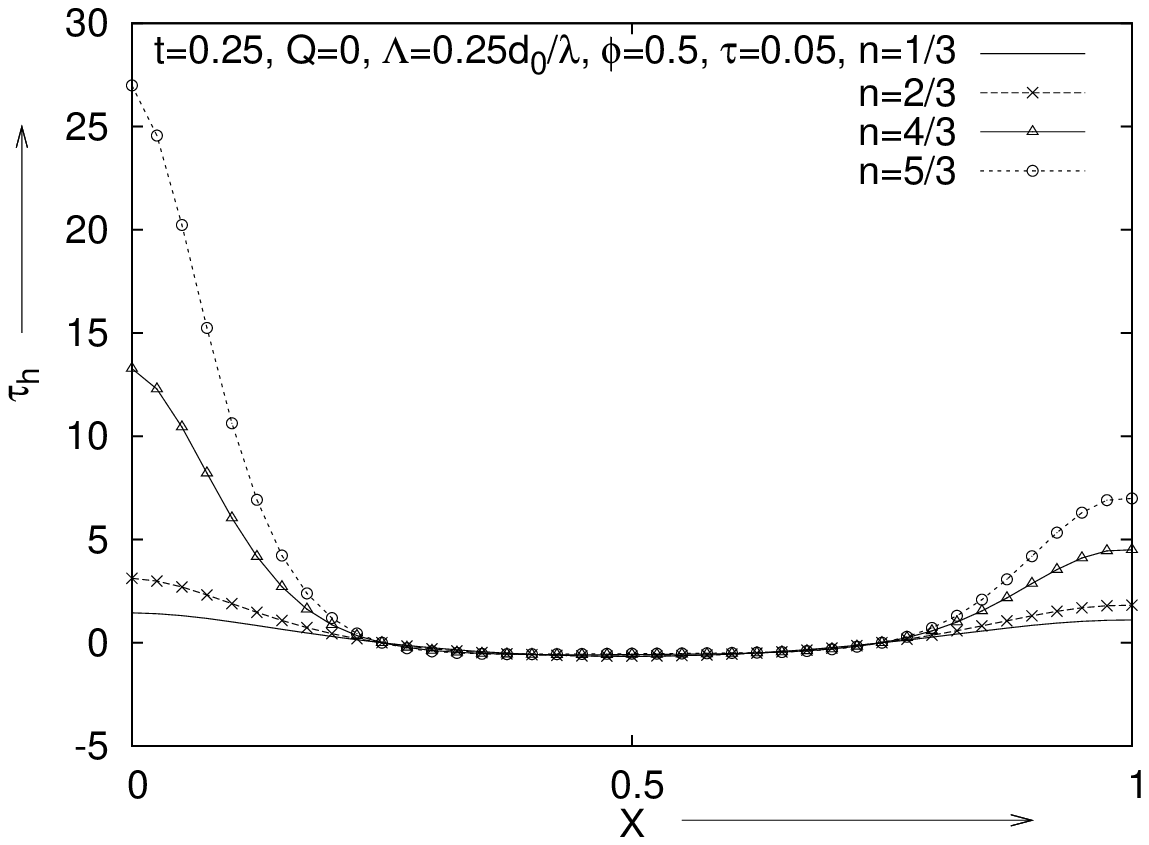}
\\$~~~~~~~~~~~~~~~~~~~~(e) ~~~~~~~~~~~~~~~~~~~~~~~~~~~~~~~~~~~~~~~~~~~~~~~~~~~~~~~~~(f)~~~~~~~~~~~~~~~~~~~~~~~~~~~~~~~~~~~~~~~~~$\\
\end{figure}

\begin{figure}
\includegraphics[width=3.6in]{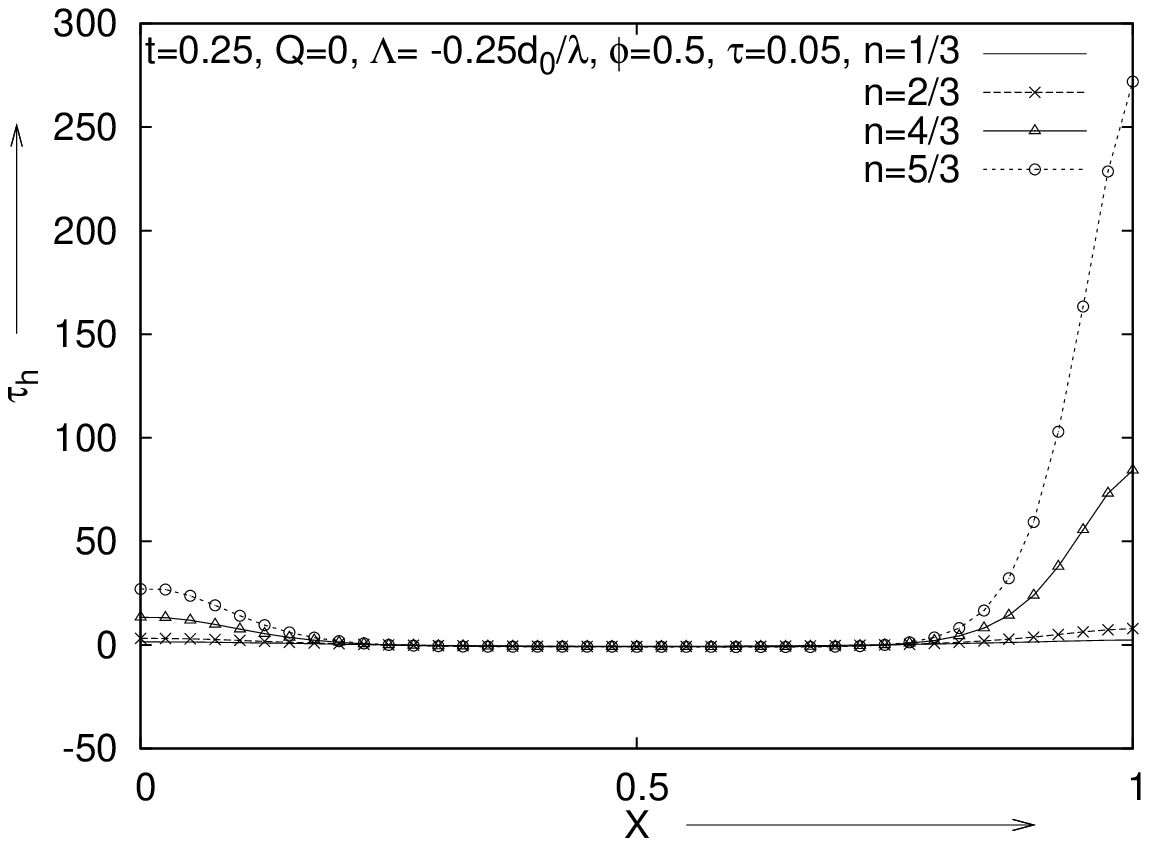}\includegraphics[width=3.6in]{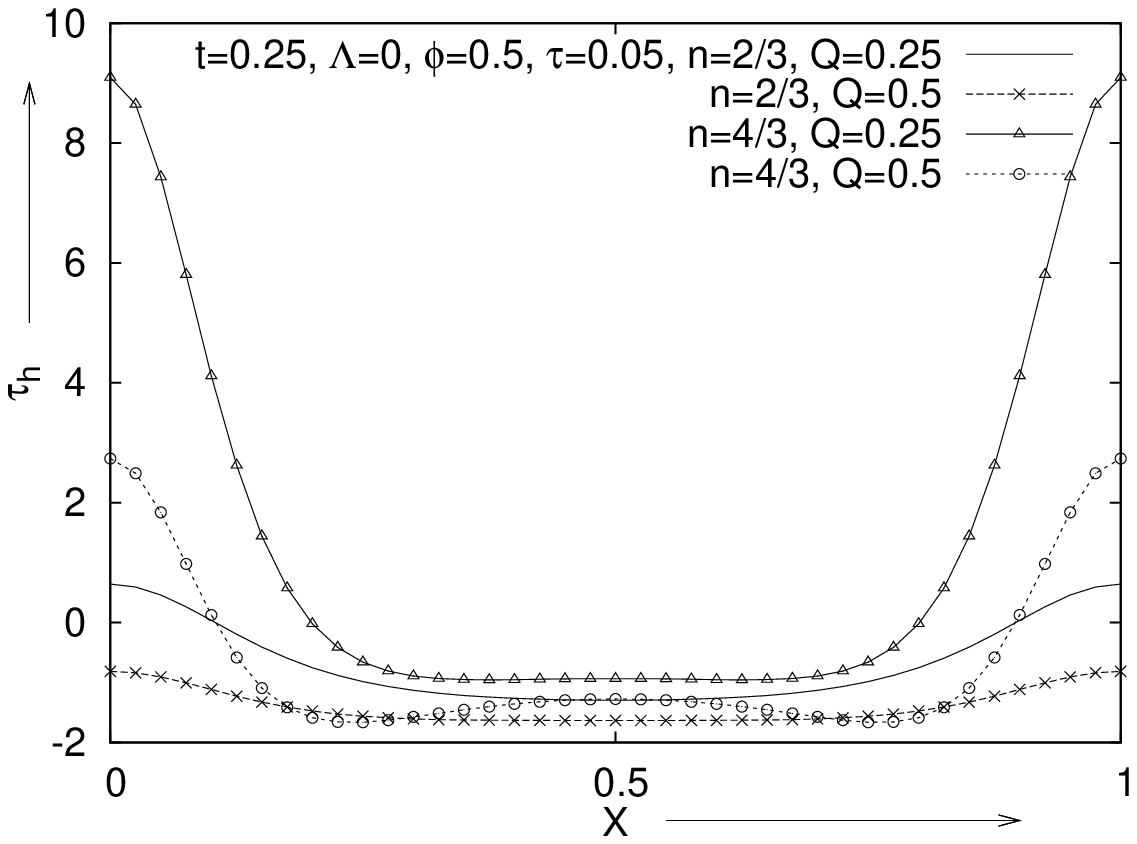}
\\$~~~~~~~~~~~~~~~~~~~~(g) ~~~~~~~~~~~~~~~~~~~~~~~~~~~~~~~~~~~~~~~~~~~~~~~~~~~~~~~~~(h)~~~~~~~~~~~~~~~~~~~~~~~~~~~~~~~~~~~~~~~~~$\\
\includegraphics[width=3.6in]{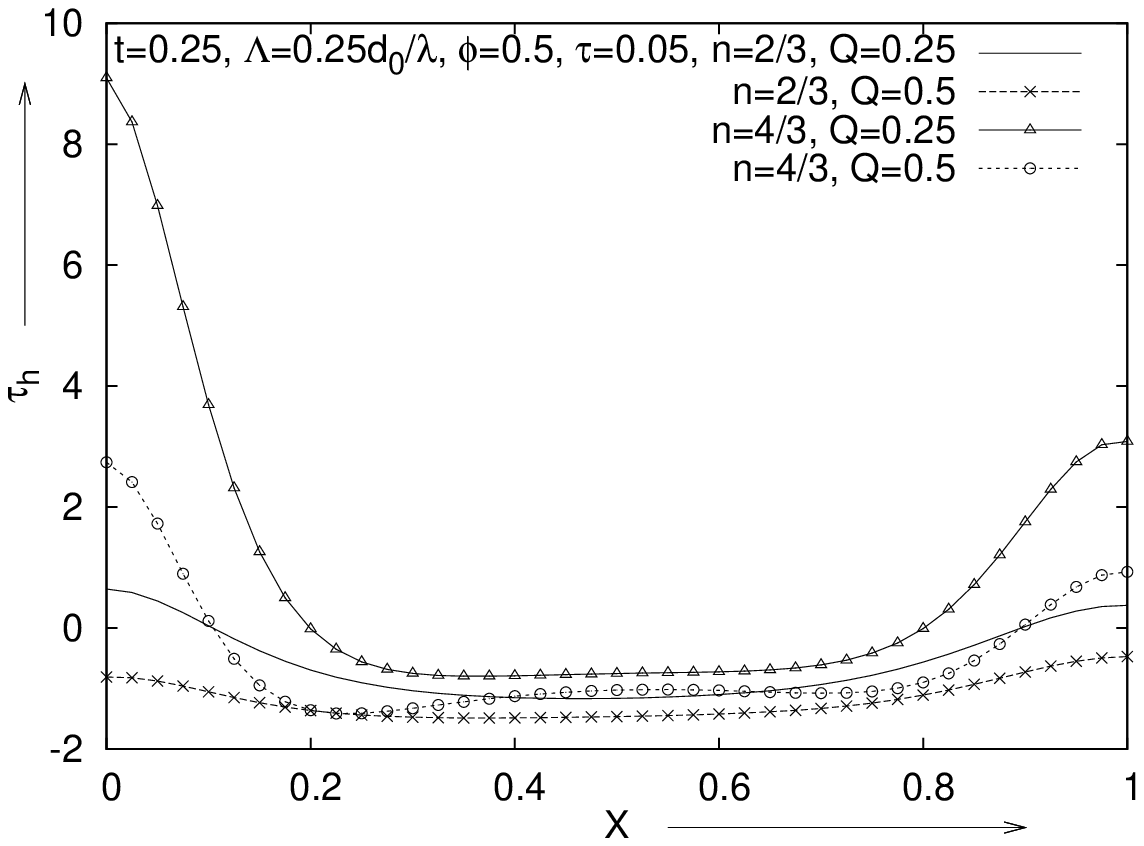}\includegraphics[width=3.6in]{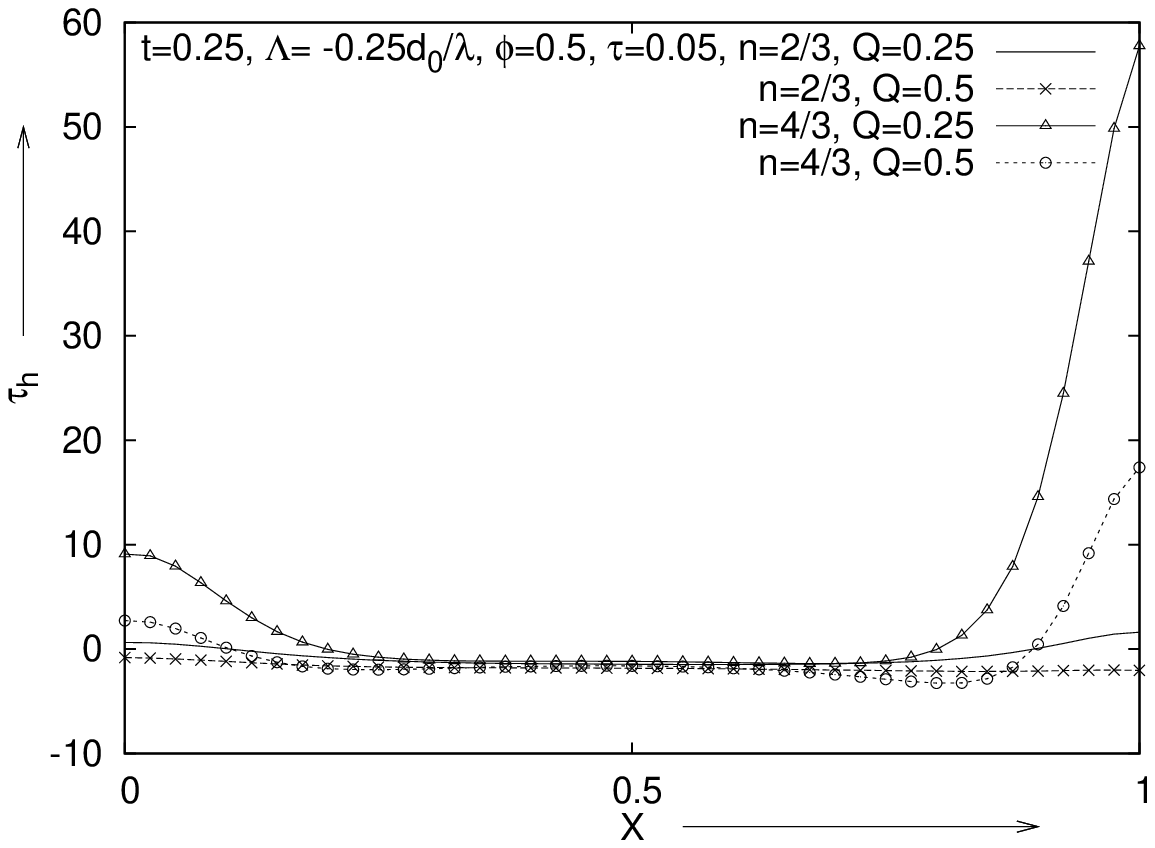}
\\$~~~~~~~~~~~~~~~~~~~~(i) ~~~~~~~~~~~~~~~~~~~~~~~~~~~~~~~~~~~~~~~~~~~~~~~~~~~~~~~~~(j)~~~~~~~~~~~~~~~~~~~~~~~~~~~~~~~~~~~~~~~~~$\\
\caption{Wall shear stress distribution for different cases}
\label{jam_shear4.1-4.5.9}
\end{figure}

\subsection{Wall Shear Stress}
It is known that when the shear stress generated on the wall of a
blood vessel exceeds a certain limit, there is a possibility that it
can cause damage to the constituents of blood. Moreover, the magnitude
of wall shear stress plays an important role in the molecular
convective process at high Prandtl number/Schmidt number
\cite{Higdon}. Because of this reason, investigation of shearing
stress deserves special attention in the hemodynamical flow of blood
in arteries. Figs. \ref{jam_shear4.1-4.5.9} depict the wall shear
stress distribution under varied
conditions. Fig. \ref{jam_shear4.1-4.5.9}(a) gives the distribution of
wall shear stress at four specially chosen instants of time during one
complete wave period. This figure shows that at each of these time
instants, there exist two peaks in the wall shear stress distribution,
with a gradual ramp in between; however, negative peak of wall shear
stress, $\tau_{min}$ is not as large as the maximum wall shear stress,
$\tau_{max}$. It may be observed that the transition from $\tau_{min}$
to $\tau_{max}$ takes place in the zone between the maximal height and
the minimum height of the channel
(cf. Fig. \ref{jam_shear4.1-4.5.9}(a)). At the point of maximum
occlusion, the wall shear stress as well as the pressure is
maximum. The pressure gradient to the left of this point is positive
and hence the local instantaneous flow will take place towards the
left of $\tau_{max}$. This may be responsible for a number of
consequences. For example, if the rate of shear at the crest is quite
high, a dissolving wavy wall will have a tendency to level out. Also,
some chemical reaction between the wall material and the constituents
of blood is likely to set in. Owing to the deposition of the products
of the chemical reaction , wall amplitude increases at a rapid
rate. This may lead to clogging of the blood vessel.
\begin{figure}
\includegraphics[width=3.6in]{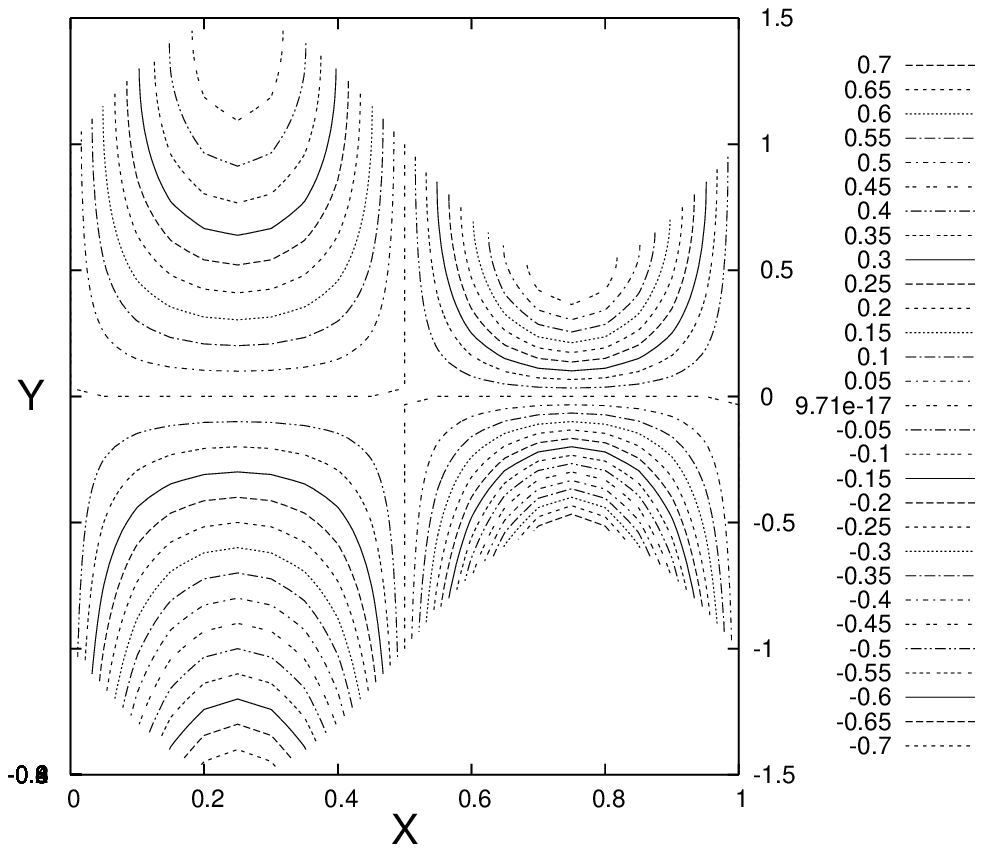}\includegraphics[width=3.6in]{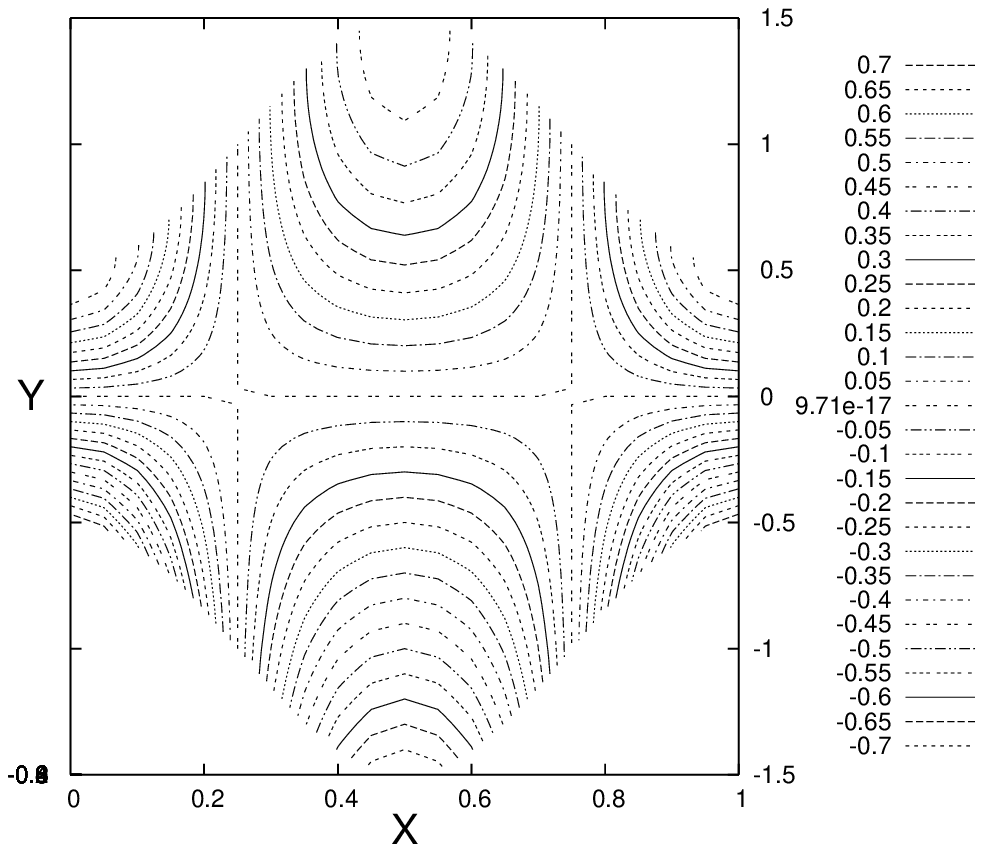}
\\$~~~~~~~~~~~~~~~~~~~~(a)~~\rm{for}~t=0.0~~~~~~~~~~~~~~~~~~~~~~~~~~~~~~~~~~~~~~~~~~~~~(b)~~\rm{for}~t=0.25~~~~~~~~~~~~~~~~~~~~~~~~~~~~~~~~$\\
\includegraphics[width=3.6in]{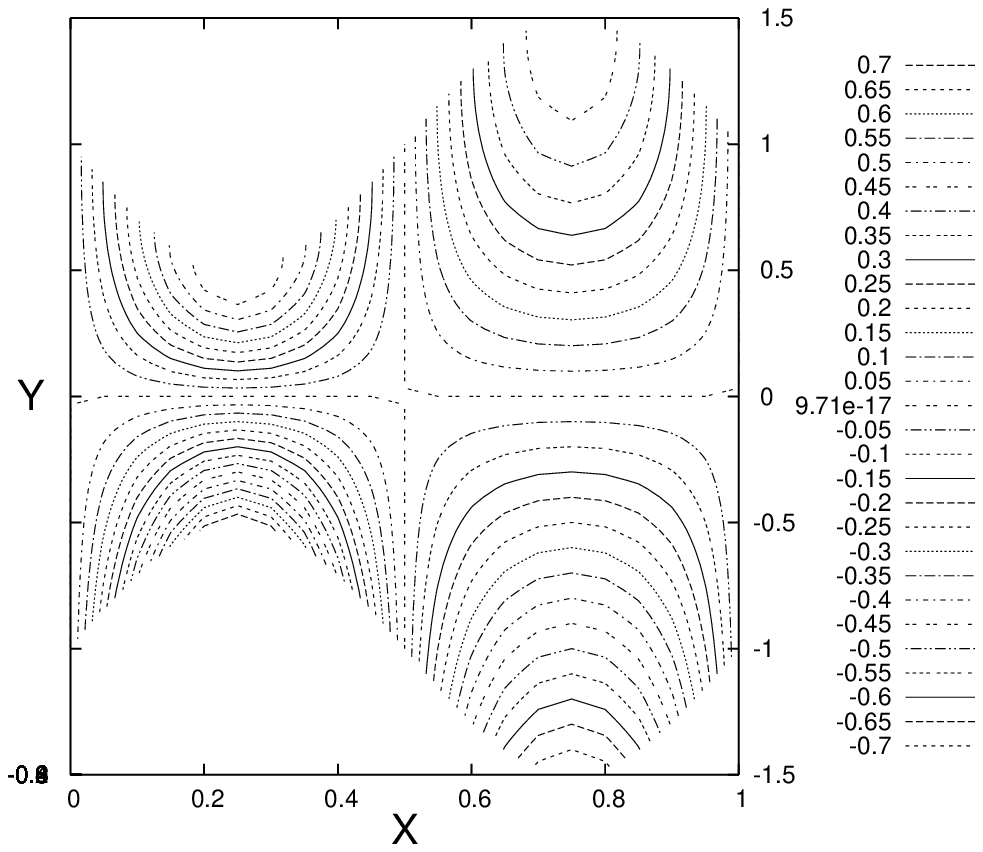}\includegraphics[width=3.6in]{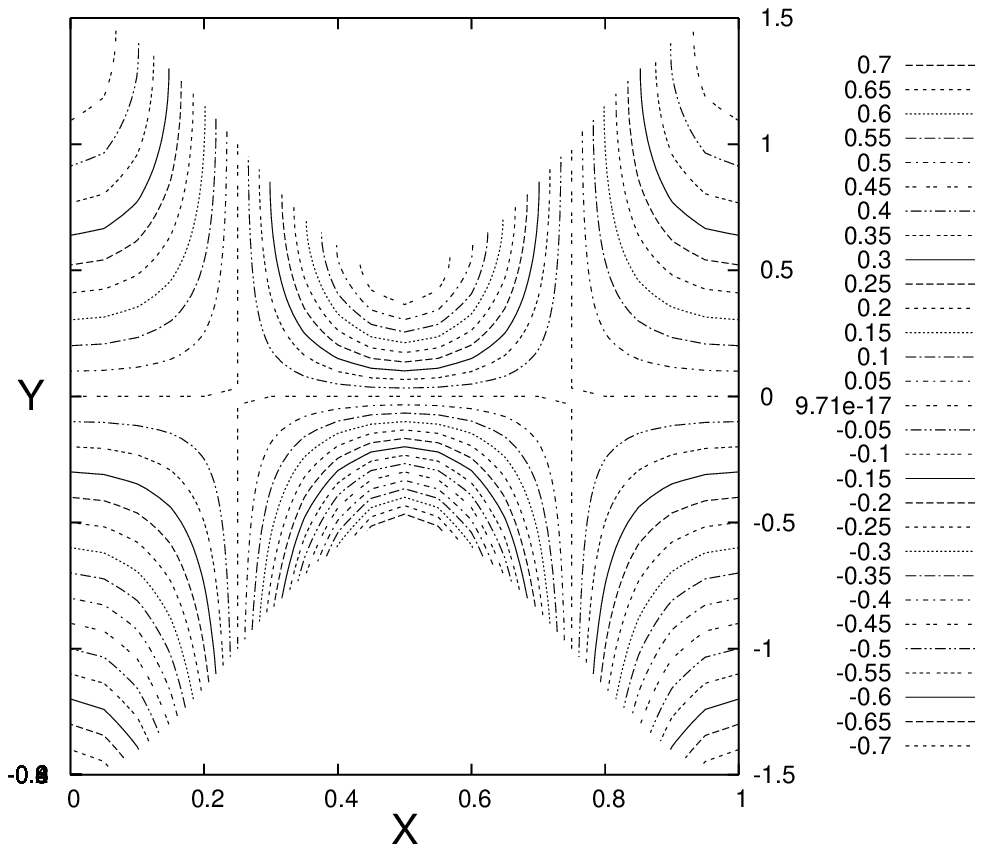}
\\$~~~~~~~~~~~~~~~~~~~~(c)~~\rm{for}~t=0.5~~~~~~~~~~~~~~~~~~~~~~~~~~~~~~~~~~~~~~~~~~~~~(d)~~\rm{for}~t=0.75~~~~~~~~~~~~~~~~~~~~~~~~~~~~~~~~$\\
\caption{Streamline patterns for peristaltic flow of a Newtonian fluid at different instants of time ($Q=0$, $\Lambda=0$, $\tau=0$, $\phi=0.5$).}
\label{jam_stline4.1.1-4.1.4}
\end{figure}

\begin{figure}
\centering
\includegraphics[width=3.4in]{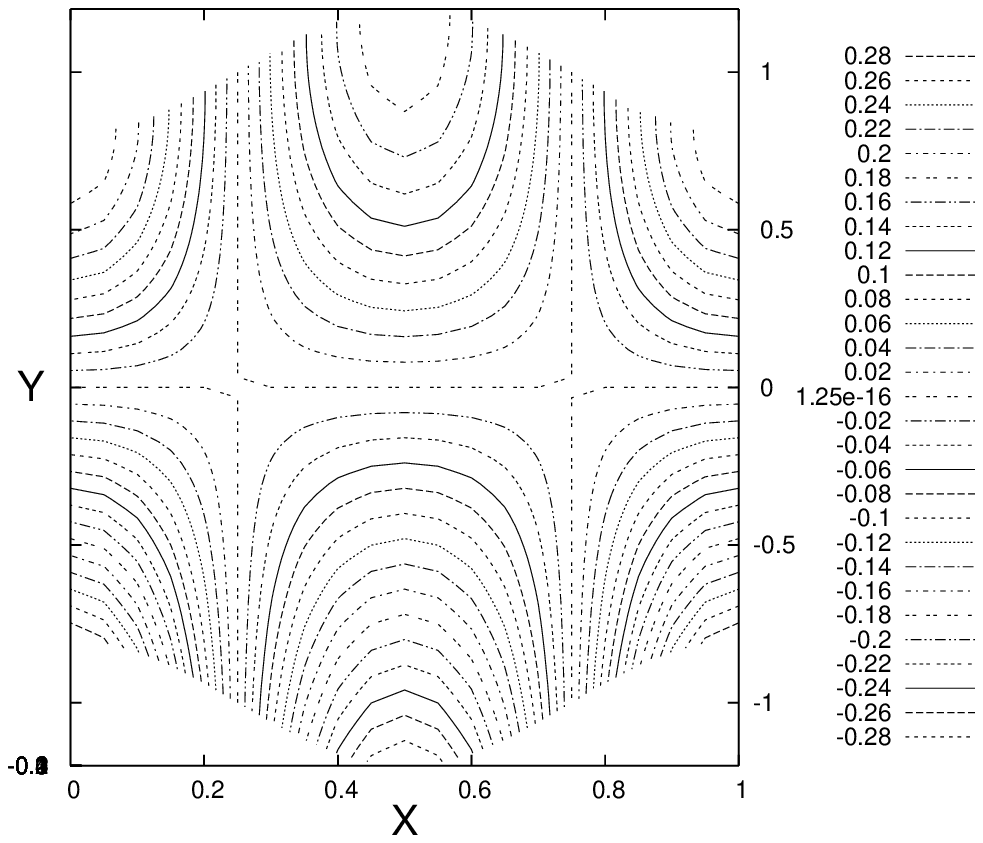}\includegraphics[width=3.4in]{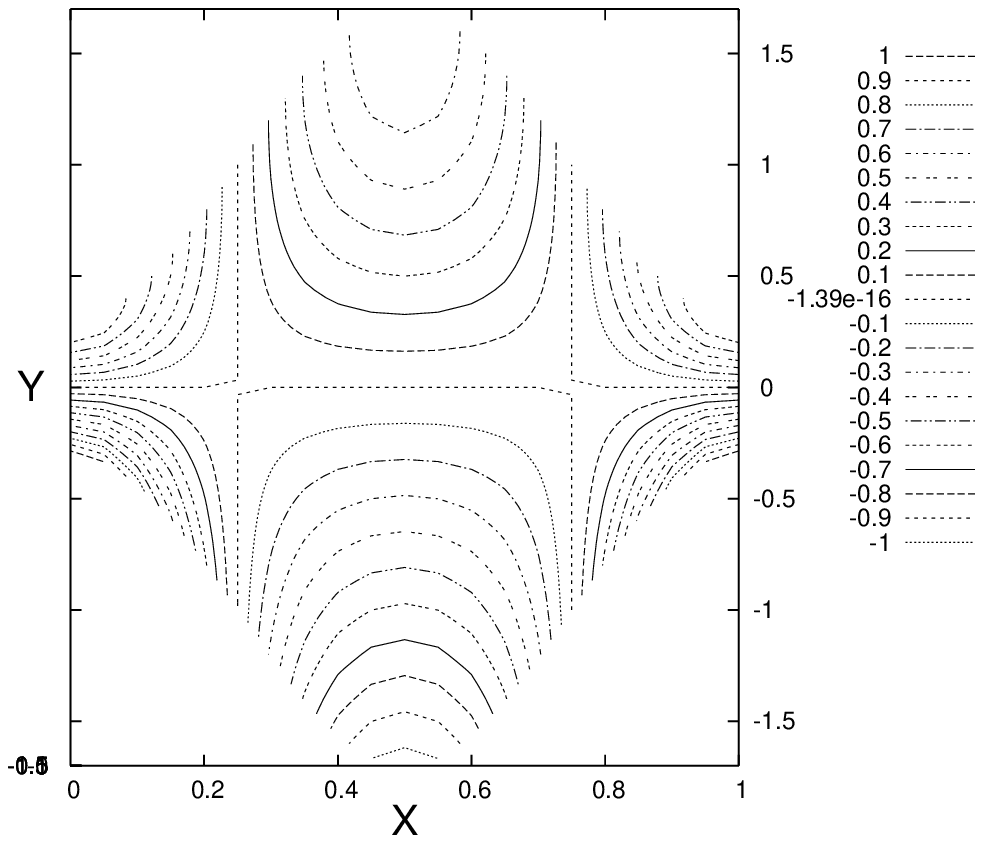}
\\$~~~~~~~~~~~~~~~~~~~~(a)~~~\phi=0.2~~~~~~~~~~~~~~~~~~~~~~~~~~~~~~~~~~~~~~~~~~~~~(b)~~~\phi=0.7~~~~~~~~~~~~~~~~~~~~~~~~~~~~~~~~$\\
\includegraphics[width=3.4in]{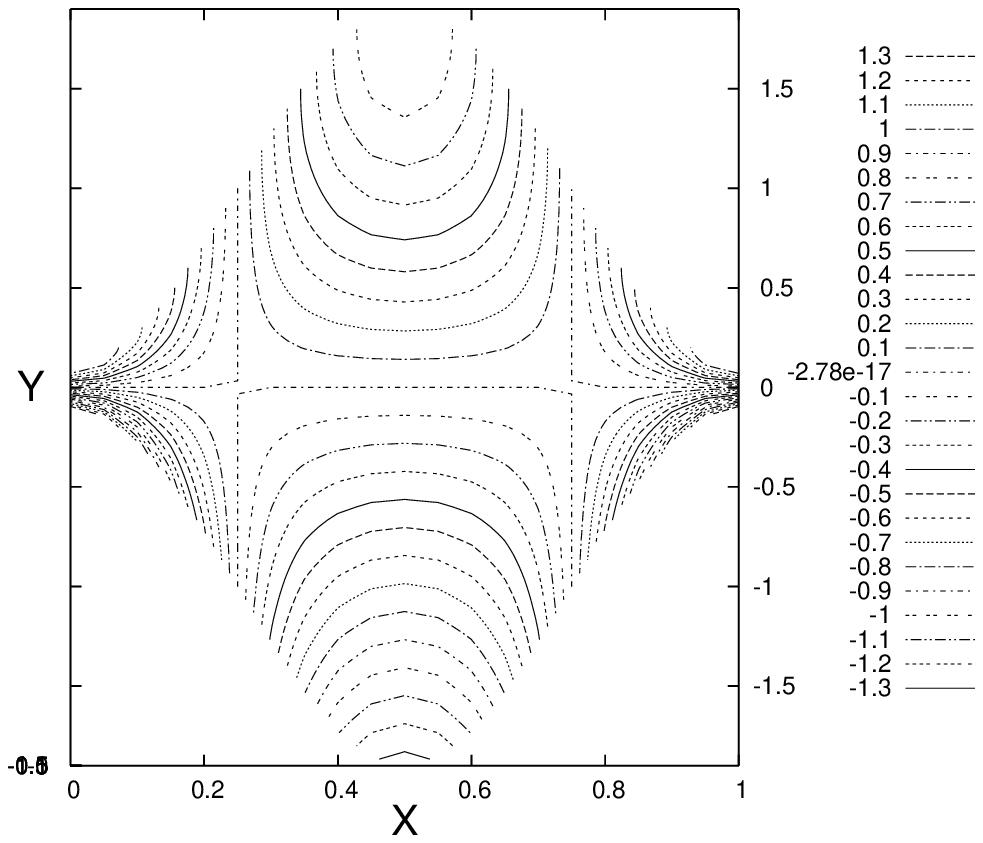}
\\(c)~~$\phi=0.9$
\caption{Streamline patterns in the case of peristaltic flow for
different values of $\phi$ when $n=1,~t=0.25~,\tau=0,~\Lambda=0,~Q=0$}
\label{jam_stline4.2.1-4.2.3}
\end{figure} 
In other regions, for wall shear stress distribution curves, the peaks
on both sides of $\tau_{max}$ are small. Thus the local instantaneous
flow will occur in the direction of the peristaltic wave
propagation. Of course, when averaged over one wave period, net effect
may be looked upon as peristaltic transport in the direction of wave
propagation. It may be observed from
Figs. \ref{jam_shear4.1-4.5.9}(b-c) that in the contracting region
where occlusion takes place, there is a significant increase in the
wall shear stress due to an increase in the value of $\phi$. This
applies for both shear thinning and shear thickening cases. However,
in the expanding region, with the increase in $\phi$, $\tau_{min}$
increases in magnitude for a shear thinning fluid; for a shear thickening
fluid, the effect is relatively
little. Fig. \ref{jam_shear4.1-4.5.9}(d) shows that $\tau_{max}$
increases with increase in $\tau$, but for $\tau_{min}$ the effect is
minimal.
\begin{figure}
\includegraphics[width=3.5in]{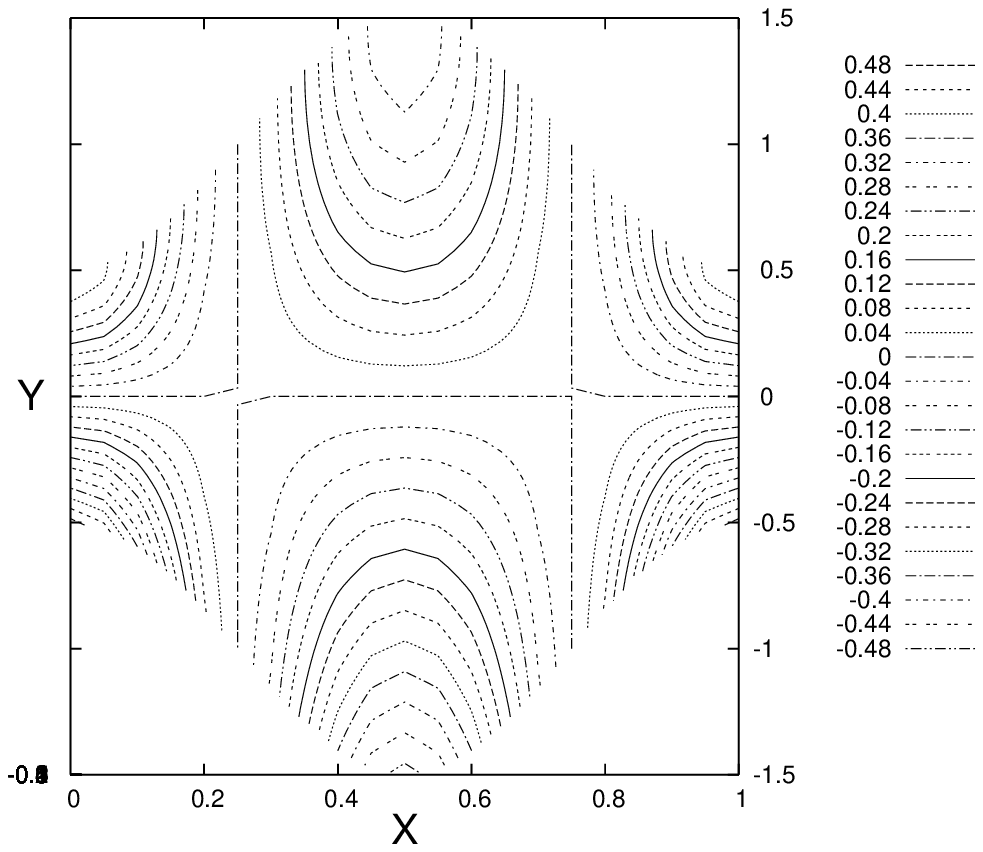}\includegraphics[width=3.5in]{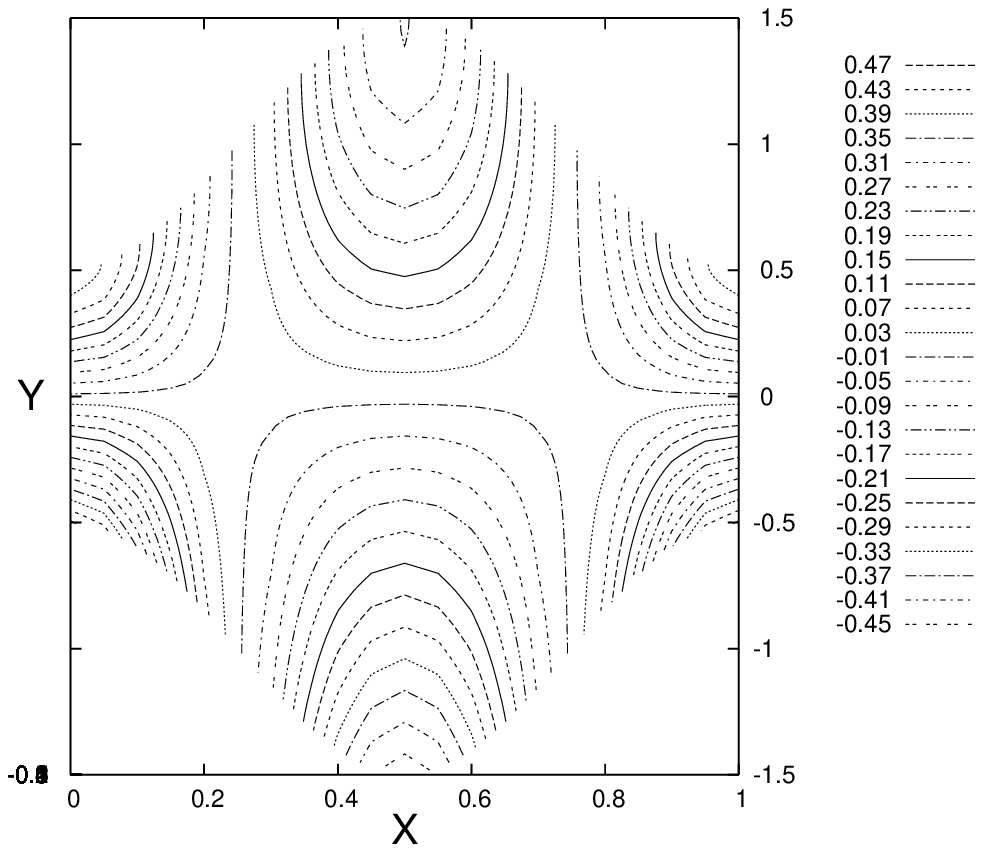}
\\$~~~~~~~~~~~~~~~~~~~~(a)~~n=2/3,~\tau=0.0~~~~~~~~~~~~~~~~~~~~~~~~~~~~~~~~~~~~~~~~~~~~~(b)~~n=2/3,~\tau=0.1~~~~~~~~~~~~~~~~~~~~~~~~~~~~~~~~$\\
\includegraphics[width=3.5in]{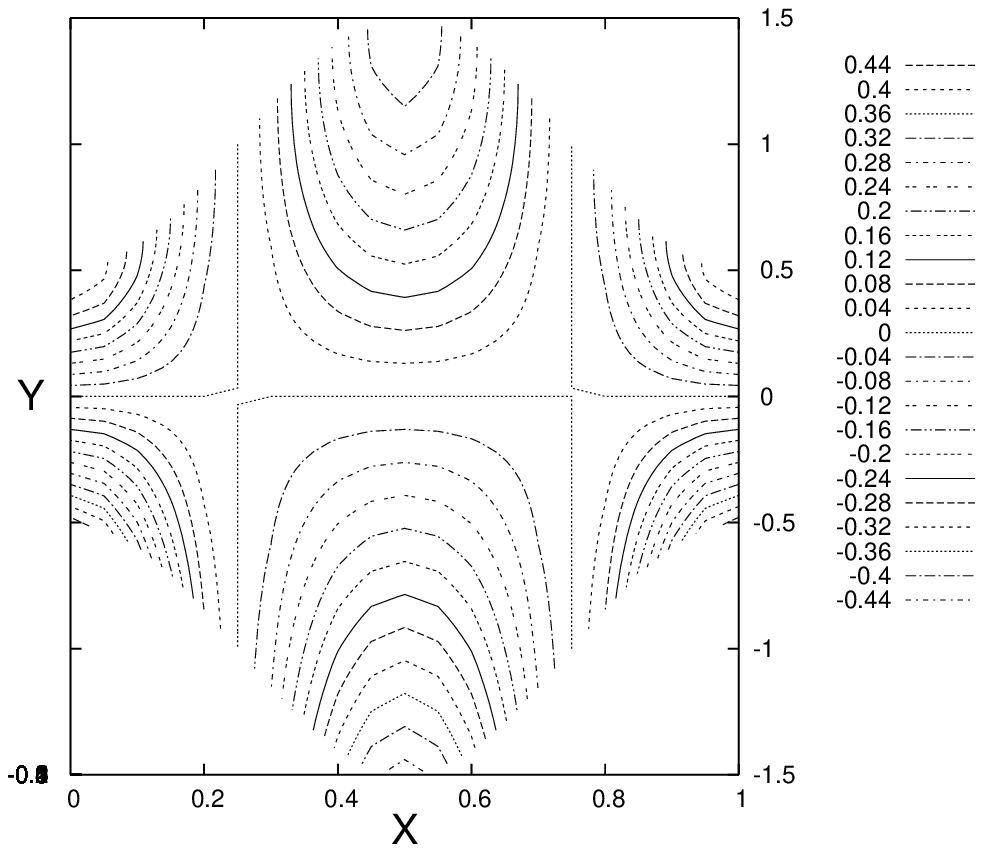}\includegraphics[width=3.5in]{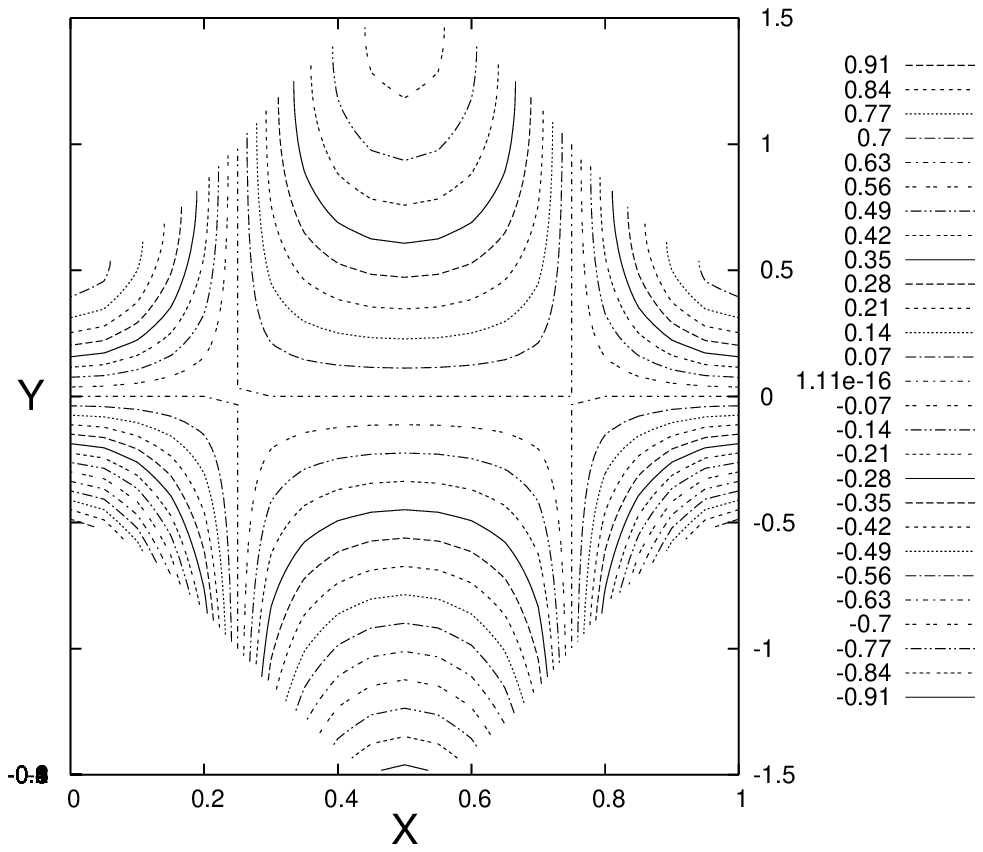}
\\$~~~~~~~~~~~~~~~~~~~~(c)~~n=2/3,~\tau=0.2~~~~~~~~~~~~~~~~~~~~~~~~~~~~~~~~~~~~~~~~~~~~~(d)~~n=4/3,~\tau=0.0~~~~~~~~~~~~~~~~~~~~~~~~~~~~~~~~$\\
\includegraphics[width=3.5in]{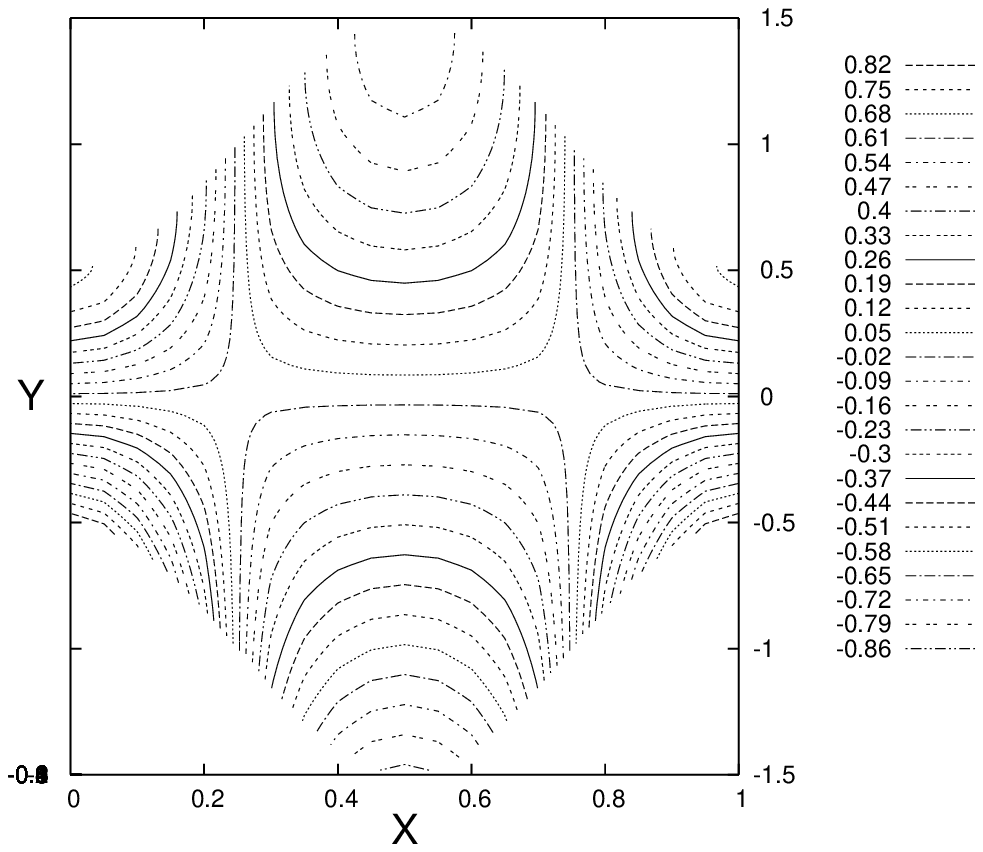}\includegraphics[width=3.5in]{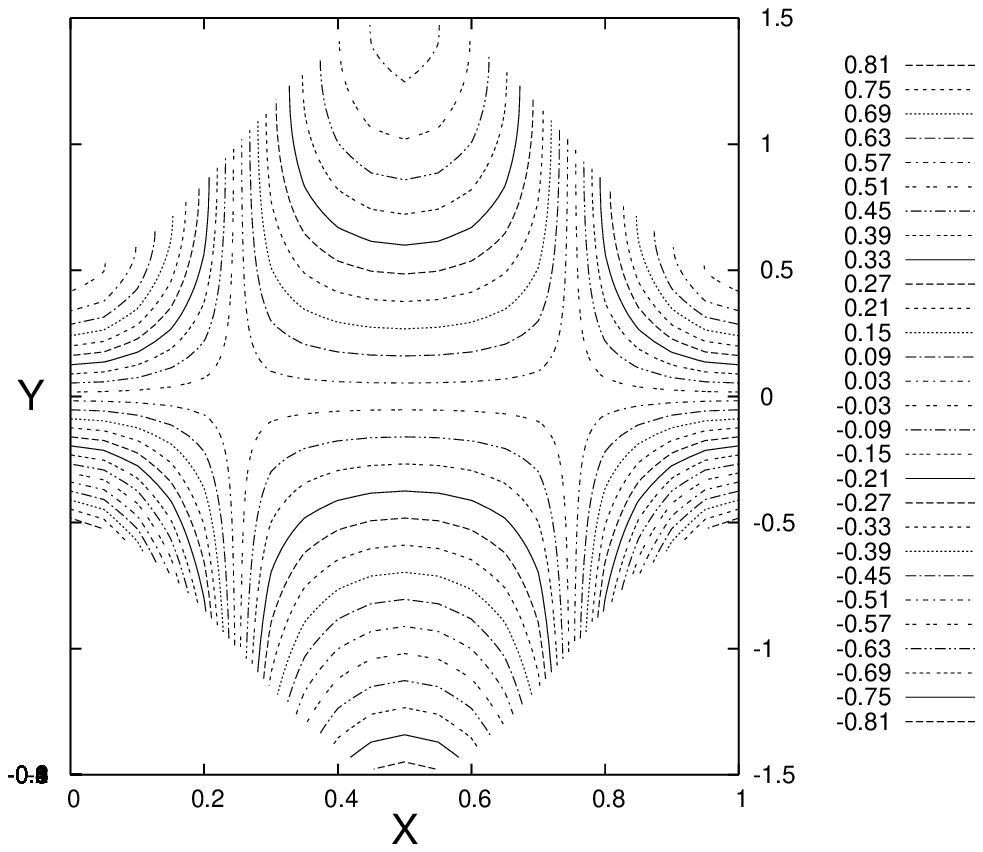}
\\$~~~~~~~~~~~~~~~~~~~~(e)~~n=4/3,~\tau=0.1~~~~~~~~~~~~~~~~~~~~~~~~~~~~~~~~~~~~~~~~~~~~~(f)~~n=4/3,~\tau=0.2~~~~~~~~~~~~~~~~~~~~~~~~~~~~~~~~$
\caption{Streamline patterns in the case of peristaltic flow of
  rheological fluid for different values of $\tau$ when $Q=0$,
  $t=0.25$, $\Lambda=0$, $\phi=0.5$}
\label{jam_stline4.3.1-4.3.6}
\end{figure} 
The effect of the rheological fluid index `n' on the distribution of
wall shear stress has been shown in
Figs. \ref{jam_shear4.1-4.5.9}(e-g) for uniform/non-uniform
channels. It may be noted that in all types of channels studied here,
values of $\tau_{max}$ enhance with an increase in the value of the
fluid index n. Moreover, one may observe that the shear stress
difference between the outlet and the inlet in the case of converging
channel is exceedingly large in contrast to the case of a diverging
channel. Figs. \ref{jam_shear4.1-4.5.9}(h-j) indicate that as the time
averaged flow rate increases, the wall shear stress decreases. This is
so for all the cases examined here. It is also to be noted that the
observation regarding the shear stress difference between the outlet
and the inlet for $Q>0$ is similar to that in case $Q=0$
discussed earlier.
\begin{figure}
\centering
\includegraphics[width=3.5in]{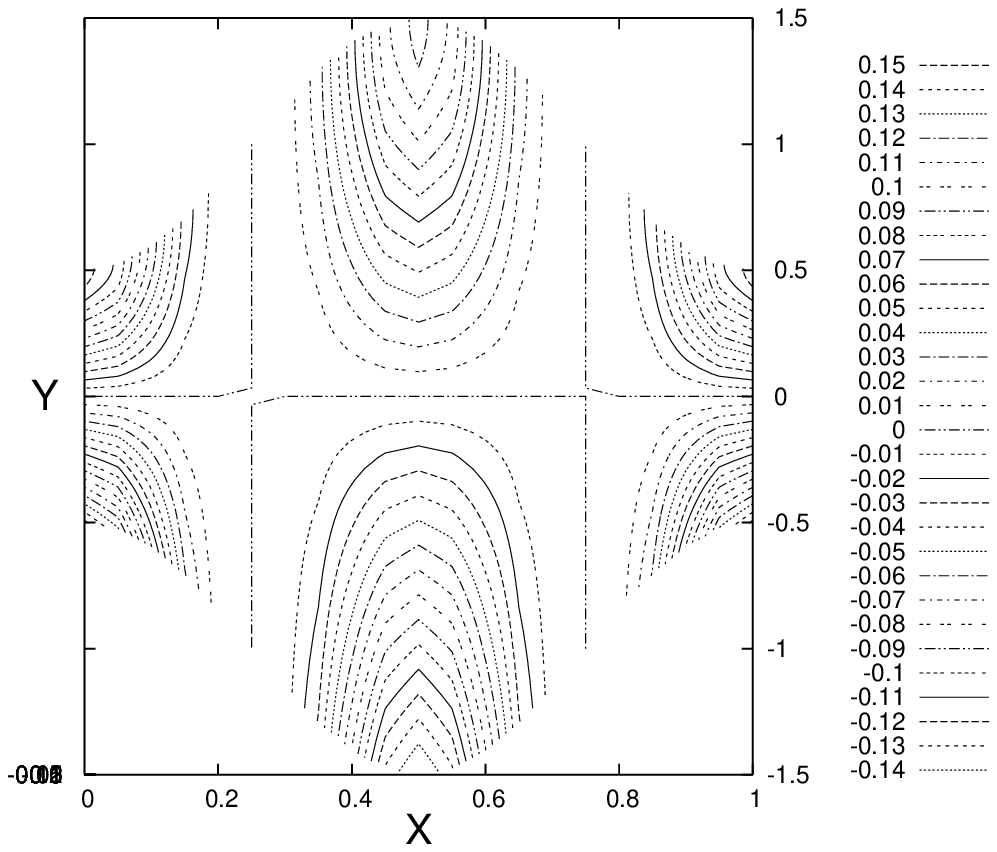}\includegraphics[width=3.5in]{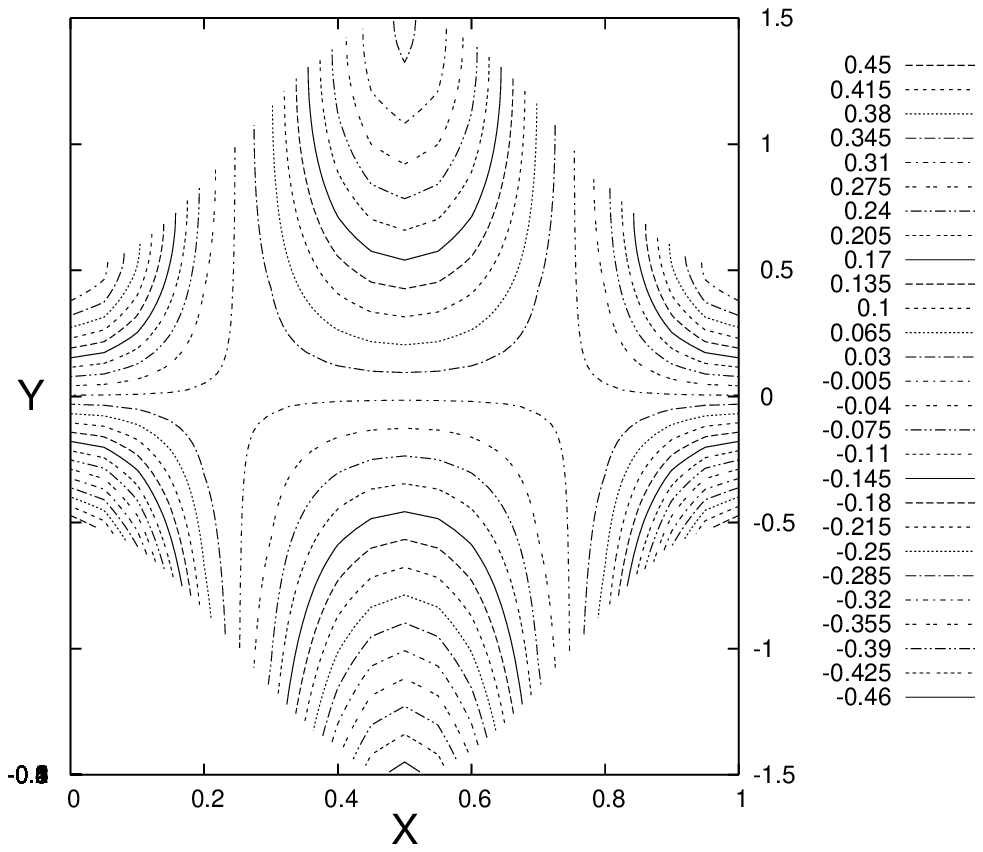}
\\$~~~~~~~~~~~~~~~~~~~~(a)~~~n=1/3~~~~~~~~~~~~~~~~~~~~~~~~~~~~~~~~~~~~~~~~~~~~~(b)~~~n=2/3~~~~~~~~~~~~~~~~~~~~~~~~~~~~~~~~$\\
\includegraphics[width=3.5in]{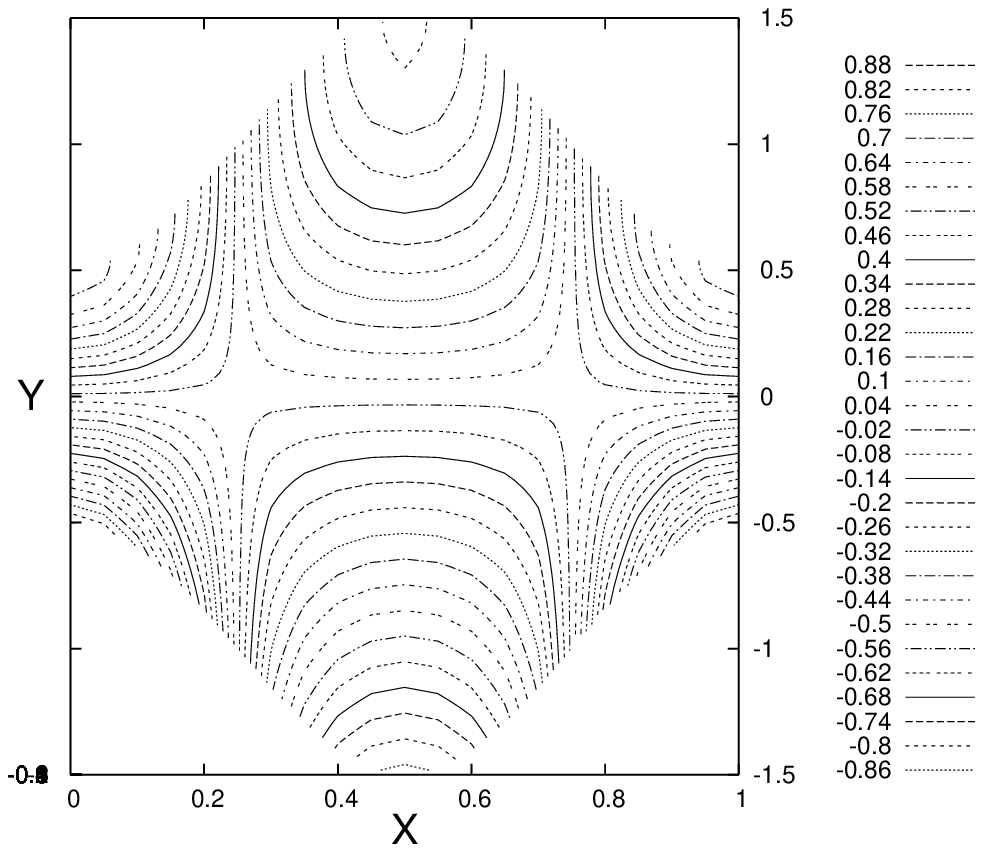}\includegraphics[width=3.5in]{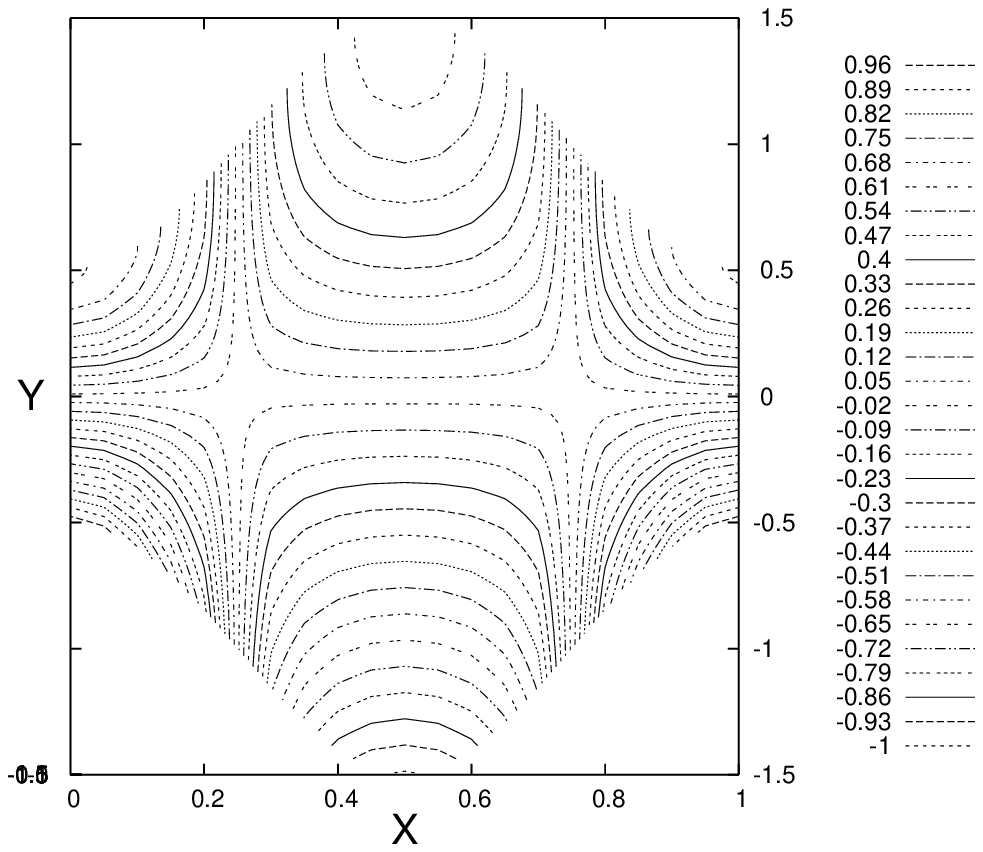}
\\$~~~~~~~~~~~~~~~~~~~~(c)~~~n=4/3~~~~~~~~~~~~~~~~~~~~~~~~~~~~~~~~~~~~~~~~~~~~~(d)~~~n=5/3~~~~~~~~~~~~~~~~~~~~~~~~~~~~~~~~$\\
\includegraphics[width=3.5in]{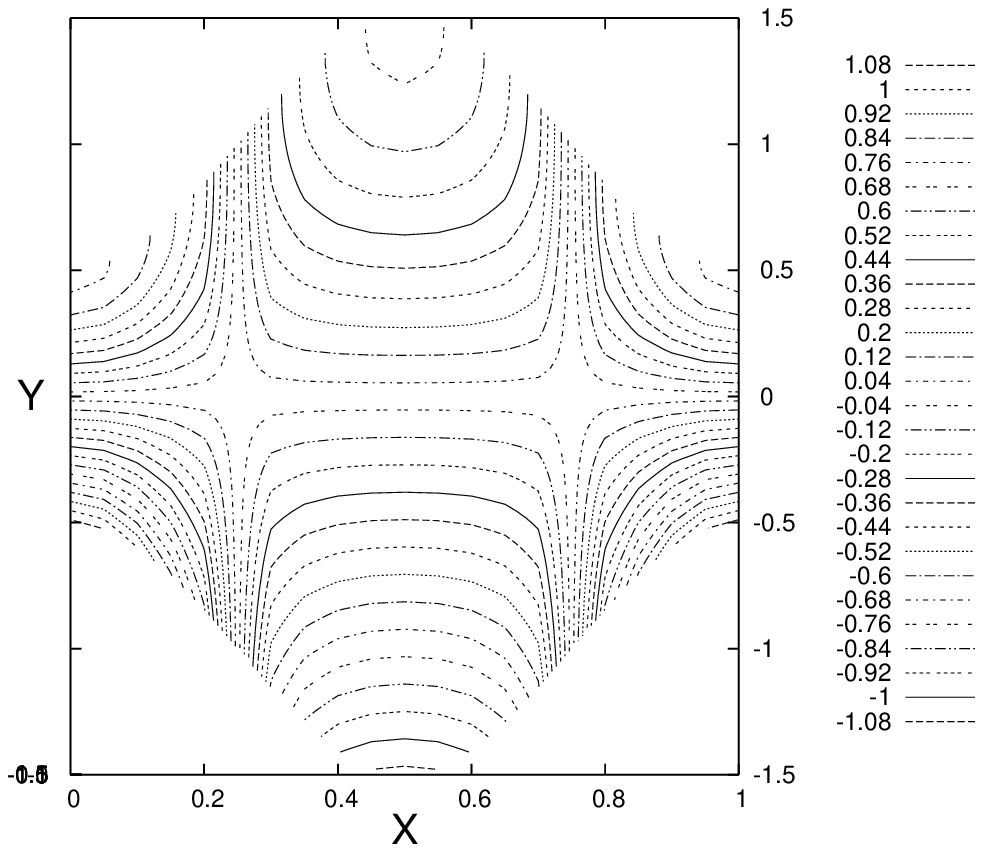}
\\(e)~~~$n=2$
\caption{Streamline patterns in the case of peristaltic flow for
  different values of n when $Q$=0, t=0.25,
  $\tau=0.1,~\Lambda=0,~\phi=0.5$}
\label{jam_stline4.4.1-4.4.6}
\end{figure}

\subsection{Streamlines}
\begin{figure}
\includegraphics[width=3.6in]{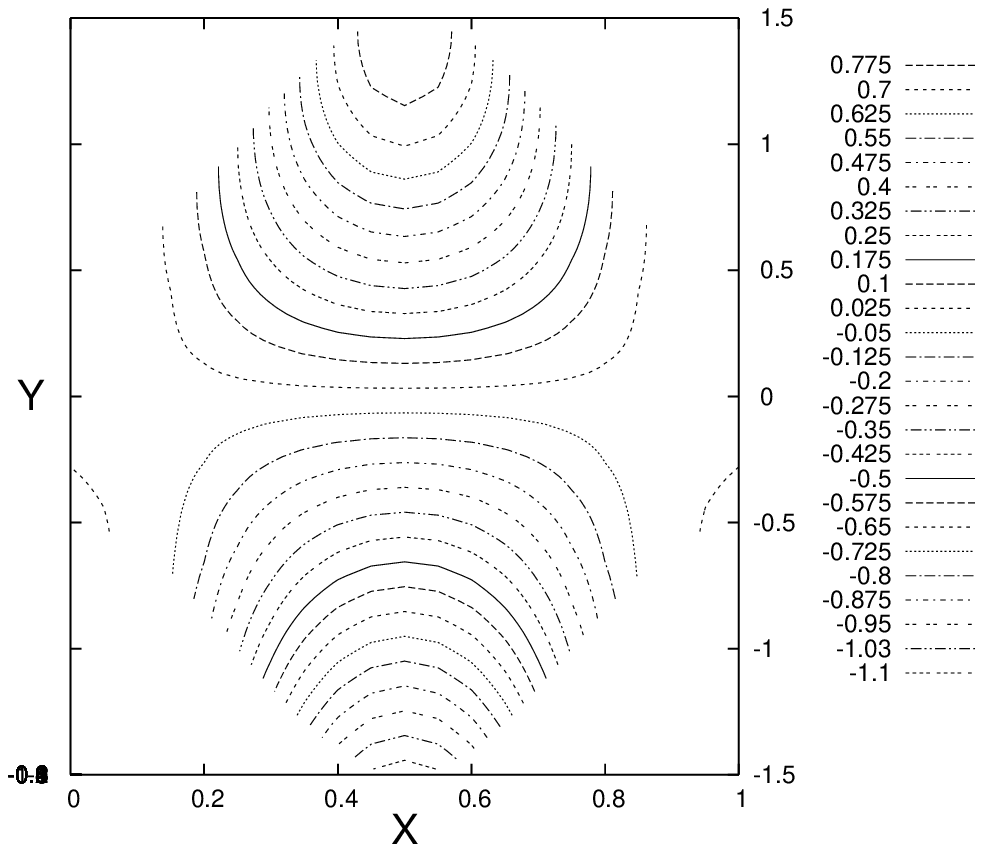}\includegraphics[width=3.6in]{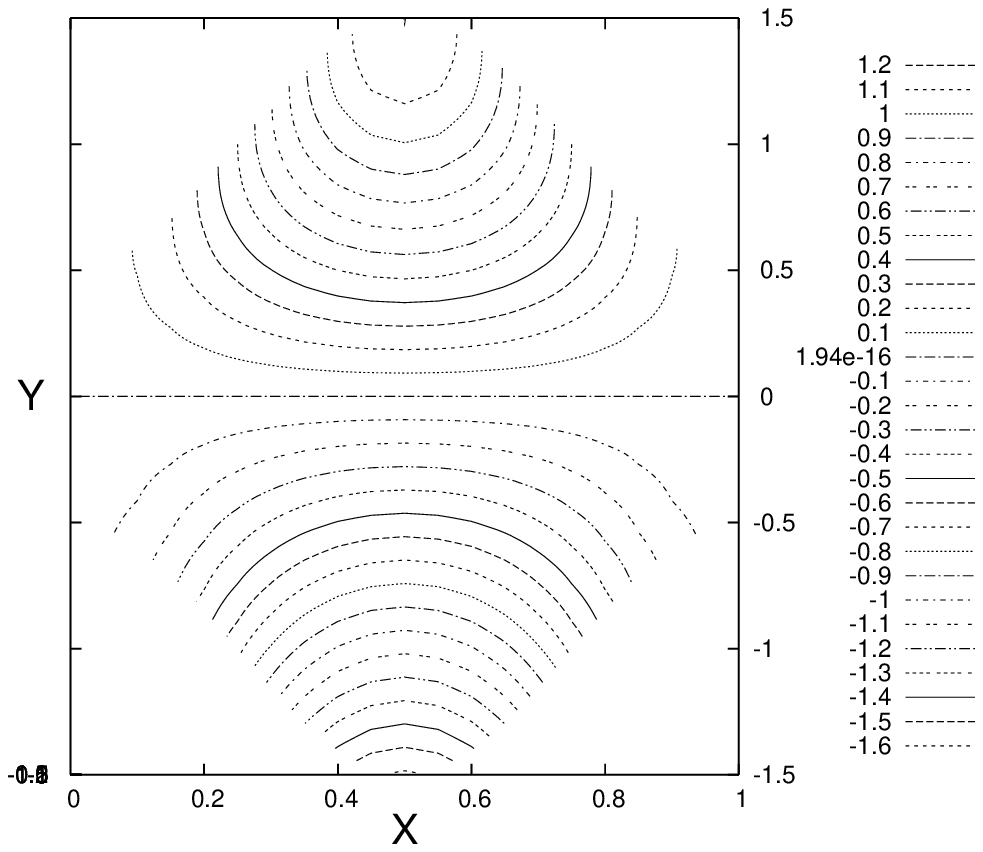}
\\$~~~~~~~~~~~~~~~~~~~~(a)~~\rm{for}~Q=0.25~~~~~~~~~~~~~~~~~~~~~~~~~~~~~~~~~~~~~~~~~~~~~(b)~~\rm{for}~Q=0.5~~~~~~~~~~~~~~~~~~~~~~~~~~~~~~~~$\\
\includegraphics[width=3.6in]{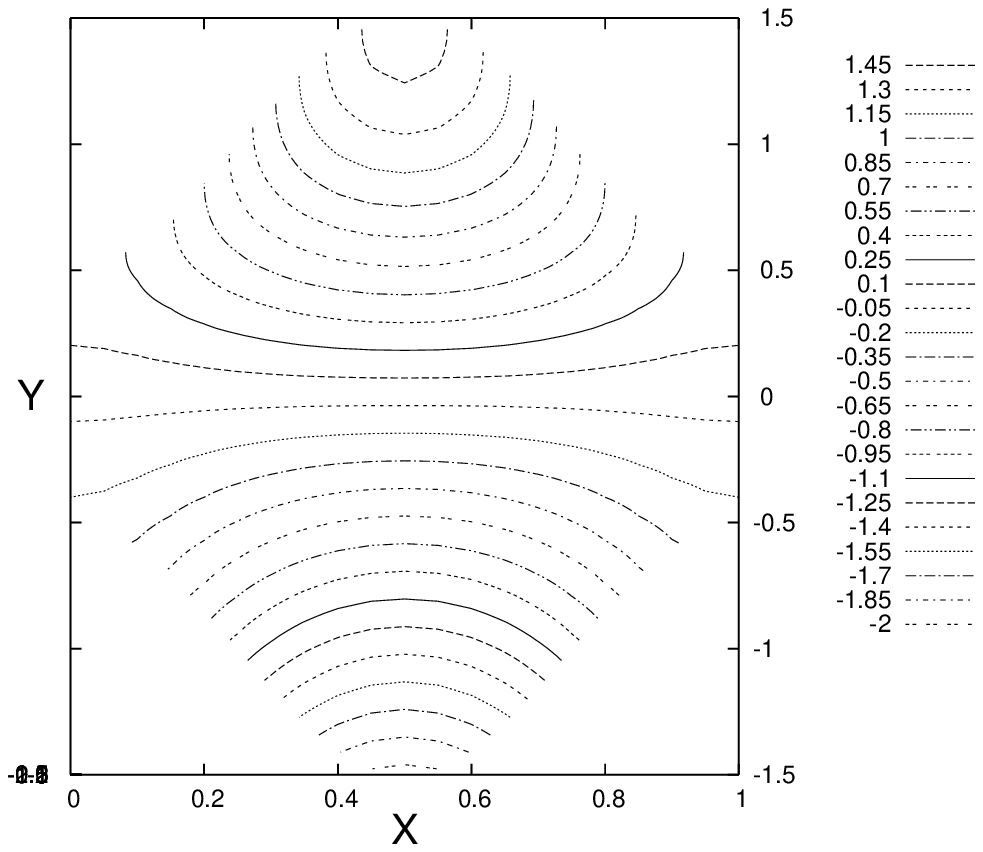}\includegraphics[width=3.6in]{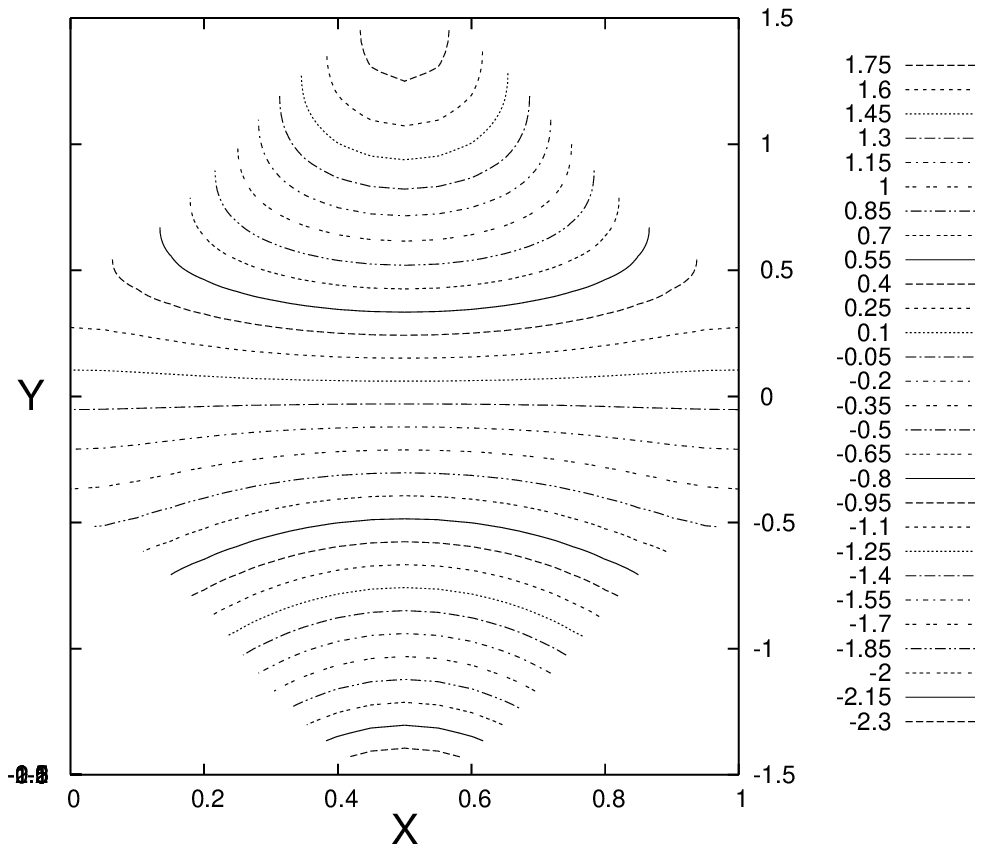}
\\$~~~~~~~~~~~~~~~~~~~~(c)~~\rm{for}~Q=0.75~~~~~~~~~~~~~~~~~~~~~~~~~~~~~~~~~~~~~~~~~~~~~(d)~~\rm{for}~Q=1.0~~~~~~~~~~~~~~~~~~~~~~~~~~~~~~~~$\\
\caption{Streamline patterns for the peristaltic flow of a shear thinning fluid (n=2/3) for different values of $Q$ when t=0.25, $\tau=0.1,~\Lambda=0,~\phi=0.5$}
\label{jam_stline4.5.17-4.5.20}
\end{figure}

\begin{figure}
\includegraphics[width=3.6in]{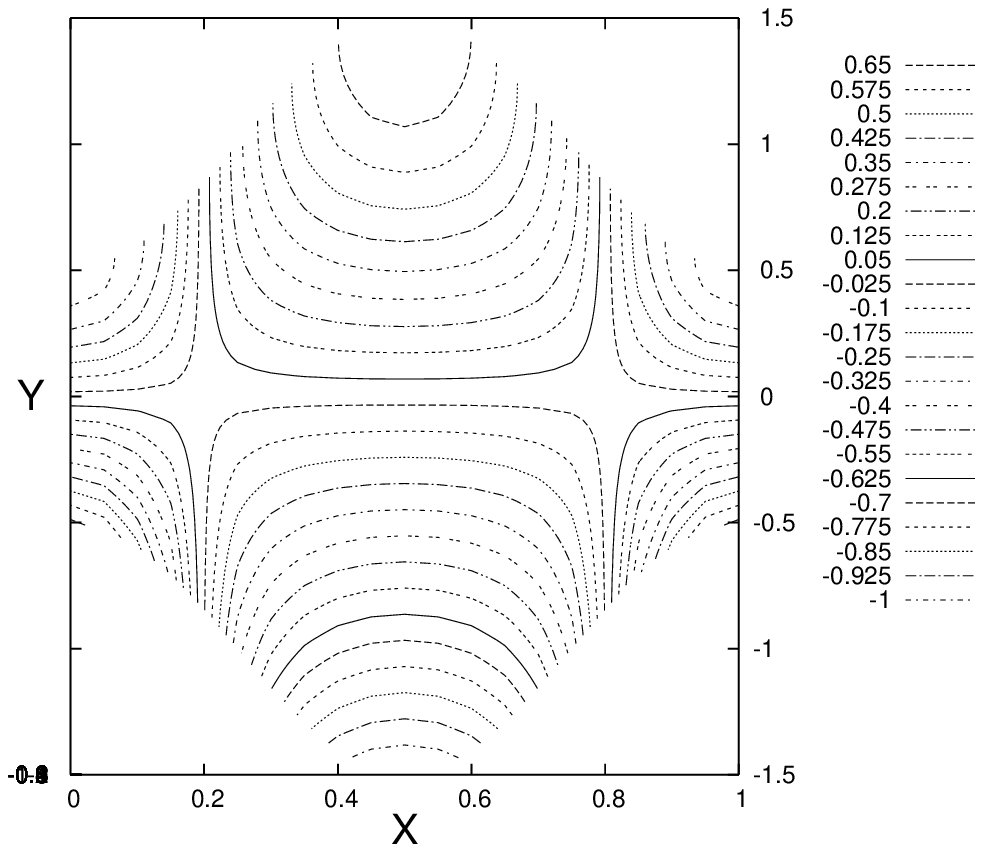}\includegraphics[width=3.6in]{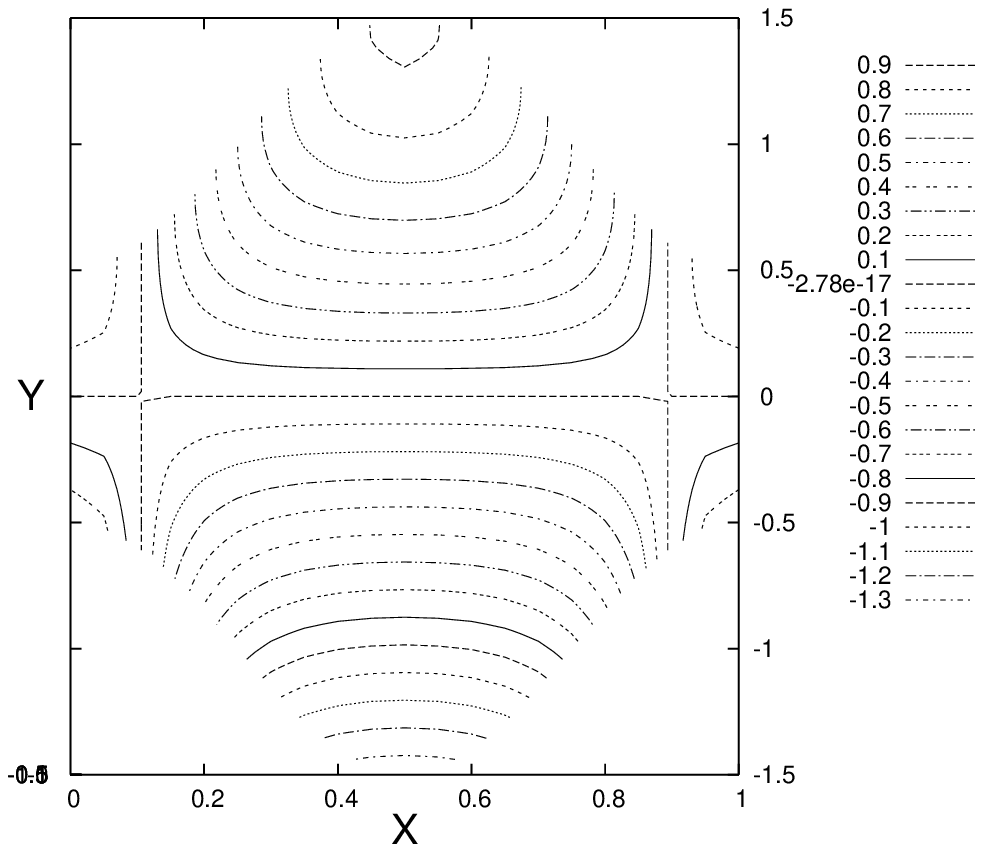}
\\$~~~~~~~~~~~~~~~~~~~~(a)~~\rm{for}~Q=0.25~~~~~~~~~~~~~~~~~~~~~~~~~~~~~~~~~~~~~~~~~~~~~(b)~~\rm{for}~Q=0.5~~~~~~~~~~~~~~~~~~~~~~~~~~~~~~~~$\\
\includegraphics[width=3.6in]{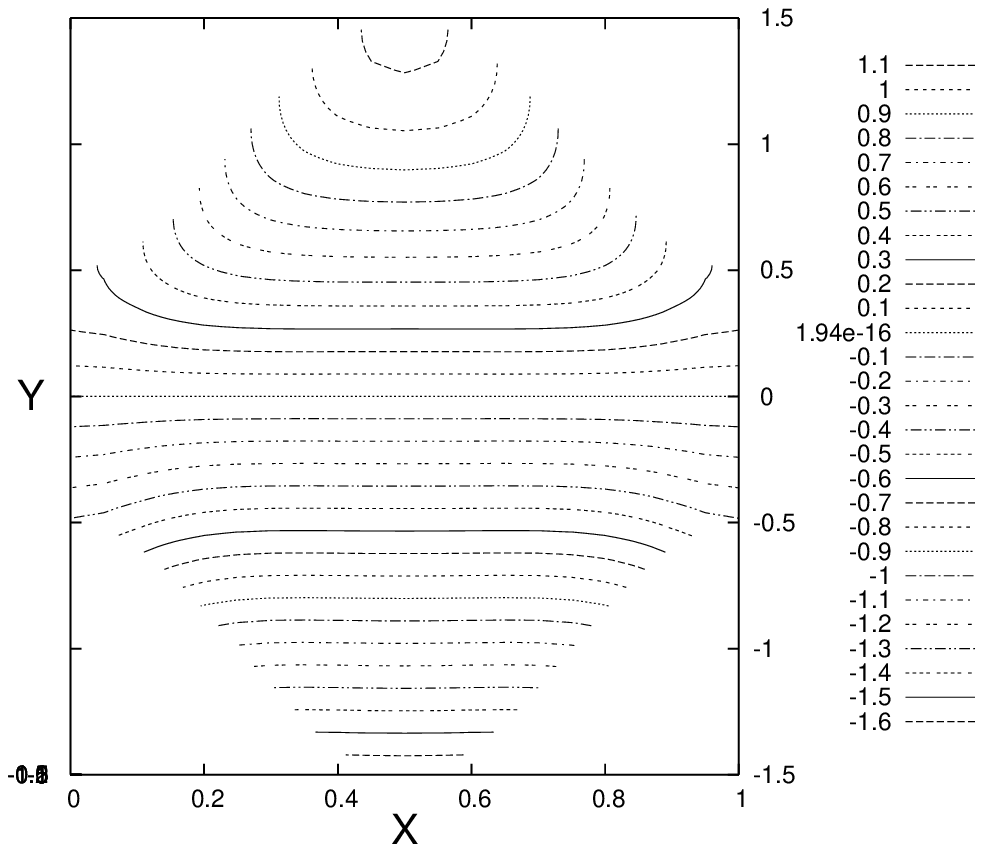}\includegraphics[width=3.6in]{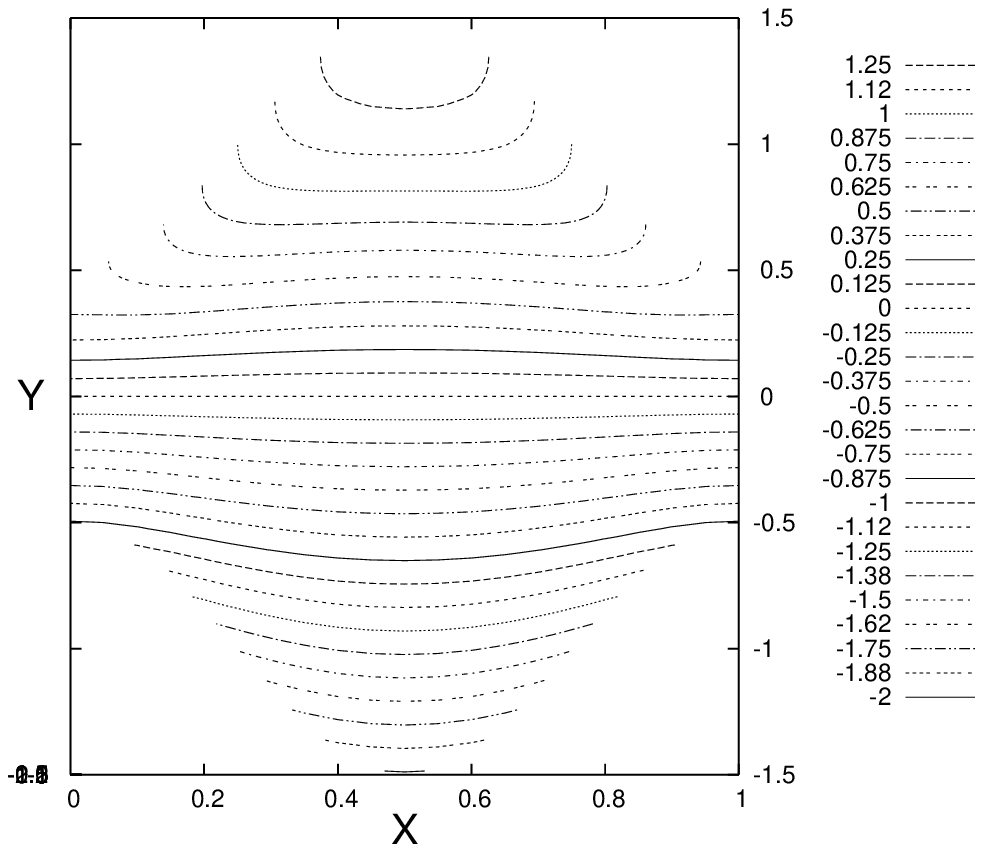}
\\$~~~~~~~~~~~~~~~~~~~~(c)~~\rm{for}~Q=0.75~~~~~~~~~~~~~~~~~~~~~~~~~~~~~~~~~~~~~~~~~~~~~(d)~~\rm{for}~Q=1.0~~~~~~~~~~~~~~~~~~~~~~~~~~~~~~~~$
\caption{Streamline patterns for the peristaltic flow of a shear thickening fluid (n=4/3) for different values of $Q$ when t=0.25, $\tau=0.1,~\Lambda=0,~\phi=0.5$}
\label{jam_stline4.5.21-4.5.24}
\end{figure}

Figs. \ref{jam_stline4.1.1-4.1.4}-\ref{jam_stline4.5.21-4.5.24}
gives an insight into the changes in the pattern of streamlines that
occur due to changes in the values of various parameters that
govern the flow of blood under the purview of the present
study. Streamlines for different flow governing parameters are
depicted in Figs.
\ref{jam_stline4.1.1-4.1.4}-\ref{jam_stline4.5.21-4.5.24}. These
figures indicate that in the portion of the channel where it is
dilating, the flow is pulled by the wall, where as in the contracting
portion, the flow is pushed away from the wall. It may be noted that
since the flow behaviour is unsteady in the fixed frame of reference,
at different times, streamlines are of different nature. Typical
nature of streamlines for the problem under the present consideration
at different instants of time is shown in
Figs. \ref{jam_stline4.1.1-4.1.4}.

In the wave frame of reference, the formation of an internally circulating bolus of fluid that moves along with the same speed as that of the wave is a very interesting phenomenon from the view point of fluid dynamics. This physical phenomenon is usually referred to as 'trapping'. However, in the fixed frame of reference it does not appear. It may be mentioned that investigation of the streamline patterns is quite important, particularly for some type of problems, because of the fact that the difference between the values of the stream function at any two points can be used to calculate the volumetric flow rate/flow flux through a line connecting the two points.

\subsection{Trajectory of Particles and Reflux}
In the micro-circulatory system, the nature of trajectories of fluid
particles play an important role in the functioning of arterioles. The
reason is that bio-chemical reactions that take place between blood
and the vessel constituents are effected by convective transport of
the fluid particles. Owing to the said bio-chemical reactions there
may be a fast increase in the wall amplitude. This may lead to
clogging of blood. From the plots of the velocity distribution
(cf. Fig. \ref{jam_velo4.1}) it is clear that at a given
cross-section, flow of blood takes place alternately in/opposite to
the direction of wave propagation in different phases. The functioning
of arterioles may then be affected by the confluence of the convective
transport in the longitudinal direction. By investigating the
pathlines of massless particles moving in the direction opposite to
that of the peristaltic wave propagation in the Lagrangian frame of
reference, it is possible to have an insight into the reflux
phenomenon.

By resorting to appropriate numerical methods for solving the
simultaneous differential equations
\begin{eqnarray}
\frac{dX}{dt}=U,~~\frac{dY}{dt}=V
\end{eqnarray}
successively starting from the initial location of the particles, it has been possible to determine the trajectories of the particles. In the above equations, (X,Y) are the non-dimensional coordinates of a particle at time t.

\begin{figure}
\includegraphics[width=3.6in]{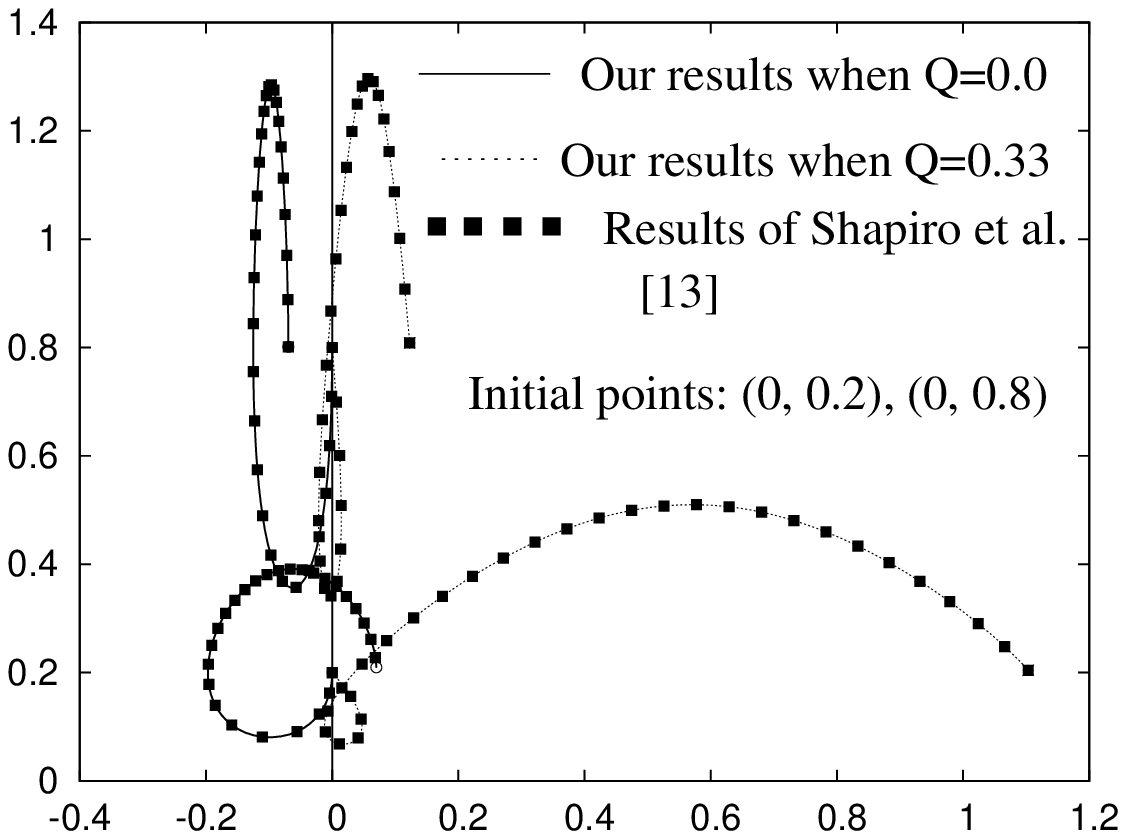}\includegraphics[width=3.6in]{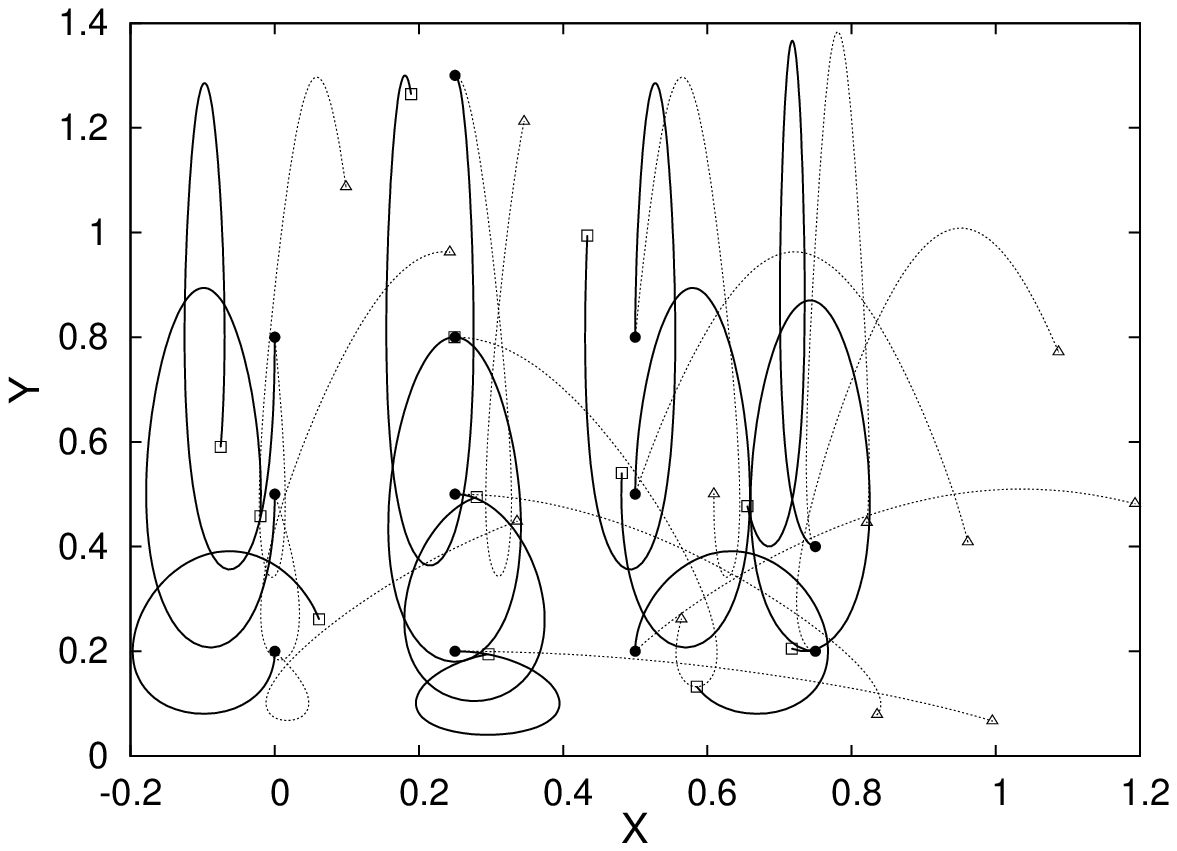}
\\$~~~~~~~~~~~~~~~~~~~~(a)~~\rm{for}~n=1~:~\rm{comparison}~\rm{with}~[13]~~~~~~~~~~~~~~~~~~~~~(b)~~\rm{for}~n=1~~~~~~~~~~~~~~~~~~~~~~~~~~~~~~~~~$\\
\includegraphics[width=3.6in]{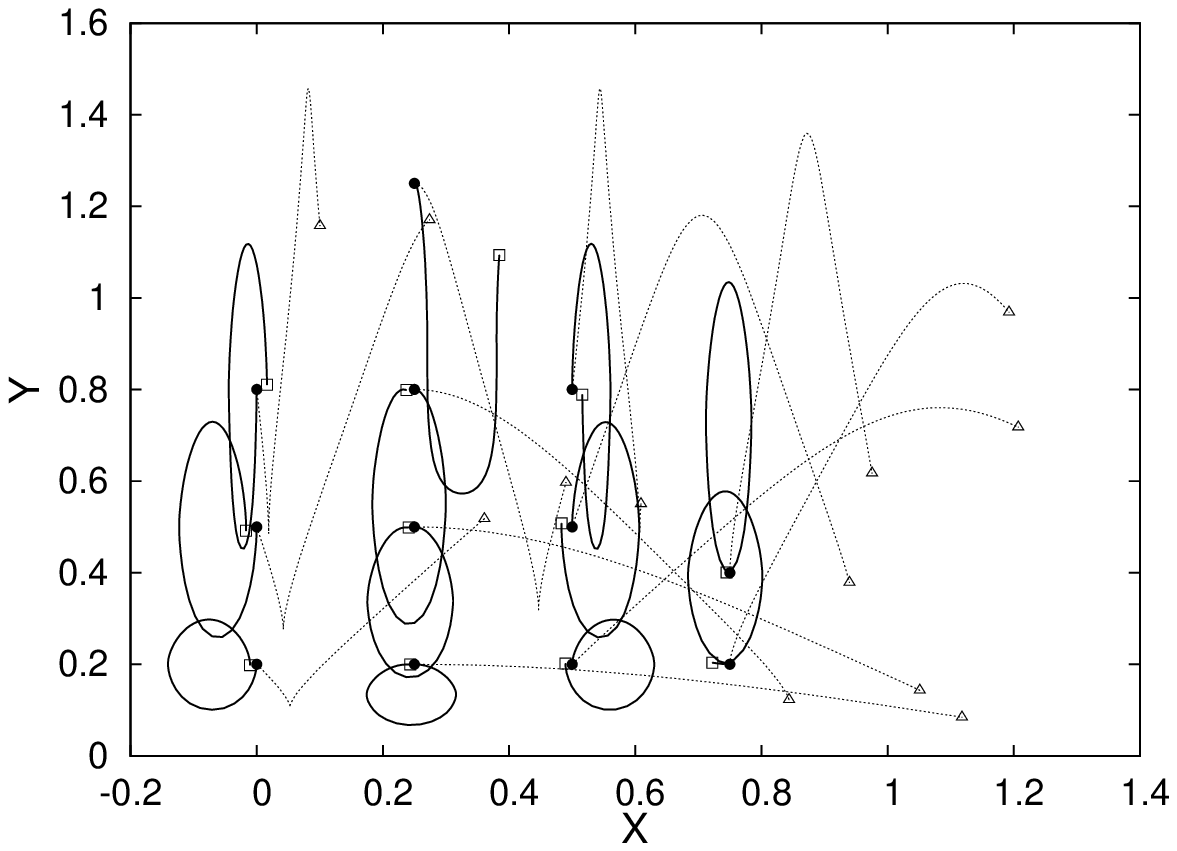}\includegraphics[width=3.6in]{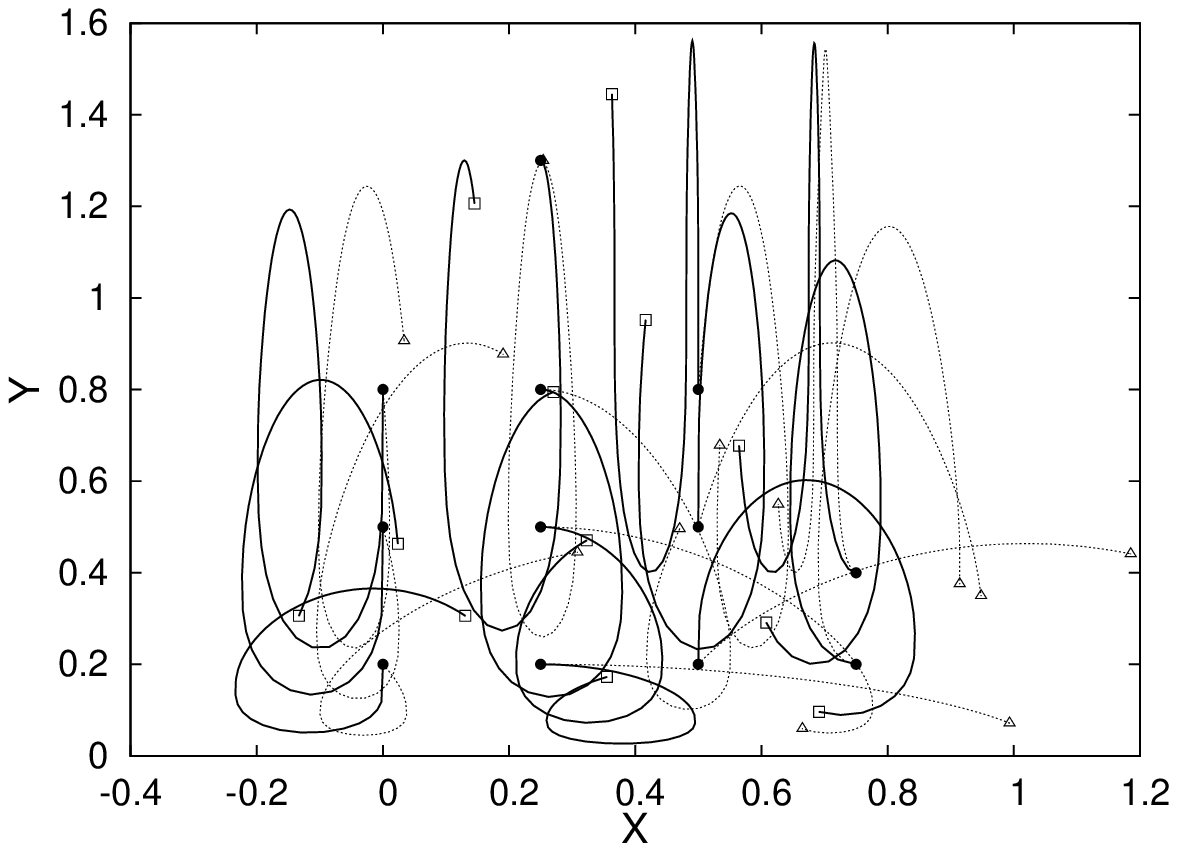}
\\$~~~~~~~~~~~~~~~~~~~~(c)~~\rm{for}~n=2/3~~~~~~~~~~~~~~~~~~~~~~~~~~~~~~~~~~~~~~~~~~~~~(d)~~\rm{for}~n=4/3~~~~~~~~~~~~~~~~~~~~~~~~~~~~~~~~$\\
\caption{Trajectories of massless particles for
  Newtonian/non-Newtonian fluids at different locations
  ($Q=0,~\phi=0.5,~\tau=0,~\Lambda=0$); $\line(1,0){20}$ for $Q$=0;
  $\cdots$ for $Q=3\phi^2/(2+\phi^2)$; $\bullet$
  initial locations;  $\Box$ at the end of one wave period if $Q=0$;
  $\triangle$ at the end of one wave period if $Q=3\phi^2/(2+\phi^2)$;
  $\circ$ at the end of one particle period if $Q$=0; $*$ at the end
  of one particle period if $ Q=3\phi^2/(2+\phi^2)$. In Fig. (a)
$\blacksquare$ stand for the corresponding results reported by Shapiro
et al. [13]}
\label{jam_traj4.1.1-4.1.4}
\end{figure}

One of the novel features of this study is the examination of the
particle trajectories for a non-Newtonian fluid. To the best of our
knowledge and observation, no previous investigator has ever studied
the trajectories for a non-Newtonian fluid. It is very important to
observe that the convergence of the pathlines is very sensitive even
to a moderate deviation from n=1. Integration of the above-written
differential equations has been carried out by applying the
Runge-Kutta $4$ method. Figs. \ref{jam_traj4.1.1-4.1.4}(a-b) present
the trajectories for the Newtonian case, while
Figs. \ref{jam_traj4.1.1-4.1.4}(c-d) show the trajectories for the
non-Newtonian case. The trajectories displayed in
Fig. \ref{jam_traj4.1.1-4.1.4}(a) are exactly similar to those
reported in \cite{Shapiro}. It is worthwhile to observe that the
period of the particle is different from the wave period and that when
$Q=0$, particles in the vicinity of the axis of the channel in
most cases undergo a net positive displacement, whereas displacement
of particles in the neighbourhood of the boundary is negative. These
observations tally with those reported in
\cite{Shapiro,Takabatake}. However, at the point (0.75,0.2) where the
velocity profile of the particle at x=0.75 and time t=0.0 is minimum,
the particle experiences a negative axial displacement
(cf. Fig. \ref{jam_traj4.1.1-4.1.4}(b)). Previous authors
(\cite{Takabatake}) also reported that even for Reynolds number 10, at
certain points near the axis , the longitudinal displacement of the
fluid particle can be negative.  Fig. \ref{jam_traj4.1.1-4.1.4}(c)
shows that the particle period is nearly equal to the peristaltic wave
period for a shear thinning fluid with $n=2/3$. A comparison of
Fig. \ref{jam_traj4.1.1-4.1.4}(d) with
Fig. \ref{jam_traj4.1.1-4.1.4}(c) and
Fig. \ref{jam_traj4.1.1-4.1.4}(b) reveals that within one wave period,
in the case of a shear thickening fluid, a fluid particle traverses
more distance than in the case of a shear thinning/Newtonian fluid.

\section{Validation of the Results}
The last paragaph of Sec. 2 serves a validation of the analytical
expressions derived in the present analysis. In order to validate the
numerical results of the present study, we have compared our numerical
results with those reported earlier by Takabatake and Ayukawa
\cite{Takabatake} on the basis of their numerical study of
two-dimensional peristaltic flows. The comparison given in
Fig. \ref{jam_pump4.1.2.1-4.3.2.4}(a) shows that our results are in
good agreement with those given in \cite{Takabatake}. More
particularly, one may observe from Table 1 that our results presented
in Figs. \ref{jam_velocompare}(b,c) for the axial velocity of blood in
peristaltic movement match well with the corresponding results
reported in \cite{Takabatake}. Also, a comparison between
Figs. \ref{jam_velocon4.1.1-4.1.4}(b) and
\ref{jam_velocon4.1.1-4.1.4}(e) shows that our results for the
velocity contour are similar to those presented earlier by Selvarajan
et al. \cite{Selvarajan}. Moreover, as shown in
Figs. \ref{jam_velocompare}(a) the axial velocity variation in our
case is in perfect matching with that reported by Shapiro et
al. \cite{Shapiro}. Lastly, our results presented in
Fig. \ref{jam_traj4.1.1-4.1.4}(a) are in excellent agreement with
those reported in \cite{Shapiro}.

\section{Summary and Conclusion}
In this study, an attempt has been made to investigate the peristaltic
motion of blood in the micro-circulatory system, by taking into
account the non-uniform geometry of the arterioles and
venules. Treating blood as a Herschel-Bulkley fluid, the effect of
amplitude ratio, mean pressure gradient, yield stress and the
rheological fluid index n on distribution of the velocity and wall
shear stress, pumping phenomena, streamline patterns and pathlines are
examined under the purview of the lubrication theory. The salient
observations are as follows:

(i) At any instant of time, there is a retrograde flow region for both
Newtonian and non-Newtonian fluids, when $\Delta P=0$ and also for
some negative values of $\Delta P$.

(ii) With an increase in the value of n and $\phi$, the regions of
forward/retrograde flow advance at a faster rate.

(iii) The parabolic nature of the velocity profiles are significantly
affected by the value of the rheological fluid index `n'.

(iv) In case of a shear thinning fluid, flow reversal does not occur
in the case of a channel having uniform geometry; it transforms to forward
flow in the case of a diverging channel. For a shear thickening fluid,
flow reversal reduces, but it does not vanish altogether when $\Delta
P$ changes from 0 to $-1$.

(v) Non-uniform geometry affects quite significantly the distribution of
velocity and wall shear stress as well as the pumping
phenomena and other flow characteristics.

(vi) Peristaltic pumping characteristics as well as the distribution of
velocity and wall shear tress are strongly influenced by the amplitude
ratio $\phi$ and the rheological fluid index `n'.

The study bears the potential of significant application to the field
of biomedical engineering and technology, because it is known that in
roller pumps, the fluid elements are quite prone to significant damage
of fluid elements; also, in the process of transportation of fluids in
living organisms, by using arthro-pumps, the fluid particles are
likely to be appreciably damaged. Moreover, the qualitative and
quantitative aspects of the present investigation for the wall shear
stress have a significant bearing on extra-corporeal circulation,
where the heart-lung machine is usually used. In this case, there is a
possibility that the erythrocytes of blood may get damaged.

{\bf Acknowledgment:} {\it The authors wish to express their deep
  sense of gratitude to all the reviewers for their learned comments
  and suggestions based upon which the manuscript has been thoroughly
  revised. One of the authors (S. Maiti) is thankful to the Council of
  Scientific and Industrial Research (CSIR), New Delhi for their
  financial support towards this study.}

\end{document}